\documentclass[fleqn,usenatbib]{mnras}

\usepackage{newtxtext,newtxmath}
\usepackage{natbib} % For citation commands
\usepackage[version=4]{mhchem}
\usepackage{booktabs}
\usepackage{array}
\usepackage{float}
\usepackage[T1]{fontenc}

\DeclareRobustCommand{\VAN}[3]{#2}
\let\VANthebibliography\thebibliography
\def\thebibliography{\DeclareRobustCommand{\VAN}[3]{##3}\VANthebibliography}

\usepackage{graphicx}	% Including figure files
\usepackage{amsmath}	% Advanced maths commands

\title[Ecological modelling of hycean worlds]{Ecological modelling of hycean worlds}

\author[G. J. Cooke et al.]{
Gregory J. Cooke,$^{1}$\thanks{E-mail: gjc53@cam.ac.uk}
Nikku Madhusudhan,$^{1}$\thanks{E-mail: nmadhu@ast.cam.ac.uk}
Emily G. Mitchell$^{2,3}$
\\
$^{1}$Institute of Astronomy, University of Cambridge, Cambridge, CB3 0HA, UK.\\
$^{2}$Department of Zoology, University of Cambridge, Cambridge, CB2 3EJ, UK.\\
$^{3}$University Museum of Zoology, Cambridge, CB2 3EJ.
}

\date{Accepted XXX. Received YYY; in original form ZZZ}

\pubyear{\the\year{}}

\begin{document}
\label{firstpage}
\pagerange{\pageref{firstpage}--\pageref{lastpage}}
\maketitle

\begin{abstract}
New observations are opening the possibility of characterising habitable environments in exoplanetary systems, with the recent example of the candidate hycean world K2-18~b. This motivates an exploration of the possible ecological conditions on such planets to better interpret biosignatures as well as understand the nature of potential life. On Earth, the Lotka–Volterra equations have been used to model  numerous coupled populations within ecosystems, from interactions between large vertebrates, to systems with multiple microbial species. In this work, we apply the Lotka–Volterra equations to the ecology of habitable exoplanets for the first time, focusing on hycean worlds. We simulate scenarios in a vertical water column with between 1--5 bacterial species that thrive in anoxic environments on Earth, i.e. similar to predicted hycean conditions. We find that a wide range of ecological diversity is possible for microbial populations under hycean conditions. We demonstrate that dominating phototrophic bacteria at the top of a water column out-compete deeper dwelling phototrophic bacteria, analogous to bacterial blooms on Earth. Incorporating microbial viruses (bacteriophages) within our models can cause ecosystem collapse depending on the time of their introduction, and such phage inclusion can be beneficial to ecological diversity. Finally, our work shows that bacterial populations inhabiting tidally locked exoplanets may be more stable due to constant illumination of the ocean, but can have lower peak population densities in such cases when compared to seasonal scenarios. Our work provides an initial step towards understanding the possible ecological diversity on habitable worlds beyond Earth. 
\end{abstract}


%%%%%%%%%%%%%%%%%%%%%%%%%%%%%%%%%%%%%%%%%%%%%%%%%%

%%%%%%%%%%%%%%%%% BODY OF PAPER %%%%%%%%%%%%%%%%%%

\section{Introduction}
\label{Introduction section}
 
Over $\sim$6000 exoplanets have been discovered to date over a wide range of planetary and host stellar properties\footnote{The \href{https://exoplanetarchive.ipac.caltech.edu/}{NASA Exoplanet Archive} lists over 6000 confirmed exoplanets.}. The diversity of these exoplanets is astounding, from Earth-sized planets around the smallest of stars, to a previously unknown class known as sub-Neptunes, and hot-Jupiters scorched by intense irradiation. Several exoplanets are considered potentially habitable because they orbit within the habitable zone of their host star \citep[the region around the star where a planet can orbit and maintain liquid water on its surface;][]{Kasting_HZ_1993, Kopparapu_HZ_2013, Madhu2021Hycean}. A primary goal of astrobiology is to determine whether any of these exoplanets do host life, which has now become potentially tangible with the arrival of the James Webb Space Telescope (JWST), albeit not through the classically proposed route.

Much of the efforts in astrobiology regarding the context of exoplanets have been applied to the habitability of a variety of exoplanets \citep{2011AsBio..11.1041S, Armstrong2014ObliqHabit, 2018ApJ...864...75K, Madhu2021Hycean}. The search for life beyond Earth has also been investigated through simulated electromagnetic spectra of habitable and inhabited planets \citep{Kaltenegger2010Spectra, Cooke2023Degenerate, Coelho2024Purple}, especially with regards to gaseous biosignatures \citep{2019AsBio..19.1075C} and more speculatively, technosignatures \citep{2022ApJ...927L..30W, 2023ApJ...943...27L}. The evolution of life \citep{2019IJAsB..18..204S, 2025MNRAS.tmp...96M}, niche habitats, and self-sustaining non-traditional habitats \citep{2024AsBio..24.1187W}, have also been investigated, in addition to the Gaia hypothesis, where life self-regulates the planet's climate \citep{2018MNRAS.477..727N}. On the practical side, laboratory studies of microbes under possible exoplanetary conditions have been performed \citep{Seager2020Lab, Abrevaya2024, Kuzucan2025Ecoli, Abrevaya2025_EXO_UV}.

Historically, terrestrial (rocky) exoplanets have been extensively studied because of their similarities with Earth. They will eventually offer opportunities to determine whether they host life in a variety of planetary systems, but a thorough characterization of existing rocky habitable zone planets is currently impractical with JWST because of the required photometric precision. A further complication is that terrestrial M dwarf exoplanets may not have atmospheres due to atmospheric erosion in the early active phase of a M dwarf's lifetime \citep{Zahnle_Cosmic_2017, Modi_2023_Impact, 2025arXiv250419872J}. Preliminary observational and modelling evidence from the `Hot Rocks Survey' \citep{Valdes_2025_HotRocks, Allen_2025_HotRocks, fortune2025hot} and the TRAPPIST-1 system \citep{Ducrot_2025_TP1, Gillon_2025_TP1} show that several of these close in rocky exoplanets are consistent with airless worlds. Revealing the atmospheres of Earth-like exoplanets remains a future goal that will likely require next-generation observatories such as the Extremely Large Telescope \citep{Cirasuolo_ELT_2024} and the Habitable Worlds Observatory \citep{Feinberg_HWO_2024}. 

In recent years, sub-Neptunes (with radii between 1.5--4 $R_\oplus$) have been at the forefront of exoplanetary research, both observationally and theoretically \citep{Espinoza_highlights_2025, Madhu_PNAS_2025_Sub_JWST}. Examples of possible sub-Neptunes include planets with rocky cores and thick hydrogen dominated envelopes (gas dwarfs), Neptune-like planets with smaller radii (mini-Neptunes), and water worlds encompassing ocean worlds with high mean-molecular weight atmospheres \citep{Leger_2004_Ocean}, steam worlds \citep{Piaulet-Ghorayeb_Steam_2024}, and hycean worlds that have hydrogen-rich atmospheres \citep{Madhu2021Hycean}. Importantly, ocean worlds and hycean worlds have been theorized to sustain a habitable global liquid water ocean \citep{leger2004new, Madhu2021Hycean, Madhu_PNAS_2025_Sub_JWST}, with the latest hycean simulations performed with a 3D global climate model \citep{Barrier_2025_Hycean}.

% Schematic system diagram figure
\begin{figure*}
	\centering
	\includegraphics[width=1\textwidth]{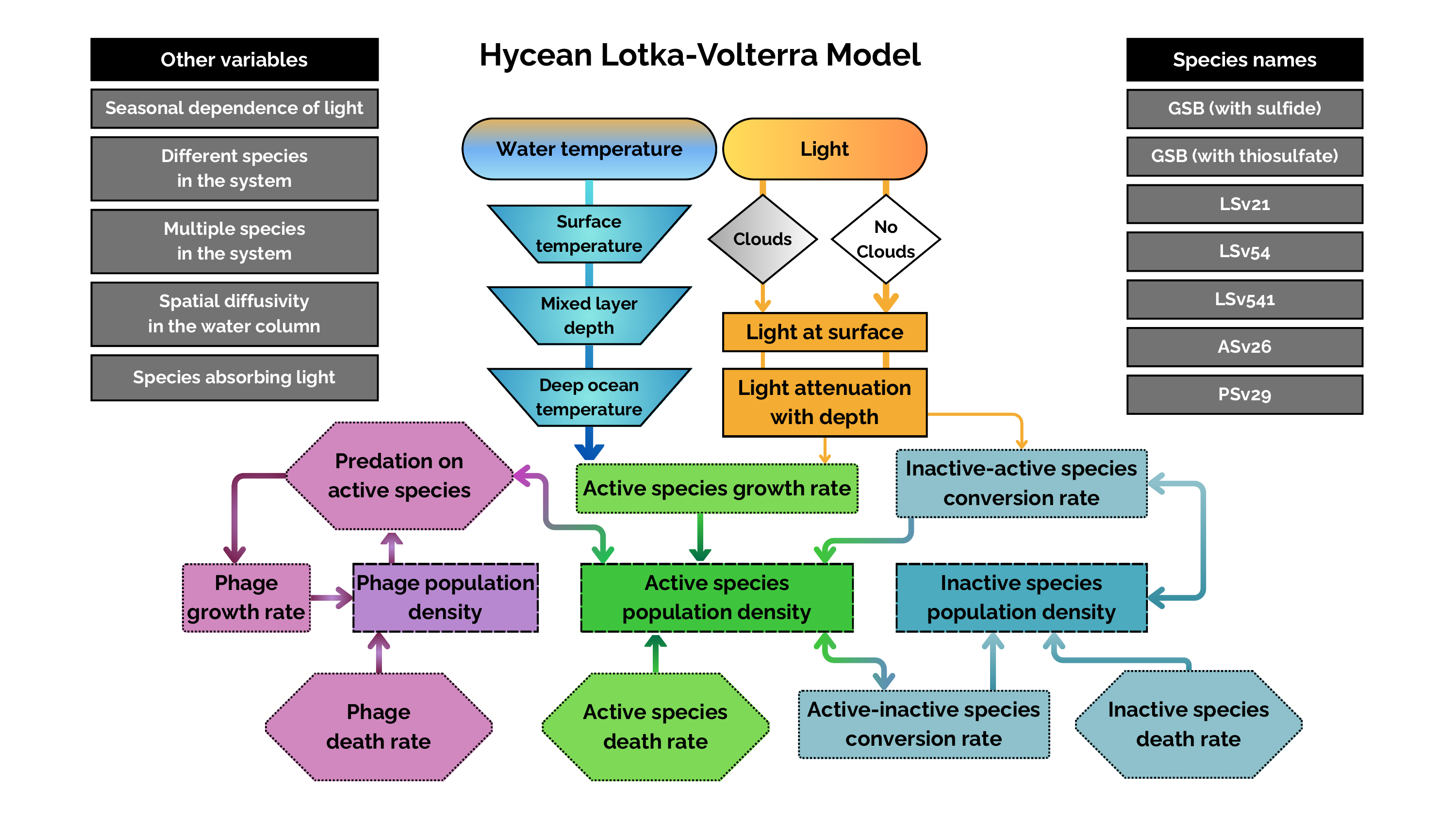}
    \caption{A schematic diagram of the hycean Lotka-Volterra model we implement in this work. The schematic shows how various parameters in our model are connected in a vertical, one-dimensional water column. Input parameters include water temperature and light. The water temperature (orange-blue oval) depends on Eq.~\eqref{Temperature profile equation}. The incident light (orange oval) can be modified by clouds (grey diamond) to reduce the relative intensity of surface light, and then the light varies with depth due to attenuation by absorption and scattering in water. Light and temperature set the growth rate for the active population, and light also influences the rate of conversion from the inactive species to the active species. The active bacteria population (green dashed rectangle), the inactive bacteria population (blue dashed rectangle), and the phage population (purple dashed rectangle) depend on various parameters which vary with depth in the water column. Growth processes for bacteria are in dotted rectangles, while death rates are in elongated hexagons (active and inactive species in green and blue, respectively, and phages in pink). Both growth and death rates act to modify the population densities. Arrows show the direction of influence, with double-sided arrows showing feedback between the parameters/processes. The species names are given by grey boxes at the top right of the schematic, with GSB standing for Green Sulfur Bacteria. At the top left of the schematic, other variables which influence our model are shown, and these variables are either turned on or off between sets of simulations, or modified.} 
    \label{Schematic system diagram figure}
\end{figure*}

In particular, hycean worlds are promising candidates in the search for life that offer a viable pathway to make significant progress with the current available technologies. Due to the H$_2$-rich atmosphere and their larger size when compared to terrestrial exoplanets, hycean worlds are conducive to atmospheric characterization and biosignature detection with JWST. Indeed, two hycean candidates, K2-18~b and TOI-270~d, have been observed by JWST and the observations are currently consistent with hycean conditions \citep{Madhu2023CarbonBearing, Madhu_DMS_2025, Holmberg2024hycean, Hu_2025_K218b}, although not conclusively so \citep{Madhu2023CarbonBearing, Madhu_DMS_2025, Benneke2024TOI270d, Hu_2025_K218b}. A biosignature candidate molecule, dimethyl sulfide, has weakly constrained posteriors (implying a tentative detection requiring more observations) in multiple atmospheric retrievals from independent JWST observations of the atmosphere of K2-18~b \citep{Madhu_2023_Faraday, Madhu_DMS_2025, Schmidt2025_K2_18b, luque2025insufficient, Hu_2025_K218b}. Biosignature molecules in hycean atmospheres have been explored in photochemical and observational simulations \citep{Madhu2021Hycean, Madhu_2023_Faraday,  Tsai2024BiogenicWaterworlds}, with a possible abiotic mechanism suggested by \cite{Hu_2025_K218b} for the production of atmospheric DMS. Note that the presence of DMS has not yet been confirmed or ruled out for K2-18~b.

Chemicals and molecules within planetary environments do not exist in isolation, so to understand and fully interpret atmospheric signatures, it is key to understand how the dynamics of chemical species are likely to coexist. Life may change these dynamics, which necessitates the use of ecological modelling techniques to elucidate the interactions between ecosystems and life in extraterrestrial settings \citep{Mendonca_2014_AstroEco, Meurer_2024_Astroecology}. Thus far, a handful of ecological models have been applied to astrobiology, which include: infection dynamics and bio-dispersal \citep{Kova_2022_Astrobio}; panspermia \citep{Ginsburg_Panspermia_2018}; various models for the dispersal of organisms in different settings \citep{Mendonca_2014_AstroEco}; biogeochemical models \citep{Schwieterman_Biosignatures_2022}; and predator-prey competition for self-replicating interstellar probes using the Lotka-Volterra (LV) equations \citep{Forgan_SRP_2019}. How life develops in novel, anoxic, and likely tidally locked environments \citep{Madhu2021Hycean, Livesey_2025_Tides}, and evolves in response to constant illumination and a lack of seasons is unknown because this situation is not present on Earth. Therefore, understanding the dynamics of these novel ecological niches with an array of organisms is key to understanding the (bio)spheres of exoplanets.

There are a huge range of different approaches for modelling ecosystem dynamics on Earth. Of these, the Lotka-Volterra predator-prey model \citep{Lotka1925, Volterra1926} has been widely used and continues to be developed \citep{2024RSPSA.48030284A, 2025NonDy.11312351G}. Studies have investigated different ecological systems on Earth, from interactions between large vertebrates \citep{kmet1994diffusive, de2012mean, bawari2024impact}, to environments with multiple microbial species \citep{xu2023proportional}. Application of these models includes both abiotic and biotic variables, such that it enables the modelling of many different biotic and abiotic interactions, beyond the original carnivorous predator and herbivore prey. It can therefore capture the dynamics of many different consumers and their resources.

In this work we introduce the LV model for hycean worlds, although our model is broadly applicable to other habitable worlds. In this study, we describe the LV model in section~\ref{Lotka-Volterra Model section} and then we detail our modification of the model in order to apply it to hycean worlds in section~\ref{Hycean Lotka-Volterra model section}. In section~\ref{Results section}, we present our model results in various scenarios under different environmental parameters, before summarising and discussing the implication of our work in section~\ref{Conclusions section}. 

\section{Lotka-Volterra Model}
\label{Lotka-Volterra Model section}

The Lotka-Volterra model comprises two first-order nonlinear differential equations that describe the dynamics of an ecological system involving two or more interacting species: with at least one acting as the consumer and at least one other as the resource. Note these species can be biological or chemical/abiotic species. The population sizes (numbers of individuals or biomass) evolve over time ($t$) according to the following equations:  

\begin{align}  
    \frac{dx}{dt} &= a x - b x y, \label{eq:preyLV}\\  
    \frac{dy}{dt} &= -c y + d x y. \label{eq:predatorLV}  
\end{align}  

\noindent Here, $x$ represents the resource population (e.g., the mass of detritus per liter), while $y$~denotes the consumer population (e.g., the mass of heterotrophic bacteria per liter). The parameter $a$ corresponds to the maximum per capita population growth rate of the resource population, and $b$ quantifies the impact of consumption on the resource's population. Similarly, $c$ represents the consumer’s per capita death rate, and $d$ characterizes the influence of resource population on the consumer population's growth rate. Because the growth of the consumer depends on the population of the resource, the simplest version of a stable model predicts cyclical growth rates for both the consumer and resource species \citep{wangersky1978lotka}, and have been observational tested in warbler populations \citep{macarthur1958population}, lynx and hare populations \citep{maclulich1937fluctuations, nedorezov2016dynamics}, microbial communities in aquatic environments \citep{gause1934experimental}, as well as experimental checks of the conditions within which an LV model can be valid \citep{dedrick2023does}. 

The dynamics of a diverse range of consumers can be incorporated into these LV equations. This diversity includes consumers which are large (e.g. classic carnivorous predators) and small (e.g. zooplankton) and their impact on resources, such as abiotic molecules like particulate organic carbon, or other life, namely bacteria or large prey (e.g. hares for lynx). The key to building suitable LV models is an understanding of what changes the growth rates ($\alpha$, $\gamma$) and death rates ($\beta$, $\delta$) of these consumers and their resources, including the shape of their distributions.  

In terms of growth rates, for phototrophs (organisms that photosynthesise, where trophic refers to nourishment, e.g., food, and photo refers to the energy source which is light), the key parameters governing organism growth are temperature and light availability, which both decrease with depth and change with ocean dynamics \citep{parkin1980effects, das2011enhanced, Overmann2013, princiotta2016temperature}. Marine phototrophs live and reproduce in the euphotic zone (also known as the photic zone), where light extends down to, by definition, 1\% of its strength at the Earth's surface \citep{lee2007euphotic}. 

Predators can modulate the abundance of grazers that are able to destroy their habitat if left unchecked \citep{griffin2011predator}, and removal of predators can destabilize ecosystems leading to extinctions \citep{o2009perturbations}.  In the Earth Oceans, the ``microbial loop''  of dissolved organic matter/carbon, bacteria and microbial predators \citep{Azam_MicrobialLoop_1983, pomeroy2007microbial}, such as protists, is crucial to the carbon cycle and ecosystem stability, illustrating that such predators also includes microbial consumers. Note that it is not known if predators are an inevitable outcome of biological evolution on other planets, so while we will model their presence, it is not a certainty, even if other life is present. 

A good analogue to hycean ecosystems can be found with Ace Lake, Antarctica \citep{lauro2011integrative}. It is a stratified lake hosting cyanobacteria and green sulfur bacteria (GSB) and so includes species present in different environmental conditions, some of which could be similar to Hycean worlds. There are three distinct zones in Ace Lake: an upper oxic layer in which wind forces mixing down to 11.5 m, followed by a transition zone characterized by the strong gradient in oxygen and salinity to around 12.7 m depth, and a lower anoxic layer \citep{rankin1999chemical}. All of these regions receive sunlight, but have different types of photosynthesizing organisms so provide multiple species of use for our study.

% Downwelling diffuse attenuation coefficient figure
\begin{figure}
	\centering
	\includegraphics[width=1\columnwidth]{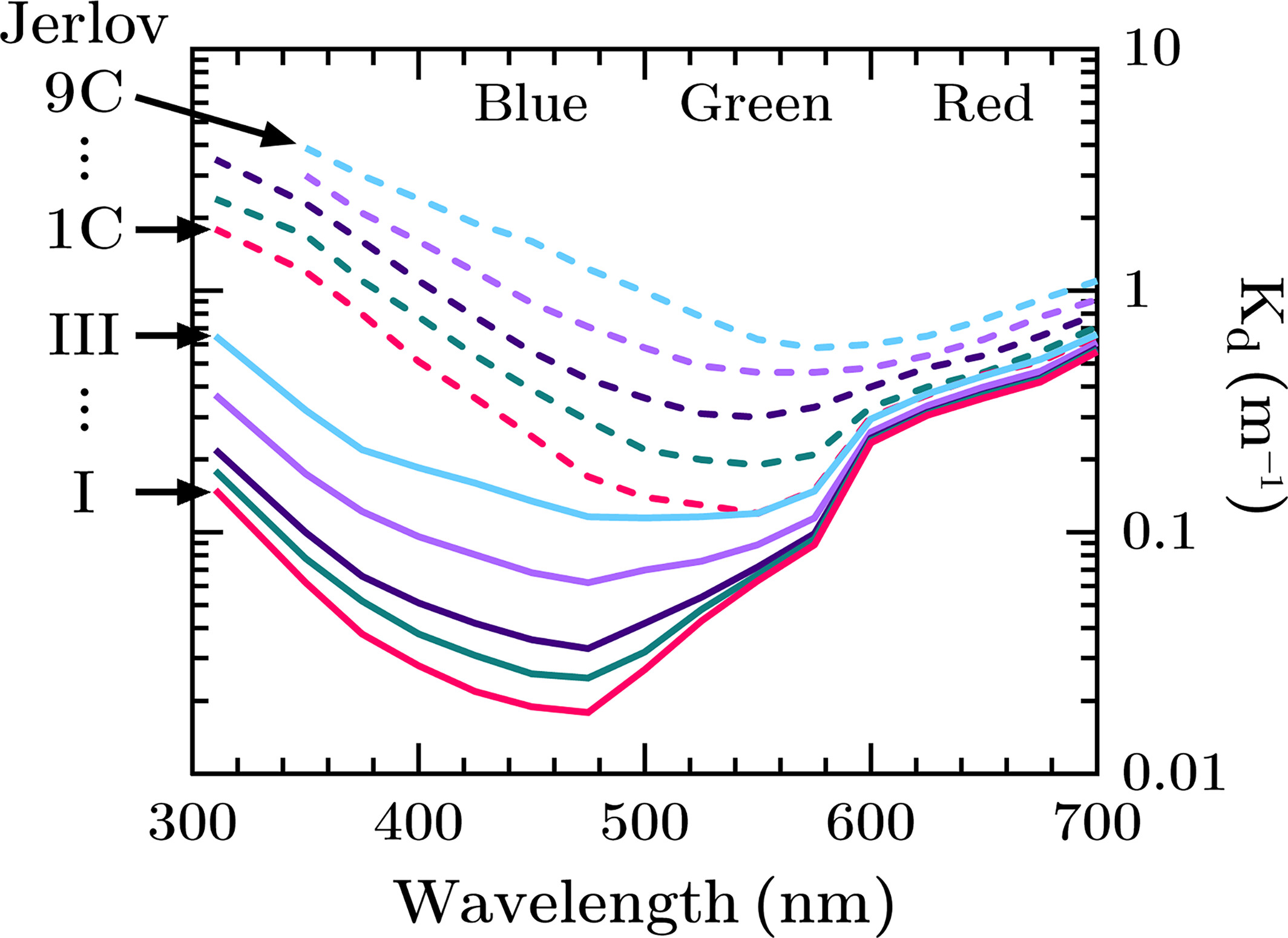}
    \caption{Jerlov water types, from open ocean (I -- III) to coastal waters (1C -- 9C). The figure shows how the downwelling diffuse attenuation coefficient, $k_d$ (m\textsuperscript{-1}) varies with wavelength, for the 10 Jerlov water types \citep{jerlov1951optical, jerlov1968optical, jerlov1976marine}. The lowest $k_d$ occurs in the visible wavelength region for all water types, with UV wavelengths and the reddest wavelengths having higher values of $k_d$. The figure is reproduced from \citet{williamson2023depth}, Limnology and Oceanography Letters, under the Creative Commons Attribution (CC BY 4.0) License. } 
    \label{Attenuation coefficient figure}
\end{figure}

In the Ace Lake model, \citet{lauro2011integrative} took microbial samples at six different depths to determine the presence of species and their densities. They modelled cyanobacteria (top 11~m of lake Ace) and green sulfur bacteria (from 12.7--14~m depth in the lake) using modified LV equations. The model of \citet{lauro2011integrative} simulated the response of both GSB and cyanobacteria to daily and seasonal light conditions, and also the introduction of phages. 

The Ace Lake model assumes that the bacteria can exist in a dormant state referred to as `inactive'. Generally speaking, bacteriophages cannot kill dormant bacteria. Dormant bacteria do not multiply, and eventually either die or undergo revival. However, predation on dormant bacteria is possible in some specific cases \citep{fischetti2008bacteriophage, butala2023unique, maffei2024phage}, or where phages prevent dormancy in the first place \citep{schwartz2022phage}. In the model, the concentrations of active bacteria cells, $A$, inactive bacteria cells, $I$, phage concentration, $P$, and total bacteria cells $G = A + I$, are tracked with a time ($t$) dependence. The availability of light affects the growth rate of the active cells and the turnover of inactive cells into active cells. The time dependent light availability, $L(t)$, is given by 

\begin{align}
    L(t) = &\begin{cases} 
    cos(2 \pi t / \omega)^\tau,& cos(2 \pi t / \omega)^\tau > 0\\
   0,& cos(2 \pi t / \omega)^\tau \leq 0, \label{Light equation}
    \end{cases}
\end{align}

\noindent where $\omega$ is the length of time within a given seasonality cycle, i.e. a period (1 year for Earth) and $\tau$ is a dimensionless modification fitted to model the seasonal changes in the flux of photosynthetically active radiation (PAR) reaching Ace Lake \citep{lauro2011integrative, powell2005ecology}. In Lake Ace, the photosynthetically active radiation can penetrate up to 12 m in the summer and only approximately 2 m in the winter \citep{rankin1999chemical}. 

The concentration of active bacteria cells grows with respect to time, such that

\begin{equation}
    \frac{dA}{dt} = \alpha A L \left ( 1-\frac{G}{C} \right ) - \beta A + \gamma I L - \delta A - \epsilon P A, \label{Eq Active GSB}
\end{equation}

\noindent where the rate of change of $A$, the active bacteria population is governed by $\alpha$, the growth rate of the active bacteria population. $C$ is the carrying capacity of the bacterial population, which limits its maximum size \citep{Gause_struggle_1936, rosenzweig1963graphical, turchin2013complex} and is related to both space and nutrient availability. \citet{lauro2011integrative} implemented a carrying capacity of $10^9$ cells ml\textsuperscript{-1}.  $\epsilon$ and $\beta$ are the death rates of the active bacteria population with and without predation due to phages, respectively. $\gamma$ is the rate at which inactive cells are converted to active cells in the presence of light, whilst $\delta$ is the rate at which active cells become inactive. The following equation describes the time dependence of the inactive population  $I$:

\begin{equation}
    \frac{dI}{dt} = -\gamma I L + \delta A - \eta I,  \label{Eq Inactive GSB}
\end{equation}

\noindent where $\eta$ is the death rate of the inactive cells. $\eta$ should be greater than $\beta$ to simulate biological ageing \citep{lauro2011integrative}. Finally, the time dependence of the phage population ($P$) is given as

\begin{equation}
    \frac{dP}{dt} = \theta P A - \mu P,  \label{Eq Phages}
\end{equation}

% Stellar fluxes figurecoefficient figure
\begin{figure*}
	\centering
	\includegraphics[width=0.85\textwidth]{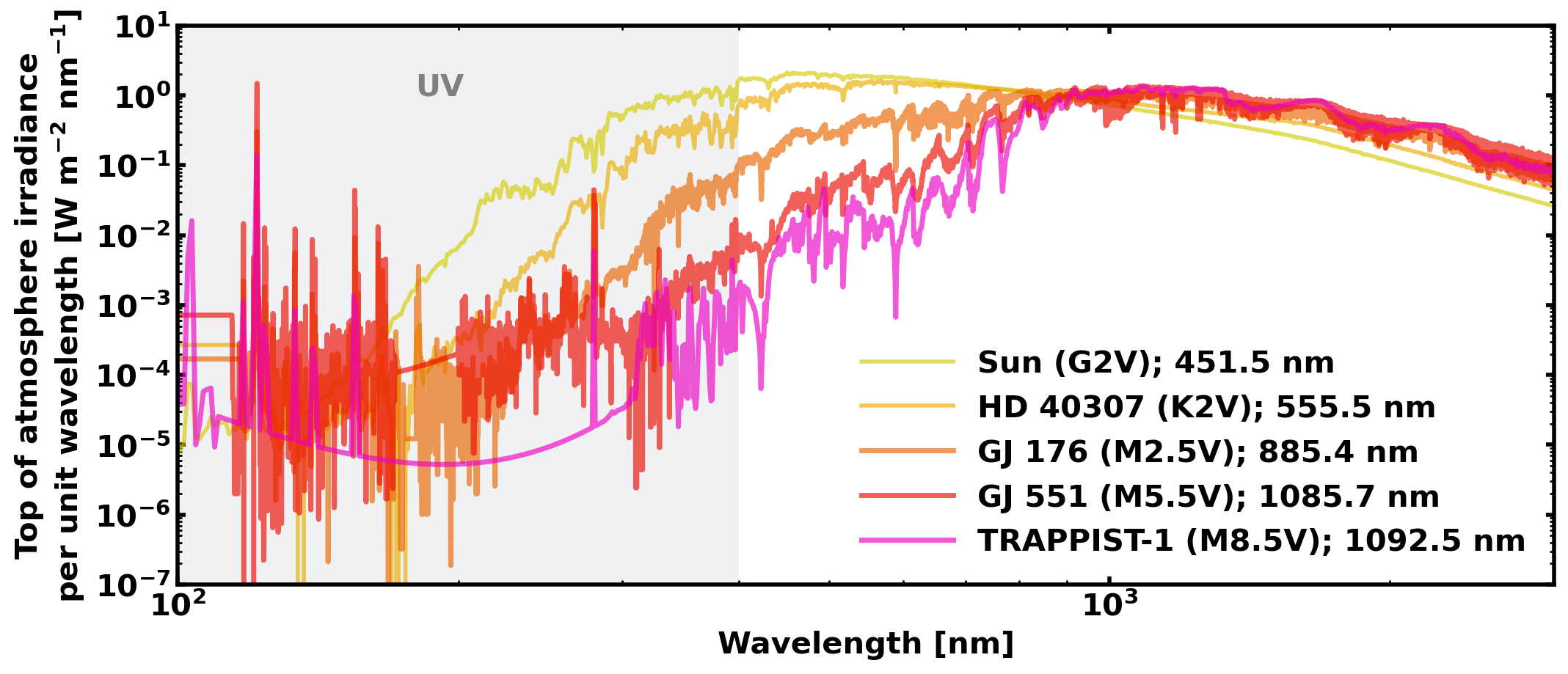}
    \caption{Spectral enegery distributions of GKM dwarf stars. The top of atmosphere irradiance is shown for several stars, all scaled to the flux that the Earth receives (1360 W m\textsuperscript{-2}), plotted against the wavelength of light in nm. The stars shown are the Sun (G2V), HD 40307 (K2V), GJ 176 (M2.5V), GJ 551 (M5.5V; also known as Proxima Centauri), and TRAPPIST-1 (M8.5V). Their stellar spectral type and peak wavelengths are given in nm. Whilst hycean candidates are only known to exist around M dwarfs, there is no reason to expect why they cannot exist around other spectral types, including G dwarfs. The UV regions (100--400 nm) is shaded in grey.} 
    \label{Stellar fluxes figure}
\end{figure*}

\noindent where $\theta$ is the growth rate of the phages due to predation and $\mu$ is the inactivation rate of the phages. Bacteriophages are not considered living organisms, but they can be inactivated by UV radiation \citep{martino2021virus}, thermal disinfection, high-pressures, as well as several artificial mechanisms \citep{kim1980mechanism, duran2002removal, muller2005inactivation, karczewska2023tackle}. 

Equations \eqref{Light equation}--\eqref{Eq Phages} describe the model used in \citet{lauro2011integrative} which was spatially dimensionless. In this work, we extend the LV model of \citet{lauro2011integrative} and integrate it with a 1D model inspired by \citet{kmet1994diffusive} and \citet{lopez2005introduction}, who both studied fish in the Danube river using a 1-dimensional (1D) diffusive LV model, and where \citet{lopez2005introduction} introduced predators at different times. We focus here on GSB (rather than cyanobacteria) because they exist in anoxic conditions which are relevant for hycean worlds.

\section{Hycean Lotka-Volterra model}
\label{Hycean Lotka-Volterra model section}

In this section, we describe our LV model devised for a hycean world scenario, which is modelled as an aquaplanet with anoxic waters in our implementation. We model the ocean as a 1D column down to 200 m deep, where the depth affects both the temperature and light availability in different layers. We investigate the inclusion of several microbial species (up to five at a time that are interacting with each other) that live in anoxic environments on Earth, in possible aquaplanet conditions, specifically those hypothesized for hycean worlds. These are green sulfur bacteria, and five psychrophilic sulphate-reducing bacteria, strains ASv26, LSv21, PSv29, LSv54, and LSv514, which are anaerobic bacteria \citep{knoblauch1999effect} and therefore relevant to hycean world conditions.

In particular, we include a light attenuation model which specifies water clarity/turbidity and light absorption by biological material. The water temperature also varies with depth, mimicking different scenarios for open ocean water temperature layers. Bacterial growth rates then depend on the ocean temperature, and we use measured growth rates for multiple bacterial species that live in anoxic environments. We end the section by providing an overview of the simulations performed in this work.

\subsection{Light availability}
\label{Light availability section}

Stellar radiation reaches the top of a planetary atmosphere and traverses through various atmospheric layers, where it is absorbed and scattered, affecting the intensity of the radiation reaching the surface. In our Lotka-Volterra model extension, Eq.~\eqref{Light equation} specifies the light availability at the ocean surface $L_{z=0}(t)$, but we modify the light availability by a factor which depends on depth ($z$, given in metres). In the ocean, light intensity decreases exponentially with depth due to absorption and scattering by water and suspended particles. This follows Beer’s Law, given by:

\begin{equation}
L_z(t) = L_0(t) e^{-k_d z}, \label{Beer's law}
\end{equation}

\noindent where $I_0$ is the initial light intensity at the surface, and $k_d$ is the attenuation coefficient in m$^{-1}$ that depends on water clarity. We use several values for $k_d$, which is the diffuse attenuation coefficient, and accounts for absorption and scattering \citep{williamson2023depth}. These $k_d$ values are based on the Jerlov water types \citep{jerlov1951optical, jerlov1968optical, jerlov1976marine}, where the attenuation varies with wavelength. Clear open ocean water has the lowest attenuation coefficients, with coastal water having higher attenuation coefficients. We will ignore coastal waters in our model as hycean exoplanets by definition do not have coasts; thus, we only consider open ocean waters (Jerlov types I--III). 

Microbial life on Earth can photosynthesize at wavelengths $>700$ nm \citep{Claudi2020Life}, including in very low light conditions \citep{Jung2023Light}, but phototropic organisms are ultimately depth limited because water is not transparent to photosynthetic active radiation. Based on measurements of $k_d$ for Earth's ocean waters in Fig.~\ref{Attenuation coefficient figure}, we consider $k_d$ to vary between 0.01 m$^{-1}$ and 0.7 m$^{-1}$. In addition to the turbidity of water, the light penetration depth will depend on the host star's spectral type. The peak emission wavelength for F dwarf stars to M dwarf stars is between 400 -- 1200 nm, and we show some example stars between G2V and M8.5V in Fig.~\ref{Stellar fluxes figure}. The light intensity is assumed to be equal to unity at the surface, i.e. $L_{z=0}(t)=1$, and we show the relative light intensity profiles with depth in Fig.~\ref{Light and temperature figure}a. Redder wavelengths tend to have higher attenuation coefficients compared to bluer wavelengths (see Fig.~\ref{Attenuation coefficient figure}). This red absorption means that light penetration depth in the oceans of planets around M dwarfs may be small compared to Earth and thus limit phototrophs to very shallow depths. Whilst this idea has been discussed before \citep{Heath_Mdwarf_1999, Kaltenegger2019DarkWaters}, to our knowledge, its impact has not been explored in relation to biological models. Among some of the hycean candidate host stars, the spectra of TOI-1468, TOI-270, and K2-18 all roughly peak at 825 nm, TOI-1231 peaks at 815 nm, LTT 3780 at 870 nm, and TOI-904 peaks at 770 nm. 

The absorption of light not only depends on the turbidity of the ocean water and the wavelength of light, but also on the biomass in the water column, including living organisms, dissolved organic compounds, and detritus, since all of these can reduce the intensity of light propagating to organisms below them \citep{huisman1995competition, bricaud1998variations, bengil2016bio, kostakis2020development, bi2023bio}. Living organisms use a variety of compounds for photosynthesis, such as Chlorophyll a and b \citep{lichtenthaler2001chlorophylls, virtanen2022chlorophyll}, and green sulfur bacteria utilise bacteriochlorophylls (BChl) a, c, d, and e \citep{otte1991comparative, kharcheva2016quantification, zhiltsova2022comparative}. In such a case where the water column is inhabited, the attenuation coefficient, $k_d$, becomes

\begin{equation}
    k_d = k_w+k_\text{bio}+k_\text{det},
\end{equation}

\noindent where $k_w$, $k_\text{bio}$, and $k_\text{det}$, are the attenuation coefficients for water, biological organisms, and detritus, respectively.

We do not aim to employ a full radiative transfer model and account for scattering as that is beyond the scope of our model. However, we test cases where $k_d = k_w$ and where $k_d = k_w+k_\text{bio}$, to investigate how the bacteria in the water column respond to self-shielding of light (the model does not track detritus). $k_\text{bio}$~depends on the concentration of cells, how much bacteriochlorophyll (BChl) each cell contains, and the absorbance of the BChl compound at a specific wavelength, $\alpha_\text{BChl}$. The wavelength absorbance peaks for different strains of bacteria depend upon which BChl compound they are utilising, and there can be more than one peak across the photosynthetic active radiation range \citep{otte1991comparative, manske2005physiology, vogl2012bacteriochlorophyll, bi2023bio, nelson1993phytoplankton, zhiltsova2022comparative, chew2007bacteriochlorophyllide, emeliantsev2020quantification}. 

For phytoplankton in the wavelength range 400--750 nm, $\alpha_\text{BChl}$ varies between 0--0.1 [Chl] m\textsuperscript{2}/mg \citep{bi2023bio}. \citet{overmann1992extremely} found $\approx 30$ -- 230 \textmu g of BCHl e per mg of cellular protein for different strains of brown phototrophic sulfur bacteria, which is 3 -- 23\% of the cellular protein in BCHl. The amount of protein in bacterial cells differs, but E. Coli have 55\% of their dry weight in proteins \citep{frigaard2003chlorobium}. A typical dry weight of a bacterium is 200 fg \citep{laufer2017physiological}. This means that in a single cell, there is approximately 1--13 fg of BCHl. \citet{bi2023bio} gives a relationship for the phytoplankton absorption coefficient: 

\begin{equation}
    k_\text{bio}=0.0237 [\text{Chl}]^{0.8987},
\end{equation}

\noindent where [Chl] is the concentration of Chlorophyll in mg/m\textsuperscript{3}. Assuming these relationships hold for sulfur reducing phototrophic bacteria, then a maximum of $10^{9}$ cells per ml (from the carrying capacity of the system) gives $k_\text{bio}=11.8$--118~m\textsuperscript{-1}.

There are also possible variations in terms of the surface intensity of light. At a particular location on a planet, light variations can be due to diurnal and seasonal cycles as well as cloud cover \citep{2003JApMe..42.1421P, 2008ACP.....8.2493S, perez2015reconstruction}. Therefore, we introduce randomness in light variations at the surface using a cloud modification factor (CMF). The CMF is defined as the ratio of radiation reaching the surface in a cloudless sky ($L_\text{Clr}$) against the radiation reaching the surface with clouds ($L_\text{Cld}$), given as

\begin{equation}
    \text{CMF} = L_\text{Cld} / L_\text{Clr}.
\end{equation}

\noindent We modify the surface light $L_{z=0}(t)$ by multiplying by the $\text{CMF}$. In Antarctica, the $\text{CMF}$ varied at two research stations by differing amounts, with medians of $\approx0.45$ and $\approx0.7$, and minimum-to-maximum ranges of $\approx 0.1$--1 and $\approx0.1$--1.15 \citep{2025AtmRe.31607940S}, respectively. $\text{CMF}$ values of greater than 1 are possible because clouds can occasionally enhance the diffuse radiation reaching the surface. To simulate temporal fluctuations in incident surface light through a $\text{CMF}$, we applied a one-dimensional Gaussian filter (the degree of smoothing is controlled by a $3\sigma$ Gaussian kernel width) to a time series of randomly generated light attenuation factors between 0.1 and 1.15 for the `Antarctica-like' cases. For tidally locked scenarios, we impose a CMF of 0.80--1, assuming that the clouds are more uniform based on global climate model simulations of exoplanets with oceans \citep{2013ApJ...771L..45Y, 2017A&A...601A.120B, 2021ApJ...913..101L, Cooke2023Degenerate}, although this depends on GCM grids, their resolution, and any sub grid-scale process parameterisations \citep{2020ApJ...894...84S, 2023NatAs...7.1070Y, 2024ApJ...970....7S}.

\subsection{Water temperature}
\label{Water temperature section}

On Earth, the temperature of a water column varies depending on geography and the time of year. At the top of the ocean sits the mixed layer, which has a roughly isothermal temperature \citep{kara2000optimal} due to turbulence from interaction with atmospheric dynamics (winds and surface stress). The thermocline is a transition layer in a body of water where the temperature decreases rapidly with depth until reaching a colder deep temperature. In terms of water temperature on exoplanets, some work has been done exploring the dynamics of tidally locked exoplanets, their oceans, and clouds \citep{2020ApJ...895...19O, 2020ApJ...896L..16S, 2023NatAs...7.1070Y, 2024ApJ...970....7S}, albeit not for hycean exoplanets. In particular, \cite{Lai_2022_Thermocline} used the MITgcm to simulate deep-ocean (up to hundreds of km deep) exoplanets and showed that the thermocline was confined to the upper 10 km and its depth varied with latitude between $\sim100$ and $\sim1000$ m. Whilst they found a global overturning circulation which could transport nutrients, they also noted that their simulations lacked high-pressure ice at the ocean floor which would affect ocean circulation - such simulations would be more representative of a hycean ocean. In the absence of hycean ocean simulations, we use a simpler mathematical model based on an exponential decay function \citep{fel1992exponential}:

\begin{equation}
T(z) = T_\text{sfc} + (T_\infty - T_\text{sfc}) \cdot e^{-\kappa_T (z-d_\text{ML})},\label{Temperature profile equation}
\end{equation}

\noindent where $T(z)$ is the temperature at depth $z$ (in meters), $T_\text{sfc}$ is the temperature at the surface (at $z = 0$), $T_\infty$ is the temperature at the deep ocean layer, $\kappa_T$ is a decay constant (in m$^{-1}$), and $d_\text{ML}$ is the depth of the mixed layer. $d_\text{ML}$ ranges between $\sim1$ to a few hundred metres on Earth \citep{treguier2023mixed} and a 200 m global average mixed layer depth was found by \citet{2020ApJ...895...19O} who varied orbital parameters for Earth-like exoplanets in 3D simulations. Additionally, modelling indicates tidally locked mixed layer depths can range from 40--100 m on terrestrial exoplanets but can possibly be more extensive \citep{2020ApJ...896L..16S}, and this can depend upon the total depth of the ocean \citep{2023NatAs...7.1070Y}. 

\begin{figure*}
	\centering
	\includegraphics[width=1\textwidth]{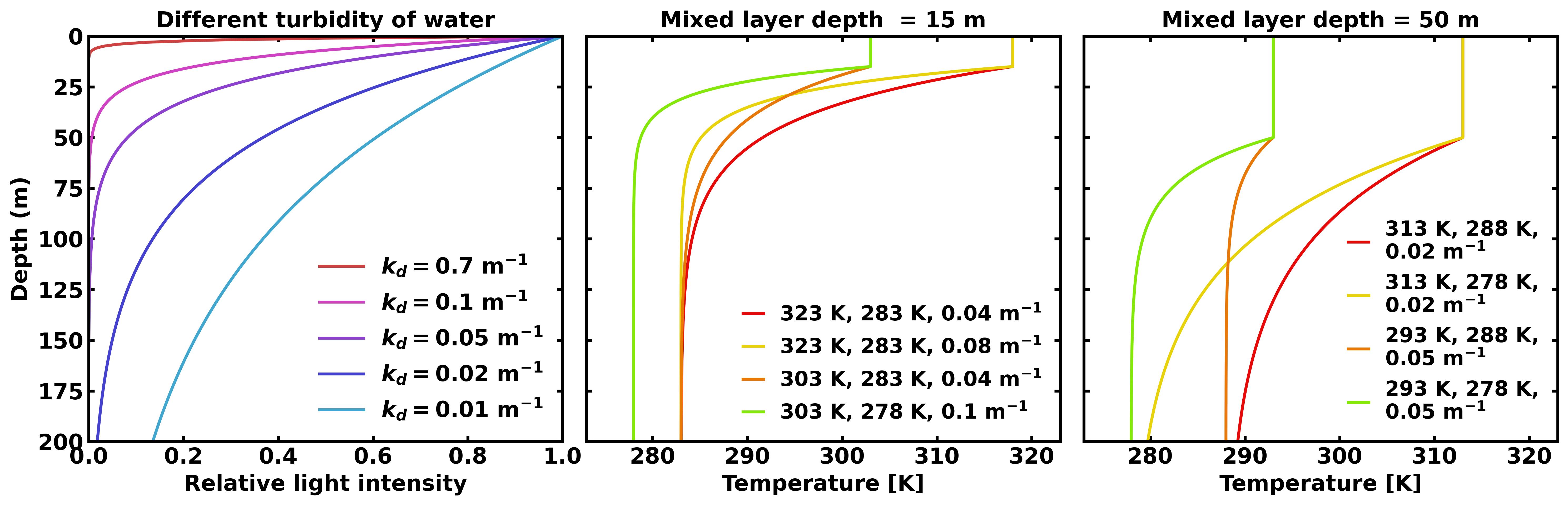}
    \caption{Light and temperature vertical profiles in the one-dimensional water column. \textbf{Left}: The attenuation coefficients ($k_d$ in m\textsuperscript{-1}) used to determine the relative light intensity which reaches different depths in the ocean column. Depth is defined as distance below the surface which is at 0 m depth. \textbf{Middle and Right}: Temperature profiles with various surface temperatures ($T_\text{sfc}$), deep ocean temperatures ($T_\infty$), and temperature decay constants ($\kappa_T$), with a mixed layer depth ($d_\text{ML}$) of 15 m. \textbf{Right}: Temperature profiles with various $T_\text{sfc}$, $T_\infty$, and $\kappa_T$, with a mixed layer depth ($d_\text{ML}$) of 50 m.} 
    \label{Light and temperature figure}
\end{figure*}

We illustrate various profiles for water temperature in Fig.~\ref{Light and temperature figure}b and Fig.~\ref{Light and temperature figure}c. The combination of light variability and water temperature ultimately influence the growth rate of the species, as will be described in section~\ref{Bacterial growth rates section}.

% Light and temperature figure

\subsection{Spatial diffusivity}
\label{Spatial diffusivity section}

Without including diffusivity, the modelled organisms would not spread out in the water column. The temperature and light would still affect each layer, but each layer would be a separate, isolated space. Inspired by  \citet{kmet1994diffusive} and \citet{lopez2005introduction}, we therefore modify Eq.~\ref{Eq Active GSB}--\ref{Eq Phages} by introducing Fick's law. This law specifies the diffusive flux in relation to the concentration gradient, in this case the concentration of active/inactive bacteria or phages. For microbial organisms, this dispersal is passive, due to oceanic currents and circulation.  

So now, our model determines the density of bacteria and phages through the following three equations:

\begin{equation}
    \frac{dA}{dt} = \alpha A L \left ( 1-\frac{G}{C} \right ) - \beta A + \gamma I L - \delta A - \epsilon P A +D_A \frac{d^2A}{dz^2}, \label{Eq Active GSB final}
\end{equation}

\begin{equation}
    \frac{dI}{dt} = -\gamma I L + \delta A - \eta I +D_I \frac{d^2I}{dz^2},  \label{Eq Inactive GSB final}
\end{equation}

\noindent and

\begin{equation}
    \frac{dP}{dt} = \theta P A - \mu P+D_P \frac{d^2P}{dz^2},  \label{Eq Phages final}
\end{equation}

\noindent where $D_A$, $D_I$, and $D_P$ are the diffusion coefficients in m\textsuperscript{2} s\textsuperscript{-1} for active bacteria, inactive bacteria, and phages, respectively.

We do not know how turbulent a water column would be on various exoplanets, and in this work we are not simulating a specific exoplanet. In Earth's oceans, the global average deep water vertical diffusivity is approximately $10^{-4}$~m\textsuperscript{2}s\textsuperscript{-1} \citep{Munk_Ocean_1998, Wunsch2004Mixing} and ranges between $10^{-6}$ to $10^{-3}$~m\textsuperscript{2}s\textsuperscript{-1} near the ocean surface \citep{Waterhouse2014_Ocean}. However, for convective regions, where the water overturns relatively rapidly, the corresponding vertical diffusivity coefficient increases and has been represented as high as 1 -- 50~m\textsuperscript{2}~s\textsuperscript{-1} \citep{Marotzke_1991_ocean, marotzke1999convective, Marshall_1999_ocean} for Earth. For exoplanets which are tidally locked, or for exoplanets with close-by satellites, increases in the vertical transport are expected due to tidal-induced mixing \citep{Si_2022_planet, Di_Paolo_2025_Ocean}. Therefore, in our column of 200 m depth, we consider diffusion coefficients of between $10^{-6}$ -- 50~m\textsuperscript{2}~s\textsuperscript{-1}. To better simulate likely biological differences, we simulate cases where the active bacteria have a greater diffusivity than the inactive bacteria and the phages.

Finally, we employ Neumann boundary conditions at the bottom and top of the water column, with $dA/dz=0$, $dI/dz=0$, and $dP/dz=0$.

% GSB growth rate figure
\begin{figure}
	\centering
	\includegraphics[width=1\columnwidth]{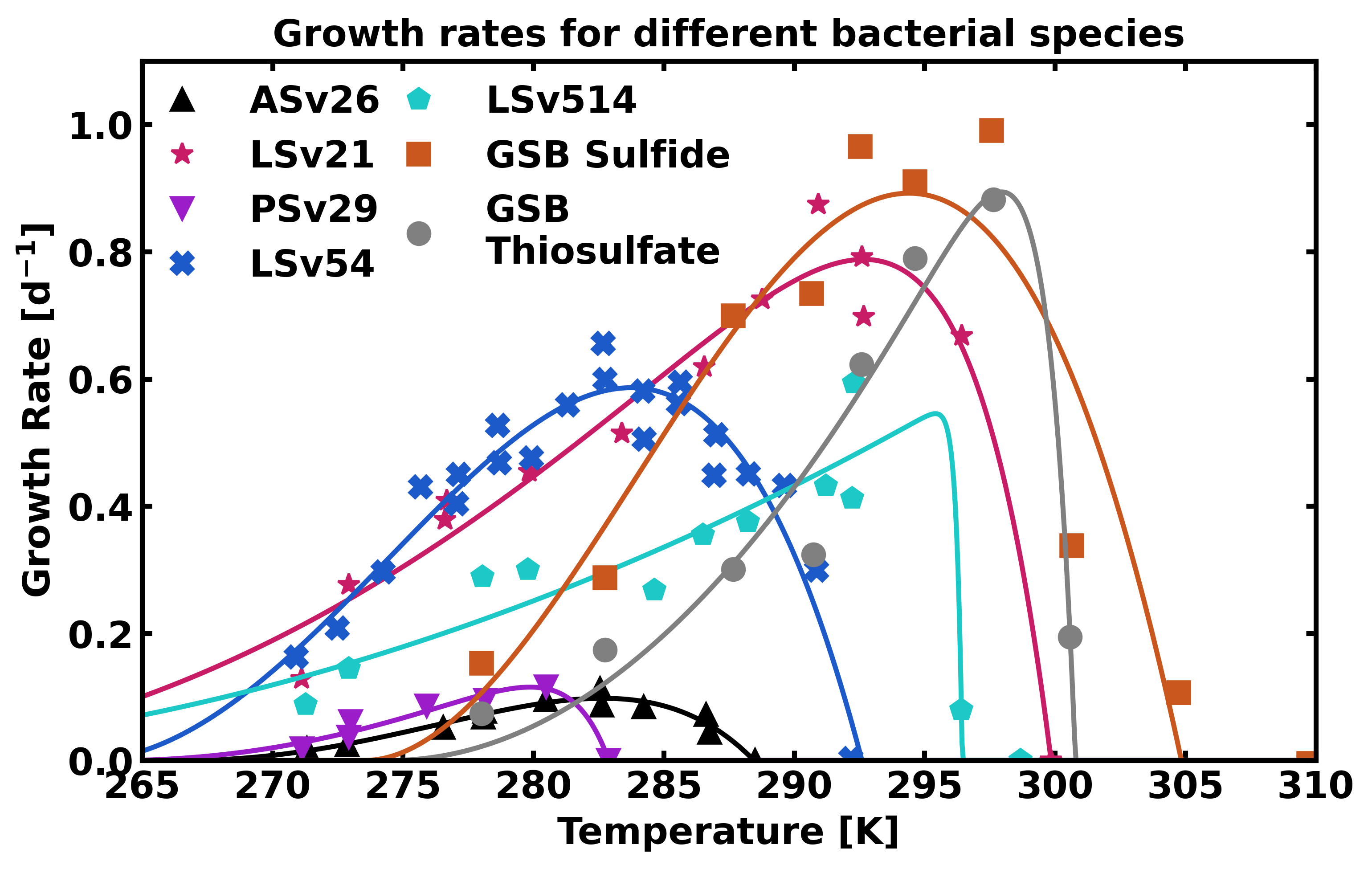}
    \caption{Biological population growth rates per day (d) of six different bacterial species. Five of these species are psychrophilic sulphate-reducing bacteria from Arctic sediments data, and the data was sourced from \citet{knoblauch1999effect}. The species shown are ASv26 (black upward triangles), LSv21 (red stars), PSv29 (purple downward triangles), LSv54 (blue crosses), and LSv514 (turquoise pentagons). Green Sulfur Bacteria grown on different electron donors, sulfide (orange squares) and thiosulfate (grey circles), with the data sourced from \citet{overmann1989pelodictyon}, are also displayed. The fits to the data use either the Ratkowsky or Square root models for bacteria growth.} 
    \label{GSB growth rate figure}
\end{figure}

\subsection{Bacterial growth rates}
\label{Bacterial growth rates section}

Several models exist to describe bacterial growth in response to temperature, with the Arrhenius model and the Ratkowsky square root model describing the growth rate of microbial populations \citep{ratkowsky1983model, zwietering1991modeling, dobric2018temperature}. The Ratkowsky square root growth model \citep{ratkowsky1983model} describes the relationship between temperature and the square root of the growth rate:

\begin{equation}
    \mu = \left [ b (T - T_{\text{min}}) \cdot \left ( 1 - e^{c (T - T_\text{max})} \right ) \right ]^2\label{Eq: Ratkowsky 2}
\end{equation}

\begin{table*}
\centering
\caption{The original parameters used for Ace Lake mixolimnion with green sulfur bacteria (GSB) and bacteriphages in \citet{lauro2011integrative}. Note that \citet{lauro2011integrative} also considered cyanobacteria, with growth rates found in \citet{powell2005ecology}. $\alpha$ is the per capita growth rate of the resource species, $\beta$ quantifies the consumer’s per capita death rate, $\gamma$ is the rate at which inactive cells become active, $\delta$ characterizes the rate at which active cells become inactive, $\epsilon$ is the death rate due to phages, $\eta$ is the mortality rate of inactive bacteria, $\theta$ is the growth rate of the phages due to predation, $\mu$ is the inactivation rate of the phages, $\omega$ is assumed to be the seasonal period (1 year for Earth), and $\tau$ is a dimensionless modification factor which was originally designated in \citet{lauro2011integrative} alongside $\omega$ to reproduce the seasonal light reaching Ace Lake in Antarctica. In this work, we extend the model of \citet{lauro2011integrative} and vary $\alpha, \omega$, $\eta$, $\mu$, and $\tau$.}
\label{Simulation Scenario Table}
\centering
\begin{tabular}
{@{}ccccccccccc@{}}
\toprule
$\alpha$ & $\beta$ & $\gamma$ & $\delta$ & $\epsilon$ & $\eta$ & $\theta$ & $\mu$ & $\omega$ & $\tau$
\\ \midrule
0.34 & 0.009 & 0.7 & 0.07 & $8\times10^{-10}$ & 0.14 & $4\times10^{-9}$ & 0.06 & 365 & 1.44\\
\bottomrule
\end{tabular}
\end{table*}

\begin{table*}
\centering
\caption{Five sets of simulations are presented in Section~\ref{Results section} and several of the parameters for each simulation are listed below. Those that are not listed are unchanged, which include $\gamma$, $\delta$, $\epsilon$, $\theta$, and $\mu$. The growth rate ($\alpha$) varies depending on the temperature as per Fig.~\ref{GSB growth rate figure}, so each species in the simulation are listed whilst the parameters required to specify the water temperature with depth (Eq.~\ref{Temperature profile equation}) are also given. Other species are included for the same simulation parameters shown in the Supplementary Material. Each $\tau$ value corresponds to the same $\omega$ value in the listed order for the `Orbital period simulation'. The initial concentrations of $A$, $I$, and $P$ are given as $A_0$, $I_0$, and $P_0$, respectively. If bioattenuation of light is included in the simulation, $k_d = k_w + k_{\textrm{bio}}$ and the value for $k_w$ is given. At the bottom, whether the cloud modification factor is included or not in the simulations is given by a tick (\checkmark) or a cross (\text{\sffamily X}).}
\small % comment out or change to \footnotesize / \scriptsize if needed
\begin{tabular}{@{}>{\centering\arraybackslash}p{1.5cm}*{5}{>{\centering\arraybackslash}p{2.8cm}}@{}} 
\toprule
Parameter & Light shielding simulation & Invasion time simulation & Competition simulation & Resource-switching/Weakness simulation & Orbital period simulation \\ \midrule
$\alpha$ [d\textsuperscript{-1}] & GSB (sulfide) & GSB (thiosulfate) & GSB (thiosulfate), ASv26, LSv21, LSv54, LSv514 & GSB (sulfide), LSv54, LSv21 & GSB (sulfide) \\
$\omega$ [d] & 365 & 365 & 365 & 365 & $36.5$, $91$, $182.5$, $365$, $730$ \\
$\eta$ & 0.085 & 0.085 & 0.085 & 0.03 & 0.01 \\
$\tau$ & 1.44 & 1.44 & 1.44 & 1.44 & 0, 1.44, 1.44, 1.44, 1.44, 1.44 \\
$A_{0}$ [ml\textsuperscript{-1}] & $10^{7}$ & $8\times10^{8}$ & $10^{8}$ & $10^{5}$--$10^{6}$ & $8\times10^{8}$ \\
$I_{0}$ [ml\textsuperscript{-1}] & $10^{7}$ & $10^{8}$ & $10^{8}$ & $10^{8}$ & $10^{8}$ \\
$P_{0}$ [ml\textsuperscript{-1}] & 10 & 10 & 10 & 1000 & 100 \\
$t_{P,\rm in}$ [d] & 0 & Varied from 0--550 & 0 & 0 & 0 \\
$d_{P,\rm in}$ [m] & 0 & 0 \& 15 & 0 & 0-200 & 0 \\
$k_d$ [m\textsuperscript{-1}] & $k_w+k_{\rm bio}$ \& $k_w=0.05$ & $k_w=0.2$ & $k_w=0.02$ & $k_w$ \& $k_w+k_{\rm bio}$, $k_w=0.1$ &  $k_w=0.7$\\
$T_{\rm sfc}$ [K] & 296 & 293 & 298 & 298 & 298 \\
$T_{\infty}$ [K] & 278 & 278 & 284 & 285 & 281 \\
$d_{\rm ML}$ [m] & 50 & 15 & 10 & 2 & 25 \\
$k_T$ [m\textsuperscript{-1}] & 0.05 & 0.08 & 0.02 & 0.1 & 0.025 \\
$D_A$ [m\textsuperscript{2}s\textsuperscript{-1}] & $10^{-3}$ & $2\times10^{-1}$ & $10^{-3}$ & $10^{-1}$ & $1$ \\ 
$D_I$ [m\textsuperscript{2}s\textsuperscript{-1}] & $10^{-3}$ & $10^{-1}$ & $10^{-3}$ & $10^{-1}$ & $1$ \\ 
$D_P$ [m\textsuperscript{2}s\textsuperscript{-1}] & $10^{-3}$ & $10^{-1}$ & $10^{-3}$ & $10^{-1}$ & $1$ \\ 
CMF & \text{\sffamily X} \& \checkmark & \text{\sffamily X} & \text{\sffamily X} & \checkmark &  \text{\sffamily X} \& \checkmark \\ 
\bottomrule
\label{Paramater table}
\end{tabular}
\end{table*}

\noindent where $\mu$ is the bacterial growth rate (per unit time), $b$ is an empirical constant related to the slope of the curve, $T$ is the temperature (in K), $T_{\text{min}}$ is the minimum temperature for bacterial growth (in K), $T_{\text{max}}$ is the maximum temperature for bacterial growth (in K), and $c$ is an empirical parameter controlling the shape of the decline at temperatures greater than the optimal temperature for growth, $T_\textrm{opt}$. Note that Eq.~\eqref{Eq: Ratkowsky 2} is not based on biology, but is a mathematical model constructed from empirical data. A modified Ratkowsky square root model was introduced by \citet{zwietering1991modeling}, who discuss various other models for bacterial growth rates, and this is given as:

\begin{equation}
    \mu = \left [ b (T - T_{\text{min}})\right ]^2 \cdot \left ( 1 - e^{c (T - T_\text{max})} \right ). \label{Eq: Ratkowsky 3}
\end{equation}

\noindent It is this equation that we use for the growth rates of our bacterial species. Note that at temperatures ($T$) lower than $T_{\text{min}}$ or greater than $T_{\text{max}}$, growth does not occur. However, see \citet{zwietering1991modeling} for a discussion on this and some other models.

In \citet{lauro2011integrative}, their LV model was zero-dimensional and used the peak growth rate of a single cyanobacteria and a single green sulfur bacteria (GSB) species. Here we show the growth rates of several different species against temperature in Fig.~\ref{GSB growth rate figure}. Growth rate data for a single GSB species, with two different electron donors, are taken from \citet{overmann1989pelodictyon}. The ASv26, LSv21, PSv29, LSv54, and LSv514 bacteria strains have growth rates from \citet{knoblauch1999effect}. These bacteria are able to grow at cold temperatures, including the freezing point of Earth's seawater ($-1.8^\circ $C). They generally have optimum growth in the neutral pH range \citep[7--7.9;][]{knoblauch1999psychrophilic} and between the species, they can grow on a variety of substrates, including sulfide, sulfur, thiosulfate, formate, acetate, and iron(III). 

To determine a smooth function for the temperature-dependent growth rate of each species, we fitted the empirical growth rate data for each species using the modified Ratkowsky square root model, Eq.~\eqref{Eq: Ratkowsky 3}, from \citet{zwietering1991modeling}, and find lines of best fit using non-linear least squares regression. 

Because our LV model is one-dimensional, we vary the growth rates depending on the temperature of the layer the bacteria are in according to Eq.~\eqref{Temperature profile equation}. The death rate is assumed to be constant because we do not know of any temperature dependent death rate for GSB or psychrophilic sulphate-reducing bacteria. Between all six species, peak growth rates vary between $\sim0.1$--1, and $T_\textrm{opt}$ varies between 280--300~K. 

We note here that these growth rates are not exhaustive, instead representing a range of anoxic microbial organisms that we will use to illustrate how certain physical conditions can affect biodiversity. \citet{biebl1978growth} measured doubling times for GSB in mixed cultures with sulfur and sulfur reducing bacteria at a fixed temperature, under different illumination intensities. Other growth rates are available but do not have temperature dependence given \citep{schmidt1970variable, eichler1988new, folt1989field, musat2008single, costa2017potential}, although the optimum temperature for growth may be provided. Some bacterial growth rates in Arctic marine sediments are also available in \citet{scholze2021psychrophilic}.

\subsection{Numerical Simulations}

The numerical simulations we use incorporate the equations set out in Section~\ref{Lotka-Volterra Model section} with the original parameters used by \citet{lauro2011integrative} listed in Table~\ref{Simulation Scenario Table}. Our extensions are given in Sections~\ref{Light availability section}--\ref{Bacterial growth rates section}. In our Hycean LV Model extension, we simulate many different scenarios with alternative values for $\alpha, \omega$, $\eta$, $\mu$, and $\tau$. We adjust $\omega$ and $\tau$ to control for different orbital periods or constant illumination. The seasonality of the light we impose is based on light reaching Antarctica, such that the minimum in the `winter months' is zero. If we instead modelled seasonality based on a location at or close to the equator, then the minimum light during the darker periods would not be zero. We use $\tau=1.44$ for Antarctica \citep{lauro2011integrative} and 0 for tidally locked worlds receiving constant illumination. $\eta$ is varied to alter the rates of biological aging and always remains higher than $\beta$. 

Bacteriophages are likely the most abundant biological entity on Earth with an excess of $10^{30}$ phage particles, outnumbering bacteria by a factor of $\sim10$ \citep{hendrix2002bacteriophages, mushegian2020there, dion2020phage}. Additionally, phages are incredibly genetically diverse \citep{Suttle2005_Viruses_sea, dion2020phage, hatfull2015dark} and have differing levels of resistance to UV inactivation \citep{zablocki2016diversity, tom2018experimental}. Changes in temperature and UV radiation between different planetary surface environments are likely to affect the phage inactivation rate. As such, in the main manuscript, we opt to not alter $\epsilon$, $\theta$, and $\mu$, which control the phage response in our model. We do include some simulations in the Supplementary Material where these parameters are varied.

The Lotka-Volterra equations are continuous functions and allow for situations where the total number of bacteria drops below 1 (extinction) either due to lack of light for growth, or attacks from phages, but then the populations rebounds. An example of this can be seen to occur in Fig.~5i from \cite{lauro2011integrative}. Instead, when the total active population ($A$) and the inactive population ($I$) of bacteria simultaneously drop below 1, we impose a zero-growth condition and assume the population of that species to be extinct in our model. The same condition applies to phages once the total population drops below 1. 

The photic zone varies in our models depending on the ocean turbidity and population densities. At $k_d=0.02$ m\textsuperscript{-1}, the photic zone extends to $\approx 230$ m, whereas this depth is $\approx 6.5$ m for $k_d=0.7$ m\textsuperscript{-1}. Thus, the photic zone will be much more restricted for planets where the water attenuation coefficient is greater due to the water turbidity, or where the photosynthetically active radiation wavelengths occur at wavelengths $<300$ nm or $> 600$ nm. 

In Section~\ref{Temperature and light dependence for growth}, we explore the dependency of bacterial growth in the water column to light and temperature. We first demonstrate how the combination of temperature and light acts to modify the growth rate of different specific species by choosing a range of nominal values. Then, we show the time evolution of a GSB (sulfide) population responding to seasonal light cycles (1 yr period) without phages. The simulation parameters are: $\eta=0.085$ d\textsuperscript{-1}, $A_0=I_0=10^{7}$ cells ml\textsuperscript{-1}, $T_\text{sfc}=296$ K, $T_\infty=278$ K, $d_{\rm ML} = 50$ m, $k_T = 0.05$ m\textsuperscript{-1}, $k_d = 0.05$ m\textsuperscript{-1}, $D_A = D_I = D_P = 10^{-3}$ m\textsuperscript{2}s\textsuperscript{-1}, and the model is run for 2000 d. In this setup, we then include and exclude biological light shielding, with and without an applied cloud modification factor (0.2--1.15). 

We introduce phages to all scenarios to determine the response of the active and inactive populations, and present the results in Section~\ref{Phage results section} regarding the light shielding simulations and the ``Invasion time" simulations. Ten phages are introduced at 0 m and 15 m depth to compare with the light shielding results in Section~\ref{Temperature and light dependence for growth}. Then, we include a different GSB (thiosulfate) population and change the parameters of the water column to: $\eta=0.085$ d\textsuperscript{-1}, $A_0=8\times10^{8}$ cells ml\textsuperscript{-1}, $I_0=10^{8}$ cells ml\textsuperscript{-1}, $P_0=10$ phages ml\textsuperscript{-1} at $t_{P,\textrm{in}}=0, 15$ m with introduction every day varied between 0 -- 550 days, $T_\text{sfc}=293$ K, $T_\infty=278$ K, $d_{\rm ML} = 15$ m, $k_T = 0.08$ m\textsuperscript{-1}, $k_d = 0.2$ m\textsuperscript{-1}, $D_A = 2\times10^{-1}$ m\textsuperscript{2}s\textsuperscript{-1} and $D_I = D_P = 10^{-1}$ m\textsuperscript{2}s\textsuperscript{-1} and the model is run for 1400 d. The diffusivity is a factor of 2 higher for the active population compared to the phages and inactive bacteria. The phages are introduced at different times to explore how the timing of consumer invasion affects whether the resource population collapses or survives. 

Each iteration and simulation of the model discussed so far assumes that a single species, limited by the carrying capacity of the system, inhabits the ocean and grows in response to temperature and light. However, in Earth's oceans, many ecological areas are occupied by multiple species which co-exist, sometimes competing for the same resource, be it space, light, or nutrients. We modify the model to enable five different species to grow at once, each being acted upon by the same phage. The total carrying capacity is unchanged, so the species are considered to be competing for space and nutrients. Section~\ref{Competition between species} explores how five bacterial species (GSB grown on sulfide, ASv26, LSv21, LSv54, and LSv514) in a water column compete for space to find a niche with their ideal growing conditions, whilst phages are present. The parameters of the simulation are:  $\eta=0.085$ d\textsuperscript{-1}, $A_0=I_0=10^{8}$ cells ml\textsuperscript{-1}, $P_0=10$ phages ml\textsuperscript{-1} at $d_{P,\textrm{in}}=0$ m, $T_\text{sfc}=298$ K, $T_\infty=284$ K, $d_{\rm ML} = 10$ m, $k_T = 0.02$ m\textsuperscript{-1}, $k_d = 0.02$ m\textsuperscript{-1}, $D_A = D_I = D_P = 10^{-3}$ m\textsuperscript{2}s\textsuperscript{-1} and the model is run for 3000 d. 

An environmental situation pertinent to our work is a bacterial bloom \citep{eiler2004composition, paerl2013harmful, huisman2018cyanobacterial} or an algal bloom \citep{hallegraeff2003harmful, balaji2022methods}. In this scenario, the environmental conditions enable a single species to dominate in that ecological space, giving consumers a vast resource. Viruses can preferentially target the most abundant or fastest-growing species, thereby curbing their dominance. With three bacterial species we compare two situations with and without phages and where light bioattenuation is switched off, then implement a scenario with phages and bioattenutaion of light present. In a fourth scenario, phages attack only the fastest growing species (Resource Weakness model). Then, in a fifth scenario, the phages switch prey based on whichever species is most abundant (Resource-switching scenario). The parameters in the simulations are:  $\eta=0.03$ d\textsuperscript{-1}, $A_0=10^{5}-10^{6}$ cells ml\textsuperscript{-1}, $I_0=10^{8}$ cells ml\textsuperscript{-1}, $P_0=1000$ phages ml\textsuperscript{-1} at $d_{P,\textrm{in}}=0$--200 m (all model levels), $T_\text{sfc}=298$ K, $T_\infty=285$ K, $d_{\rm ML} = 2$ m, $k_T = 0.1$ m\textsuperscript{-1}, $k_d = 0.02$ m\textsuperscript{-1}, $D_A = D_I = D_P = 10^{-1}$ m\textsuperscript{2}s\textsuperscript{-1} and the models are run for 1825 d (5 years). 

Finally, in Section~\ref{Rotational period results section}, we vary $\omega$, the periodicity of the light to account for seasonality, and we explore how constant illumination (i.e. tidally locked worlds) affect the modelled bacterial population, both with and without phages.  In the other sections, our models are only applicable to asynchronously rotating planets in an anoxic Antarctica-like scenario where light returns to the poles based on a specified one-year orbital period. Thus, the simulations assume a planetary obliquity like Earth. Such a scenario could be applicable to both hycean and Earth-like worlds, as long as the water is anoxic, and the planet is not tidally locked. The 6 different scenarios differ only in the value of $\omega$ and $\tau$, which specify the seasonal period of light. One case is a tidally locked case, represented by lacking a seasonal period. The seasonal periods tested are 0.1 yr, 0.25 yr, 0.5 yr, 1 yr, and 2 yr. This corresponds to $\approx 36.5$, 91 d, $\approx 182.5$~d, 365 d, and 730 d, respectively. The shortest time period is comparable to the orbital period of K2-18~b, assuming it has seasons and is not tidally locked. The longer period of 2 years corresponds to a possible orbit in the habitable zone around an F type dwarf star \citep{Kopparapu_HZ_2013, 2017AN....338..413S}. Note that it would also be possible to consider alternative latitudes, such as the tropics or extratropics in our model by modifying the light function. The parameters of the simulation are:  $\eta=0.01$ d\textsuperscript{-1}, $A_0=8\times10^{8}$ cells ml\textsuperscript{-1}, $I_0=10^{8}$ cells ml\textsuperscript{-1}, $P_0=100$ phages ml\textsuperscript{-1} at $d_{P,\textrm{in}}=0$ m, $T_\text{sfc}=298$ K, $T_\infty=281$ K, $d_{\rm ML} = 25$ m, $k_T = 0.025$ m\textsuperscript{-1}, $k_d = 0.7$ m\textsuperscript{-1}, $D_A = D_I = D_P = 1$ m\textsuperscript{2}s\textsuperscript{-1} and the models are run for 1460 d (4 years). 

The summary of all the parameters for each simulation can be found in Table~\ref{Paramater table}.

\section{Results}
\label{Results section}

We implement our Hycean Lotka-Volterra Model by testing out various parameters and how they affect the modelled organisms, starting with a single bacterial species only. The effects of water temperature and light intensity with depth are initially examined through various temperature profiles and ocean turbidity values in Section~\ref{Temperature and light dependence for growth}. Bacteriophages are introduced as a consumer in different circumstances in Section~\ref{Phage results section}, and the introduction of competition with multiple bacterial species is then included in Section~\ref{Competition between species}. Finally, we assess the impact of rotation rate and constant illumination on the simulated scenarios in Section~\ref{Rotational period results section}, emulating asynchronous rotators and tidally locked planets, respectively.

\subsection{Temperature and light dependence for growth}
\label{Temperature and light dependence for growth}

% alpha L growth rate figure
\begin{figure*}
	\centering
	\includegraphics[width=1\textwidth]{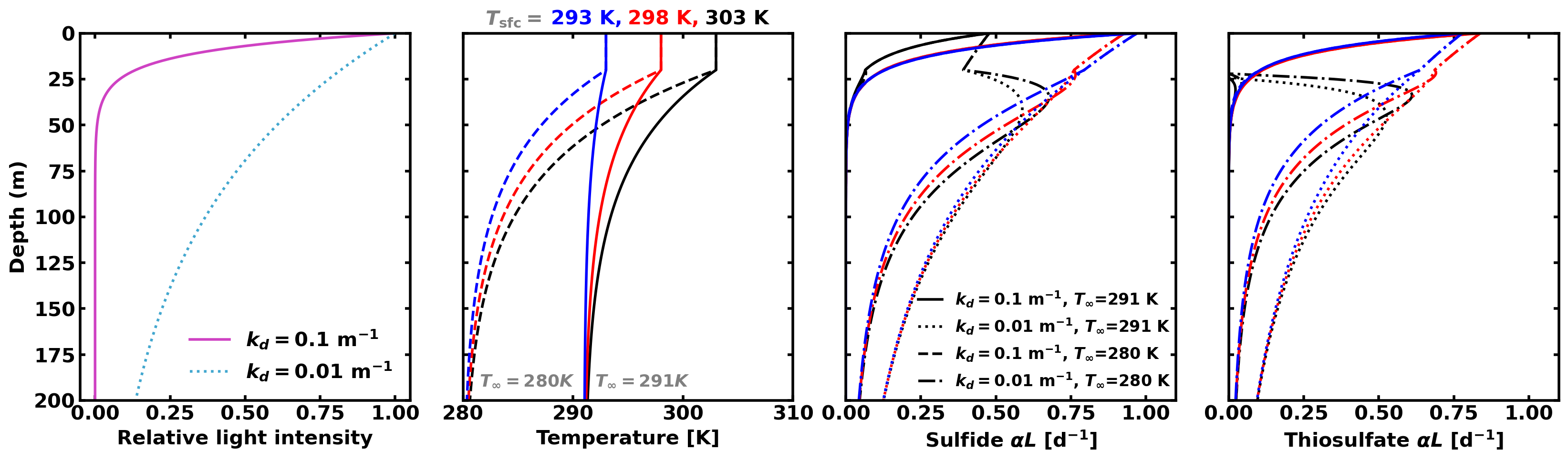}
    \caption{Light, temperature and growth rate vertical profiles in the one-dimensional water column. The availability of light ($L$, \textbf{left panel}) affects the product $\alpha L$ which specifies how fast the active Green Sulfur Bacteria can grow for two different electron donors, sulfide and thiosulfate. The growth rate per day ($\alpha$) depends on water temperature (\textbf{middle left panel}). Different surface temperatures (293 K, 298 K, and 303 K) are marked by different colours (blue, red, and black, respectively). Therefore, the light attenuation coefficient ($k_d$ in m\textsuperscript{-1}) and the temperature profiles set whether the GSB can grow, and the prodct $\alpha L$ is shown for GSB grown on sulfide (\textbf{middle right panel}) and thiosulfate (\textbf{right panel}). In the middle right and right panels, the solid line indicates a $k_d$ of 0.1 m\textsuperscript{-1} with a deep ocean temperature of 291 K, whilst dotted, dashed, and dash-dotted lines, represent 0.01 m\textsuperscript{-1} and 291 K, 0.1 m\textsuperscript{-1} and 280 K, and 0.01 m\textsuperscript{-1} and 280 K, respectively. These are illustrative examples and not necessarily the profiles used in actual simulations.} 
    \label{alpha L growth rate figure}
\end{figure*}

We first explore the dependence of species' growth on temperature and light. Even if the water temperatures are viable for bacterial growth, the light availability profile is also essential for determining whether any phototrophic organisms (e.g., GSB) grow in our Hycean LV model.  

% Light self shielding figure with phages
\begin{figure*}
	\centering
	\includegraphics[width=1\textwidth]{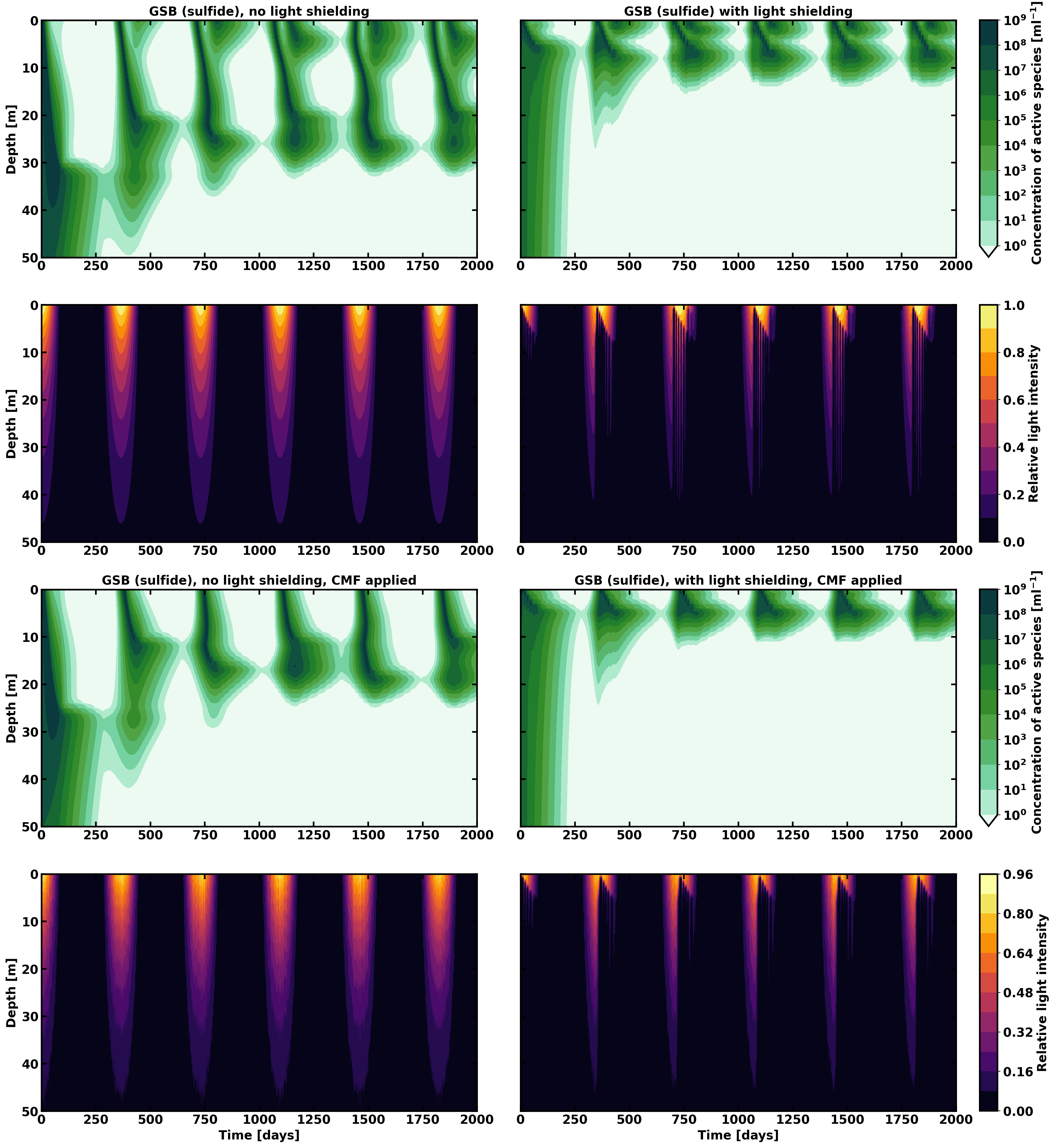}
    \caption{A comparison between the exclusion (\textbf{left}) and inclusion (\textbf{right}) of bioattenuation of light. Between depths of 0 m (surface) and 50 m, the top two panels show how the active population of Green Sulfur Bacteria responds to changes in light when there is no shielding included in the model (left) and when there is shielding included (right), plotted against time, assuming a period of 365 days. A concentration of 10 phages are included at 0 metres depth at $t=0$. The second row shows the relative light intensity with depth, assuming unity at the surface with no cloud modification factor (CMF). The third row shows how the active species responds when using the CMF varying with time. The bottom row shows the propagation of light in the column when using the CMF varying with time.}
    \label{Light self shielding figure with phages}
\end{figure*}

Fig.~\ref{alpha L growth rate figure} shows the product $\alpha L$ which sets the growth rate for the active GSB population in Eq.~\eqref{Eq Active GSB} for both electron donors, sulfide and thiosulfate, assuming various parameters for temperature and light attenuation. In some cases, it is possible that no growth occurs in the uppermost parts of the ocean because it is simply too hot. When this is coupled with higher rates of light attenuation (e.g. $k_d\geq0.1$ m\textsuperscript{-1}) then no growth occurs throughout the vast majority of the water column. However, if the light attenuation coefficient is relatively low (clear water, $k_d=0.02$ m\textsuperscript{-1}), bacterial growth can occur down to 200 m in depth regardless of the modelled species, provided that the deeper waters have conducive temperatures for bacterial growth. This figure is illustrative and actual light and temperature profiles are specified in Table \ref{Paramater table}.

At $k_d=0.1$ m\textsuperscript{-1} and $k_d=0.7$ m\textsuperscript{-1}, by 10 m depth, the relative light intensity has decreased to $\approx 0.37$ and $\approx 0.001$ its initial value, respectively. In the more extreme case of $k_d=0.7$ m\textsuperscript{-1}, if light modulation is included from particulates, detritus, or as we have included, biological organisms themselves, then the bacterial species will be confined to depths $< 10$ m.

In Supplementary Material Fig.~2, we show how the total biological population responds to the bacteria in the column blocking incoming light. Upon introducing a CMF, the effect is to reduce the ability of the bacterial population to grow deeper in the water column (now reaching $\approx25$ m), simply because the CMF generally reduces incoming light. The randomness of the CMF produces a new instance of simulation results each time the model is run, but the population of GSB remain stable in each case. Once light shielding is introduced, the bacteria begin to grow but start to block out significant portions of light upon reaching $\sim 10^{7}$ cells ml\textsuperscript{-1} (where $k_\text{bio}=0.19$). Now, the bacteria population extends to only $\approx 5$ m instead of $\approx 30$ m.

% Phages with time GSB Thiosulfate figure
\begin{figure*}
	\centering
    \includegraphics[width=1\textwidth]{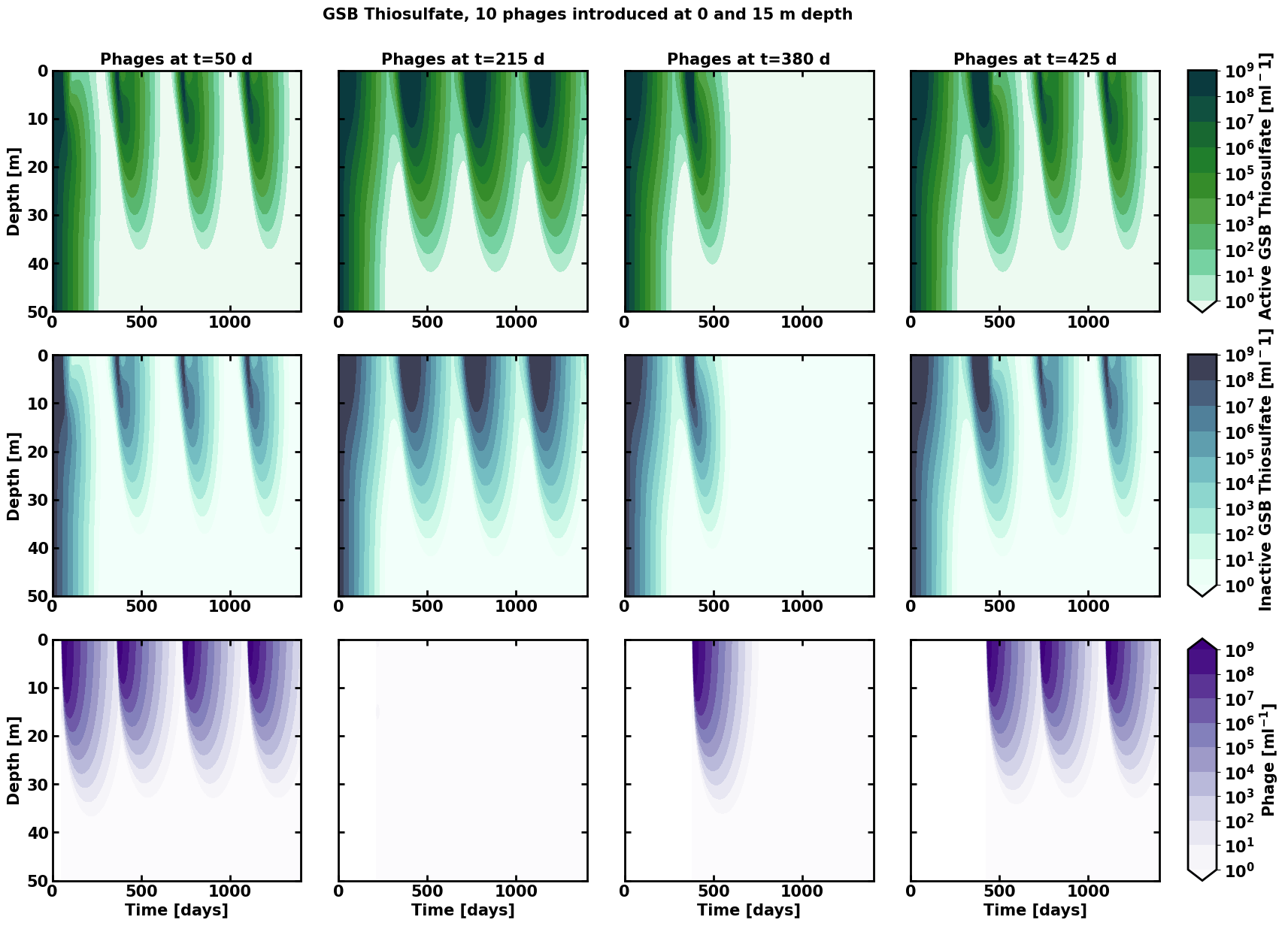}
    \caption{The effect of introducing phages at different times. A concentration of $10$ phages were introduced at $t=50$~d (first column), $t=215$~d (second column), $t=380$~d (fourth column), and $t=425$~d (fourth column) at 0 and 15 metres depth. The bacterial growth rates considered are for GSB grown on thiosulfate. In the first and fourth columns, the active population of GSB cells survive the introduction of the phages and continue to live with the phages. In the second column, the phages are introduced when there is a small amount of bacteria to consume and quickly die out. In the third column, the introduction of the phages causes the bacteria population to die out.}
    \label{Phages with time GSB Thiosulfate figure}
\end{figure*}

\subsection{Introducing consumers into the system}
\label{Phage results section}

Here we discuss the effect of including phages as consumers in the Hycean LV model. In the Ace Lake system (1 year seasonal period for returning light), once phages are introduced, they grow very rapidly and consume much, if not all of the active GSB. One reason the overall population can be saved from extinction is due to the inactive GSB population which is immune to phage attack.

Fig.~\ref{Light self shielding figure with phages} shows the introduction of ten phages ml\textsuperscript{-1} at the surface at $t=0$~d, with light shielding excluded and included in two models presented side-by-side. This is similar to Supplementary Material Fig.~2 (comparing light shielding and CMF effects) but instead with phages included. When light shielding is included, it can be seen by comparing Supplementary Material Fig.~2 to Fig.~\ref{Light self shielding figure with phages}, that the introduction of phages enables the bacterial population to survive at greater depths. In this simulation, the growth rates are maximized near the surface, and a substantial bacterial population builds up, blocking light from reaching greater depths. Introducing the phages results in the removal of a significant amount of this population near the surface, allowing light to penetrate further into the water column, enhancing the growth rates ($\alpha L$) at depth, and as the diffusivity spreads out the bacterial populations, then the bacteria can now grow and persist at greater depths.

The time at which the phages are introduced to the system can impact whether the bacterial population collapses or whether it is able to rebound. We explored the impact of introducing $10$ phages into a system at different times, with each of the different bacterial growth rates from Fig.~\ref{GSB growth rate figure}. For GSB grown on thiosulfate, we introduce $10$ phages ml\textsuperscript{-1} at 0 and 15 metres depth and at times between $t=0$~d and $t=550$~d. In Fig.~\ref{Phages with time GSB Thiosulfate figure}, we show the results for $t_{P,\rm in}=50$~d, $215$~d, $380$~d and $425$~d. For the GSB grown on thiosulfate species, if the phages are introduced at between 0--350 d, no collapse occurs. If the phages are introduced 350--390 d, the bacterial population is destroyed, and eventually the phages die too because there is no bacteria left to consume. If the phages are introduced at 90--290 d, the phages die out and the bacteria population reproduces without the presence of phages. Supplementary Material Figs.~11--13 show the same experiments but with GSB (sulfide), LSv21, and LSv514. With these simulations, the introduction of phages can also destroy the bacterial population or result in coexistence. These results are similar to previous work, where the time of introduction can affect whether a prey and predator are able to coexist or not \citep{lopez2005introduction}. 

\subsection{Competition between species}
\label{Competition between species}

% Competition model figure
\begin{figure*}
	\centering
	\includegraphics[width=1\textwidth]{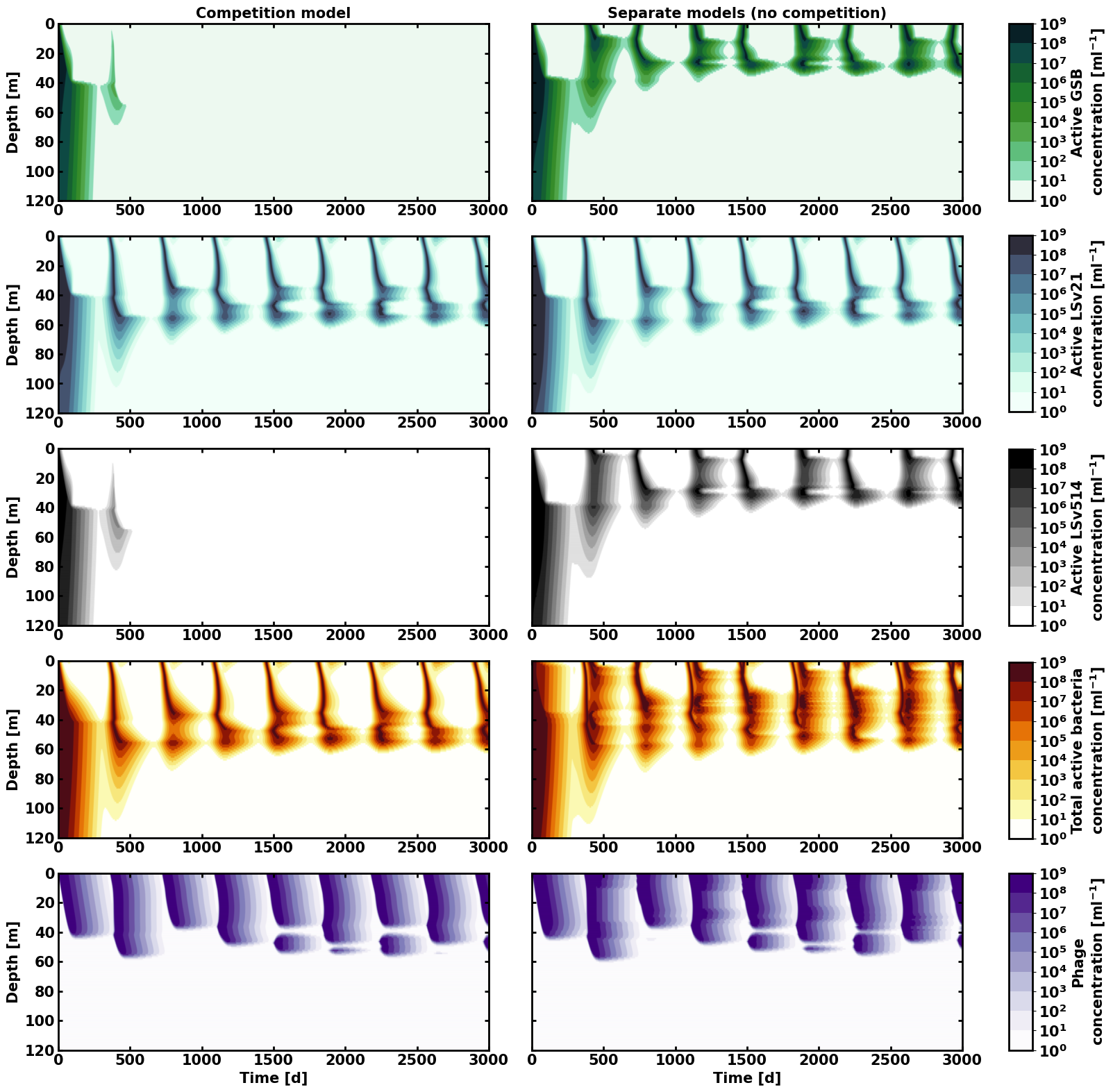}
    \caption{Models with and without competition between species are compared. The figure compares two systems, where five bacteria species are included in a competition model (left) with phages, and where each bacteria species is considered alone with no competition (right), also with phages. Included in the models are GSB grown on thiosulfate as an electron donor, ASv26, LSv21, LSv54, and LSv514. The GSB (top row), LSv21 (second row), and LSv514 (third row) species were able to survive in the multi-species system. Therefore, they are the only species shown. The summation of all active bacteria concentrations is shown in the fourth row. The phage concentration is shown in the bottom row.}
    \label{Competition model figure}
\end{figure*}

% Competition model figure
\begin{figure*}
	\centering
	\includegraphics[width=1\textwidth]{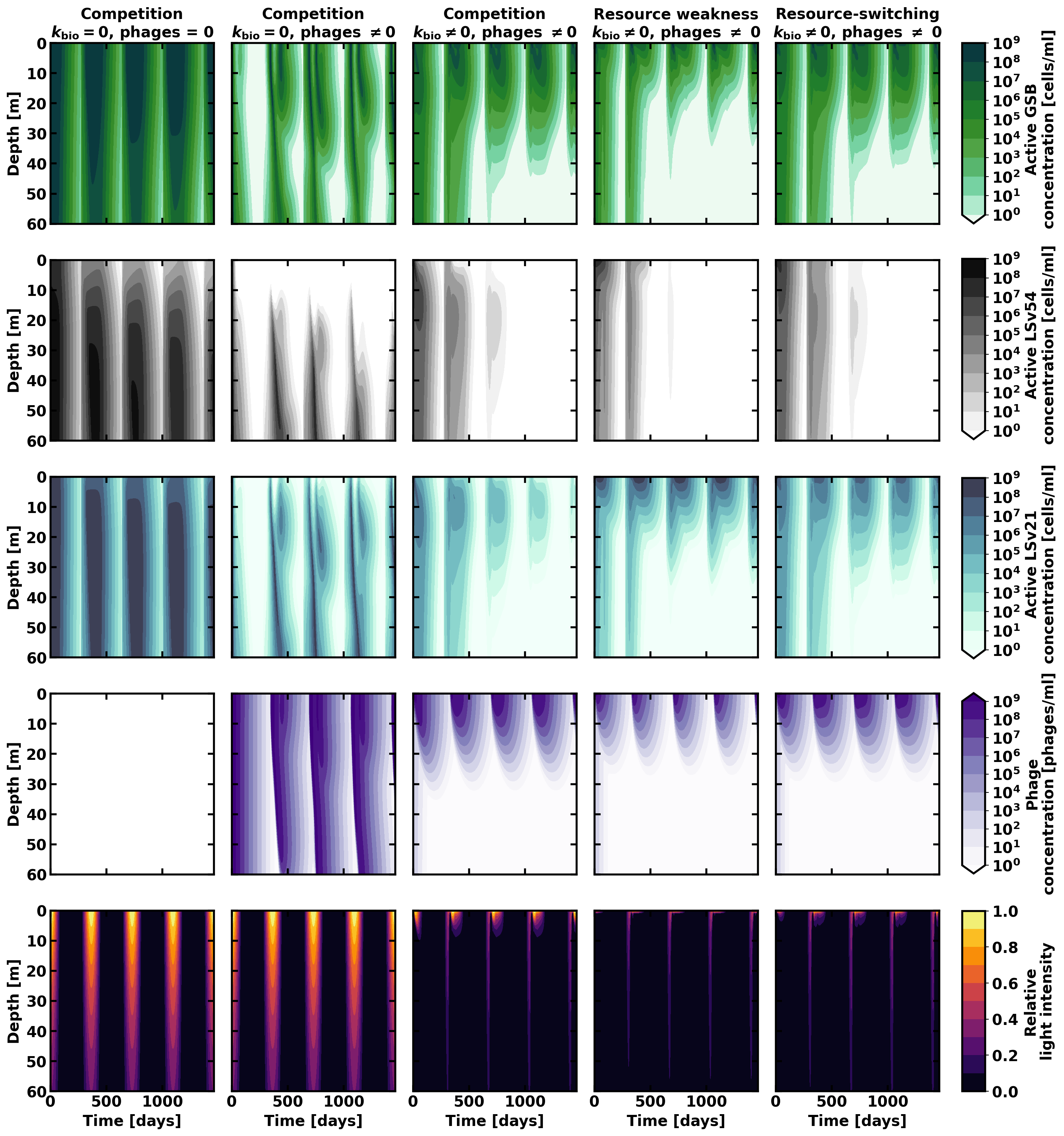}
    \caption{Five competition scenarios, each with three bacterial species, are compared. Phages and bioattenuation of light are included or excluded in a each scenario. The three bacteria species are: GSB with sulfide (top row), LSv54 (second row) and LSv21 (third row). GSB grown on sulfide are the fastest growers. Phages are not included in column one, but are included in the other columns. Bioattenuation is not included in column one and two, but is included in the other columns. In column four, the phages only attack the GSB and no other species - this is the Resource weakness model. In column five, the phages only attack the most abundant species, and switch resource depending on the column integrated abundance of active species - this is the Resource-switching model.}
    \label{Kill-the-winner figure}
\end{figure*}

Here we increase the amount of species that are inhabiting the water column in the Hycean LV model. Fig.~\ref{Competition model figure} shows a side-by-side comparison of the competition model (GSB grown on sulfide, ASv26, LSv21, LSv54, and LSv514) versus individual simulations where the bacteria are assumed to be alone, albeit predation with phages occurs. Three species, GSB, LSv21, and LSv54,  were able to survive in the water column simultaneously due to the selected temperature profile and the assumption of $k_\text{bio}=0$. When competition for space is included, LSv21 dominates and the other species die out. By changing the water parameters, it is possible to get a different amount of species growing and surviving in the same competitive system, as shown in Supplementary Material Fig.~15 and 16 where GSB and LSv21 survive. In such cases, there can be some spatial overlap between the species, but because they are competing for the same resource (space), and their growth rates are maximized at different depths in the system, each species tends to inhabit their own spatial niche in the water column. 

% GSB sulfur species orbital period figure
\begin{figure*}
	\centering
	\includegraphics[width=1\textwidth]{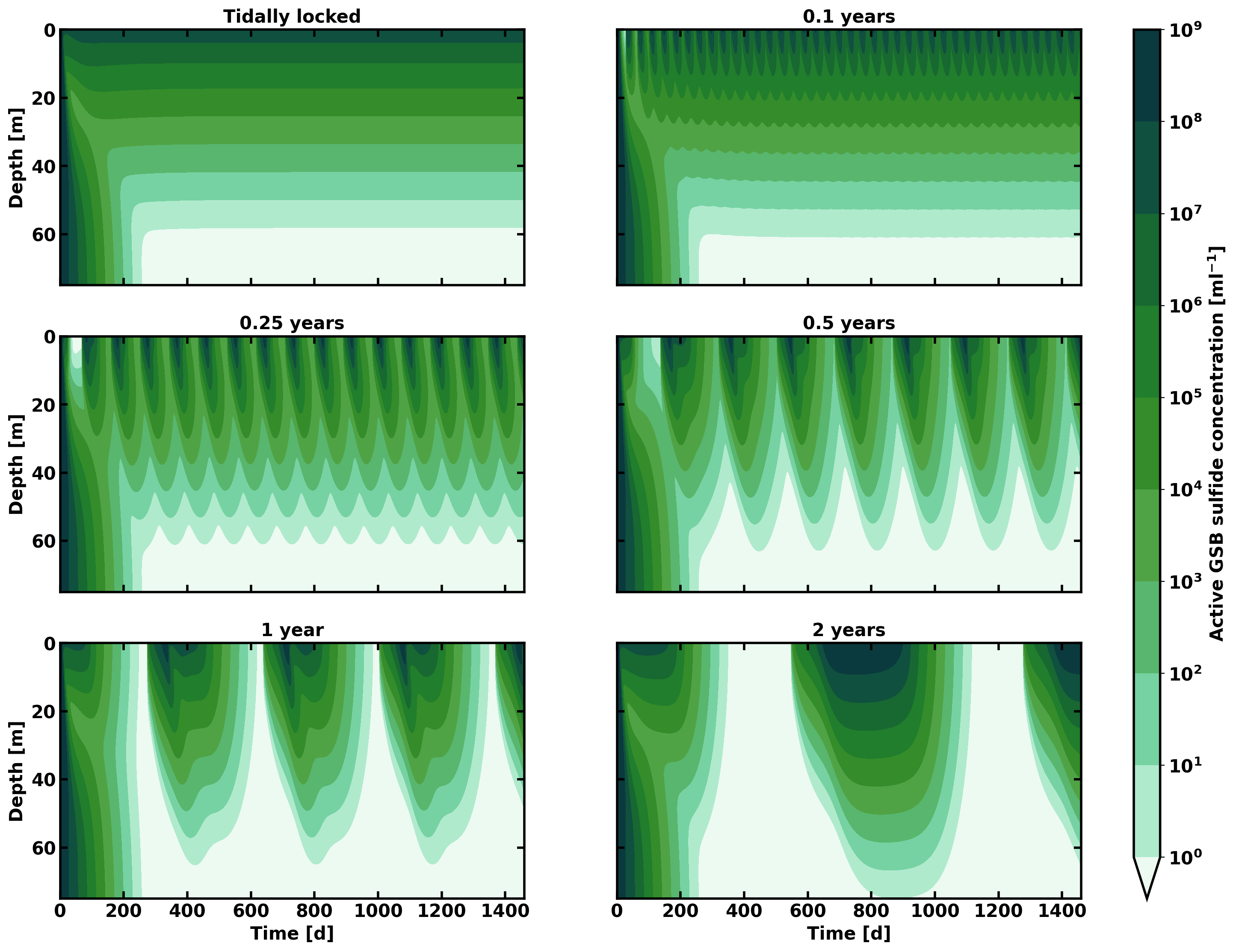}
    \caption{A comparison of varying seasonal periods with a constant illumination (tidally locked) scenario. The population density of green sulfur bacteria (GSB) grown on sulfide is shown for 6 different scenarios where the initial conditions are the same. The top left panel is a tidally locked scenario with constant illumination. The other panels show results for orbital periods ($\omega$) of 0.1, 0.25, 0.5, 1, and 2 yr orbital periods. The model is initialised with 100 phages per mililitre which are introduced at the surface (0 m depth).} 
    \label{GSB sulfur species orbital period figure}
\end{figure*}

Fig.~\ref{Kill-the-winner figure} shows five scenarios for a three-species (GSB on sulfide, LSv21, and LSv54) competition model. With $k_\text{bio}=0$ and no phages present, all three species coexist in the water column. Introducing phages allows all species to coexist, each with reduced total abundance. When biological attenuation is implemented ($k_\text{bio} \neq 0$), then only the fastest grower survives (GSB on sulfide) because it blocks out the light needed for growth lower down in the water column. Introducing phages that only predate the fastest growing population (Resource-weakness scenario) changes the competitive situation in the water column: once phages are introduced, they kill a significant portion of the population of GSB near the surface, which enables light to penetrate deeper into the water column. Now, LSv21 is able to grow and survive, although LSv54 does not survive because the light does not penetrate down to the depths where LSv54 would outcompete LSv21. A fifth scenario is simulated where the phages switch resource to whichever species is the most abundant in the water column. Similarly to the Resource Weakness scenario, LSv21 and GSB co-exist. Here, light penetrates further down because neither species reaches the carrying capacity of the system towards the top of the water column. This enables both species to have a greater vertical extent. Therefore, when comparing the five scenarios in Fig.~\ref{Kill-the-winner figure}, the presence of phages actually enables more species to survive than would otherwise be possible, providing bioattenuation of light is included. 

\subsection{Influence of planetary rotational period}
\label{Rotational period results section}

Here we consider constant illumination in a tidally locked scenario as well as various seasonal light cycles. In Fig.~\ref{GSB sulfur species orbital period figure}, we show the tidally locked scenario and the 0.1, 0.25, 0.5, 1.0, and 2.0 year orbital period simulations. Under similar water column temperature conditions with and without phages, all of these scenarios are able to sustain the GSB (grown on sulfide) bacterial population. For comparison purposes, all other species are shown in the Supplementary Material (Figs~17--29) both with and without phages. 

Phages are included in the different scenarios. As the orbital period is decreased from 2 years, the times between the peak density of bacteria consequently reduces. For periods $\geq 1$ yr, the active bacteria pass through a time period where the density is $<10$ cells ml\textsuperscript{-1} throughout the whole column. In the top 20 m of the column, for a period $\geq 0.5$ yr, the active density is $<100$ cells ml\textsuperscript{-1}.  In the shortest time period, the bacterial density is never less than $10^4$ cells ml\textsuperscript{-1}, although it is also never above $10^8$ cells ml\textsuperscript{-1} after the first 50 days at depths below the top 3 m of the water column. The tidally locked case does undergo oscillations, although they are smaller than the resolution of the plotted data, which is constantly between $10^7-10^8$ cells ml\textsuperscript{-1} in the top 4 m of the water column. This is essentially a quasi-equilibrium. Thus, decreasing orbital periods act to stabilize high GSB (sulfide) bacterial number densities in our model, yet a peak population close to the carrying capacity is never reached because phages are constantly present and consuming the resource bacterial species. The phages are consistently present in each scenario apart from in the 2 yr orbital period scenario, where they die because the active bacteria decrease in numbers in between seasonal light cycles.

Provided the water column parameters are changed to optimise species growth rates, all species can survive in the tidally locked case with and without phages. In Supplementary Material Fig.~20 demonstrates that a species can survive in the tidally locked cases with phages due to constant illumination and also survive in the $\geq 1$ yr orbital period cases, but eventually are wiped out in the periods $0.1\leq \omega\leq0.5$ yr. This is because the short orbital periods do not allow the species to reach the carrying capacity of the system before the light disappears and their growth rate reduces, such that they eventually go extinct.

\section{Summary and discussion}
\label{Conclusions section}
 
In this work, we presented the first set of dynamic ecological models for hycean worlds under a  Lotka-Volterra framework. Assuming a one-dimensional water column, we explored the influence of multiple factors on various populations of modelled anoxic bacteria. These included: temperature, light availability, seasonal periods or lack thereof, the introduction of bacteriophages, and competition for space and light. Our research opens up new avenues for considering how life may interact with its environment in distant and unfamiliar contexts.

The Lotka-Volterra equations are a powerful tool for Earth-based ecological research which we have extended to astroecology. In the model, temperature and light control the simulated biological growth, with a tendency for organisms to accumulate in distinct regions in the water column where their growth conditions are optimized. Applied to hycean conditions, our work has the following key conclusions:

\begin{enumerate}
    \item We find that a range of ecological diversity is possible for anaerobic organisms under assumed hycean world conditions.
    \item The modelled bacteria undergo less severe population oscillations in tidally locked hycean scenarios due to phages as a result of more `stable' growth conditions. Concurrently, their peak densities are smaller than those experienced when seasonal cycles are implemented because phages are constantly present and killing bacteria.
    \item Consumers (e.g. bacteriophages) can be beneficial to ecological diversity. If the parameters are optimal, then the fastest growing species are able to outcompete other bacteria which eventually die off, with this situation analogous to a bacterial bloom on Earth. In this case, phages can act to reduce the population of the fastest growers in a Resource Weakness scenario, where less competitive species which have harnessed stronger defence mechanisms begin to thrive.
    \item Bacterial population collapse is sensitive to the timing of a consumer invasion when seasonal cycles are considered. Consumers can take advantage of the oscillation present when including seasons to the point of ecological catastrophe.
    \item The photic zone around M dwarf stars will be considerably contracted compared to G dwarf stars at a given ocean turbidity if the peak emission of the star is the defining parameter for phototrophic organisms.
\end{enumerate}

This work presents a computational model of an ecosystem on an exoplanet and without a known target for comparison. Our results demonstrate some possibly important causal relationships for life beyond Earth, rather than distinct quantitative inferences. The piecewise model construction has enabled us to tease out the impact of a variety of theoretical exoplanetary parameters on the ecosystems we devised, such that our results convey how bacterial ecosystems may be hindered or bolstered by a specific set of planetary characteristics. 

We expect our work to be a springboard for future modelling studies which could expand on the complexity of the system we introduced. In what follows, we discuss our model in this context, and consider how the specified initial conditions, and the explicit resource limitations, influence the evolution of the chosen ecosystems and what the implications might be for extraterrestrial populations on a variety of habitable worlds. 

\subsection{Influence of predation}
\label{Influence of predation section}

Light attenuation during bacterial blooms can be significant \citep{moore2017bio, zhang2022influence} and affect organisms growing below the bloom, even if the bloom does not occur at the surface \citep{vadeboncoeur2021blue}. The third column of Fig.~\ref{Kill-the-winner figure} shows that when light is attenuated by the bacteria ($k_\text{bio} \neq 0$), with phages, the fastest growing species dominates at the top of the model. The high density of organisms reduces light penetration, thus restricting the populations of life in the lower water column. We also implemented a Resource-weakness model and Resource-switching model in Fig.~\ref{Kill-the-winner figure}, where the fastest growing (most dominant) species is attacked by the phages leaving the slower growing species ungrazed, and where the most abundant species are attacked, respectively. One of the reasons such scenarios can take place is that slower growers may have substituted rapid growth for greater defences against marine phages \citep{winter2010trade, vaage2014optimal, yang2023trade}.

These models are analogous to a bacterial bloom but where specific factors affect the competitive outcome. In the Resource weakness model, GSB (on sulfide) do not dominate temporally. Rather, their population reaches a maximum abundance before the slower grower LSv21 does, yet has reduced relative abundance once the phages start to graze upon the GSB, when LSv21 becomes the dominant species. This reduced abundance then occurs in a cyclical manner as light availability increases and recedes with the imposed season. Thus, additional species can survive in systems where otherwise the fastest grower would have dominated the ecosystem by absorbing all the light at the top of the column. However, we note that the fastest growing bacteria may be attacked during the onset of a bloom but resistance selection effects can still result in that species dominating the bloom regardless \citep{suttle2007marine}. 

\subsection{Initial conditions}

As with many types of numerical models, the initial conditions in a LV system are important for determining how the number density of species evolves with time \citep{wangersky1978lotka, gokhale2013lotka}.

For instance, in a simple competitive case, the initial conditions determines the outcome, whether that is coexistence of multiple species or a sole, dominant species \citep{holt1985infectious}. Furthermore, the amplitude and frequency of the LV cycles depends on the initial conditions \citep{gilpin1975limit, mougi2010evolution}, and competing species in LV systems can produce chaos \citep{vano2006chaos, bazeia2024chaotic} and Hopf bifurcations\footnote{A Hopf bifurcation is a bifurcation which results in the emergence or disappearance of a limit cycle (periodic oscillation).} \citep{lu2002two, wang2010bifurcations, li2021stability}. 

This work was a proof of concept, and as such we did not explore the full sensitivity of our results to all possible initial conditions. Initial conditions include the temperature profile of the ocean, the growth rates of various species, and the initial density of active and inactive bacteria and phages. The one-dimensional nature of our model, and the various parameter choices (e.g., carrying capacity, difussivity, etc.) means that the results are constrained by and restricted to the specified initial conditions. Using more complex models may significantly affect the results we present. Such models include the NPZD model, which tracks nutrients and detritus \citep{heinle2013internal}, or the LOBSTER \citep{Levy_Lobster_2005} and MARBL models \citep{Long2021_MARBL}, which track various nutrients (e.g., nitrates, phosphate, ammonia, etc.) and can be coupled to 3D ocean models. 3D models, in particular, will have a more accurate treatment of temporal and spatial variations for atmospheric winds and ocean currents, water temperatures, and the propagation of light to the ocean surface.

Of course, it is difficult to obtain initial conditions in the oceans of Earth, so ascertaining knowable initial conditions in oceans on other exoplanets is unfeasible. Nevertheless, there are foreseeable future scenarios which may allow us some constraints on initial conditions. Knowledge regarding surface temperatures and the presence or lack of oceans can theoretically be deduced \citep{2001Natur.412..885F, 2005AsBio...5..372S, Cowan2013Spectra, 2018haex.bookE.147C, 2022PSJ.....3...33R, 2022A&A...664A.172T, 2023MNRAS.524.5477V}, albeit requiring extremely precise and detailed observations of exoplanetary atmospheres and exoplanetary surfaces, which is currently beyond the limit of existing telescopes.

\subsection{Resource limitations}

One major possible limitation for phototrophic organisms living on habitable exoplanets (including hycean worlds) is that significant portions of light may not penetrate to the surface. For tidally locked worlds, the distribution of starlight depends on the latitude and longitude and not on the `time of day'. Light intensity is additionally modulated by the presence or lack of clouds that absorb and scatter incoming and outgoing radiation. Even if the surface irradiation is relatively high, light can be further limited at lower water column depths resulting from absorption of light as it passes through biological material.

`Hutchinson's paradox', or the `Paradox of the plankton' \citep{hutchinson1961paradox, calbet2024paradox}, a specific instance of the `Competitive exclusion principle' \citep{hardin1960competitive, armstrong1980competitive, britton1989aggregation, cropp2012mechanisms}, describes the situation where a large number of plankton species are able to survive in a habitat which has a small range of limiting resources \citep[e.g., carbon, light, nitrate;][]{hutchinson1961paradox, roy2007towards}. This paradox is likely resolved by a few different interacting factors, such as spatial-temporal variations due to water turbulence and mixing, as well as competitive species and consumer-resource interactions which introduce chaos \citep{scheffer2003plankton, roy2007towards, kozlov2013chaos}. 

The resulting restrictions on population growth due to a reduction in light intensity from shielding in a water column has been investigated before \citep{huisman1994light, weissing1994growth}, finding in some instances that at most two species could coexist \citep{huisman1995competition}. Our modified LV model finds a parameter space where three species can exist with no light attenuation from organisms, or two if light attenuation is included with phages also present. 

A further caveat to our model is assuming that light is the only required resource, and provided that the temperature is suitable, then bacteria can grow. In reality, various dissolved chemicals and nutrients are required for phototrophic organisms to survive in Earth's oceans, and this would presumably be the case in exoplanet oceans too. An extension of our model could track dissolved nutrients, influenced by the presence of an atmosphere above, and could be linked to planets where abundances have been constrained \citep[e.g., K2-18~b;][]{Madhu2023CarbonBearing, Madhu_DMS_2025, Hu_2025_K218b}.

Connected to the issue of resources is the `Paradox of Enrichment' \citep{1971Sci...171..385R, may1972limit, gilpin1972enriched}, where increasing the carrying capacity of the system enables the resource species to increase in number density, to the point where the consumer also increases exponentially until reaching instability, causing population collapse and extinction for the consumer. Whilst shown under specific laboratory conditions \citep{luckinbill1974effects, persson2001effects}, such a response to effectively unlimited resources has rarely been observed in nature \citep{vos2004inducible, jensen2005paradoxes}. Therefore, given the discrepancy between theory and observations, several theoretical resolutions have been proposed which suggest that mechanisms involving spatial effects, scrounging and social foraging, and prey switching, can play important roles \citep{jansen1995regulation, mougi2007resolution, toyokawa2017scrounging}. 

\subsection{Periodicity, seasonality and biodiversity}

Earth, which is the only known planet harboring life, has a seasonality induced by its axial tilt ($23.44^\circ$), with a small but non-zero orbital eccentricity ($0.0167$). Previous studies have found that both eccentricity and axial tilt \citep{Jernigan2023Superhabitability, 2024MNRAS.532.4511L}, as well as orbital period and rotation rate \citep[e.g., spin-orbit resonances and tidal locking;][]{2021ApJ...921...25C, 2025PSJ.....6....5B}, can affect the habitability metrics applied to simulations of exoplanet atmospheres, oceans, and surfaces.

Habitable zone hycean worlds around M dwarfs may be expected to be tidally locked, especially those closest to the inner edge \citep{2017CeMDA.129..509B, 2019AnRFM..51..275P}, whereas this might not be the case for Sun-like stars (e.g. G dwarf stars). Tidal interactions between a hycean planet on an eccentric orbit and a low mass M dwarf host star can also reduce the width of the habitable zone \citep{Livesey_2025_Tides}. To encompass the range of possibilities for habitable zone hyceans, we therefore tested a variety of orbital periods, including those longer than 2 years. There is an upper limit at which the bacterial population can grow to peak densities again when the light returns. Whether the population disappears in our model depends on the internal death rate of the species and the predation from phages, but also on the conversion rate between active and inactive populations, and the time the inactive population can survive in a dormant state.

We find that a diverse range of phototrophic organisms can be abundant around M dwarf hosts, which contrasts with potential expectations:   The peak radiation of M dwarf stars, being in the infrared, will have higher attenuation coefficients in water such that it may appear that M dwarf host stars could be unfavorable for phototrophic marine organisms as we know on Earth. In our models we have explored the range of Jerlov open ocean water types ($0.02 \leq k_d \leq 0.7$ m\textsuperscript{-1}), noting that several organisms photosynthesize in different portions of the visible electromagnetic spectrum. Despite this, phototrophs are known to be able to photoysthesize in extremely low light conditions, and healthy phytoplankton have been found below the photic zone, down to depths as far as 4,000 m \citep{platt1983photosynthetically, agusti2015ubiquitous, guo2018insight}. Provided that light propagates to the ocean surface in the first place, we also showed that species can grow and persist when $k_d = 0.7$ m\textsuperscript{-1}. Thus, phototrophic organisms perhaps won't be completely restricted to the smaller photic zone on M dwarfs and may instead thrive due to constant illumination on the dayside of any tidally locked exoplanet.

\subsection{Assumptions for biosphere modelling} 
\label{Plausibility of a hycean biosphere section}

Our work is focused on hycean exoplanets, but our model is applicable to a wide range of oceanic environments, including terrestrial worlds and super-Earths with oceans and lakes. The LV model we implement implicitly assumes that a habitable exoplanet has formed and life has emerged from abiogenesis. In the specific context of hycean worlds, our model assumes the presence of a deep planet-wide ocean under habitable conditions, due to an interior with a large \ce{H2O} reservoir beneath a \ce{H2}-rich envelope. Candidate hycean worlds have been nominally considered with \ce{H2O} mass fractions between 10-90\% \ce{H2O} \citep{Madhu2021Hycean, Rigby_Ocean_2024}, though even much smaller mass fractions, e.g. below 1\%, could lead to planet-wide oceans; for reference, the mass fraction of the Earth's oceans is below 0.1\%.

The emergence of water-rich sub-Neptunes has been predicted in formation models \citep{Alibert_2017_Formation, Mousis_2020_Ocean, Venturini_2020_Nature, Bitsch_2021_formation, Venturini_2024_Radius} without specifically targeting the plausibility of hycean exoplanet formation. However, open questions remain regarding the specific formation pathways for hycean worlds and how much water can be accreted and retained. For example, population synthesis studies show that in the presence of a magma ocean, atmospheric H$_2$ may react with oxidized species in the magma resulting in H$_2$O \citep{kimura2022predicted}. On the other hand, atmospheric loss from impacts coupled with a low delivery of volatiles for M dwarf planetary systems could significantly limit the initial water inventory \citep{Chen_2025_TRAPPIST1}. Whilst these studies are focused predominantly on rocky planets, similar effects may also be relevant for hycean worlds. \cite{Werlen_2025_Dry} use population synthesis models of sub-Neptunes with magma oceans to show that for planets starting with a H$_2$O mass fraction of 5-30\%, only 1.5\% is retained after equilibration with the magma. This work investigates the early stages of planet formation, presupposing the presence of a magma ocean without accounting for long-term cooling, differentiation and atmospheric reprocessing. However, \cite{Hu_2025_K218b} constrain an H$_2$O mass fraction in the envelope of sub-Neptune K2-18~b to be over tens of percent, which is inconsistent with this scenario. Future population synthesis models could explore more comprehensive models over a broader region of parameter space for forming hycean worlds. On the other hand, the water mass fraction may not need to be as high as originally predicted for some hycean candidates, because planets could instead accrete high quantities of carbon \citep{Li2025arXiv}, and this is able to explain the bulk density of some hycean candidates \citep{Lin_2025_Carbon}. 

Hycean atmospheres are expected to be composed primarily of \ce{H2} gas \citep{Madhu2021Hycean}, with other species such as \ce{CH4}, \ce{CO2}, and/or \ce{H2O} present in smaller quantities \citep{Madhu_2023_Faraday}. Many existing hycean candidates orbit towards the inner edge of the hycean habitable zone \citep{Madhu2021Hycean}. Due to convective inhibition, which arises from \ce{H2O} being heavier than the atmospheric mean molecular weight, hycean planets may require adequate Bond albedos to sustain a habitable ocean \citep{Innes_2023_hycean, Leconte_2024_K218b}. Recent global general circulation models (GCMs) show that for K2-18 b, a habitable ocean is permissible for a 1, 5, and 10 bar atmosphere with Bond albedos above 0.27, 0.35, and 0.48, respectively \citep{Barrier_2025_Hycean}. Such albedos may be caused by photochemical hazes on the dayside, as high albedos are unlikely to be attained from clouds alone \citep{Jordan_2025_Albedo, Barrier_2025_Hycean}. There is already observational evidence that sub-Neptunes posses hazes in their atmospheres, including for the exoplanets LP 791-18 c \citep{2025NatAs.tmp..256R} and GJ 1214 b \citep{gao2023hazy}. However, the observational data are currently not precise enough to confirm hazes or rule them out for the atmospheres of the hycean candidates of K2-18~b \citep{Madhu2023CarbonBearing, Barrier_2025_Hycean, Jordan_2025_Albedo} and TOI-270~d \citep{Holmberg2024hycean, Benneke2024TOI270d, Constantinou_2026_TOI270d}. For hycean worlds that may exist in longer orbital periods, significant bond albedos may not be required as the stellar instellation is much lower \citep{Madhu2021Hycean}.

A photosynthetic biosphere requires a sufficient flux of photosynthetically active radiation. The proportion of light reaching the ocean surface will depend on the incident spectral energy distribution (SED) from the host star, the atmospheric composition, and the atmospheric pressure. In global climate model simulations of K2-18~b down to 1~bar and 5~bar surface pressure, the shortwave radiation reaching the surface was only a small fraction of the top of atmospheric incoming radiation \citep{Barrier_2025_Hycean}. However, photosynthesis can occur under very dim conditions \citep{giordano2000photosynthesis, hoppe2024photosynthetic, Jung2023Light, futo2024ecophysiological}, so the low flux reaching the surface is unlikely to be a barrier for phototrophs. Whilst our simulations are not specifically tied to M dwarf host stars which have infrared wavelengths as their peak emission, organisms on Earth are able to photosynthesize beyond 700 nm \citep{kuhl2005niche, Mielke_2013_Photo}, with anoxygenic photosynthesis occurring up to $1015$ nm \citep{qian2018cryo}. Thus, the SED of M dwarf stars may not pose a barrier for life to thrive on orbiting exoplanets.

At the base of the global hycean ocean would be high-pressure water ice, rather than rock. This could create a nutrient-limited environment by sequestering nutrients in the rocky core, potentially resulting in low nutrient availability for ocean life. Whether transport can occur through such high-pressure ice layers through solid-state convection or melting is uncertain and the habitability of hycean exoplanets may depend on these processes \citep{Noack_2016_Water, 2018ApJ...864...75K, Seager_2021_Possibilities}. On the other hand, bioessential elements like phosphorus, sulfur, and refractory metals could be sufficiently supplied through asteroid impacts and atmospheric sedimentation \citep{Madhu_2023_Faraday}. Furthermore, entropy calculations have been performed for hycean candidates \citep{Petraccone_entropy_2024}, and bioenergetic calculations have already been performed for K2-18~b, showing that methanogenesis is thermodynamically feasible \citep{Glein_2024_K218b}. Overall, hycean exoplanets may possess the necessary physical and chemical conditions to sustain a biosphere similar to that of Earth's current anoxic waters, or that of the Archean Earth \citep{Madhu_2023_Faraday}, which existed 4 -- 2.4 billion years ago before global oxygenation \citep{catling2020archean}.

Generally, we have assumed temperate ocean conditions similar to Earth. However, higher temperatures can still result in liquid water, and many thermophilic organisms would be able to thrive at hotter temperatures \citep{fey2001thermophilic, stetter2006history, takai2008cell}, not to mention that biological evolution itself may occur more rapidly on hotter hycean worlds \citep{2025MNRAS.tmp...96M}. A few other physical parameters which can affect organism growth which we have not accounted for are pressure \citep{takai2008cell, kumar2013pressure, Michoud_2016}, gravity \citep{deguchi2011microbial, santomartino2020no, molecules31020229}, and water pH \citep{schonheit1980growth, franzmann1997methanogenium}. These are avenues for further research.

To summarize, provided hycean exoplanets can form and persist, then habitable environments for phototrophic organisms are possible, but future work using a 3D ocean model and explicitly tracking light through an atmosphere should be performed. Our models additionally operate under the assumption that the hycean planet is either sufficiently far away from its host star or that it has an adequate albedo and surface pressure to maintain a liquid water ocean. Overall, our work demonstrates the viability of microbial biodiversity in hycean worlds and ocean worlds over a wide range of host stellar environments. The ecological model we introduce for hycean worlds is the first step towards more general biological and ecological models in exoplanetary environments.

\section*{Acknowledgements}

G.J.C., N.M., and E. G. M., acknowledge support from “the Leverhulme Centre for Life in the Universe, the Leverhulme Trust, under Grant RC-2021-032.

\section*{Data Availability}

The data underlying this article can be found in the various cited publications. The data will be shared on reasonable request to the corresponding author. 

\bibliographystyle{mnras}
\bibliography{example} % if your bibtex file is called example.bib

@ARTICLE{Mousis_2020_Ocean,
       author = {{Mousis}, Olivier and {Deleuil}, Magali and {Aguichine}, Artyom and {Marcq}, Emmanuel and {Naar}, Joseph and {Aguirre}, Lorena Acu{\~n}a and {Brugger}, Bastien and {Gon{\c{c}}alves}, Thomas},
        title = "{Irradiated Ocean Planets Bridge Super-Earth and Sub-Neptune Populations}",
      journal = {\apjl},
         year = 2020,
        month = jun,
       volume = {896},
       number = {2},
          eid = {L22},
        pages = {L22},
          doi = {10.3847/2041-8213/ab9530},
archivePrefix = {arXiv},
       eprint = {2002.05243},
 primaryClass = {astro-ph.EP},
       adsurl = {https://ui.adsabs.harvard.edu/abs/2020ApJ...896L..22M}
}

@ARTICLE{Venturini_2020_Nature,
       author = {{Venturini}, Julia and {Guilera}, Octavio M. and {Haldemann}, Jonas and {Ronco}, Mar{\'\i}a P. and {Mordasini}, Christoph},
        title = "{The nature of the radius valley. Hints from formation and evolution models}",
      journal = {\aap},
         year = 2020,
        month = nov,
       volume = {643},
          eid = {L1},
        pages = {L1},
          doi = {10.1051/0004-6361/202039141},
archivePrefix = {arXiv},
       eprint = {2008.05513},
 primaryClass = {astro-ph.EP},
       adsurl = {https://ui.adsabs.harvard.edu/abs/2020A&A...643L...1V}
}

@ARTICLE{Alibert_2017_Formation,
       author = {{Alibert}, Y. and {Benz}, W.},
        title = "{Formation and composition of planets around very low mass stars}",
      journal = {\aap},
         year = 2017,
        month = feb,
       volume = {598},
          eid = {L5},
        pages = {L5},
          doi = {10.1051/0004-6361/201629671},
archivePrefix = {arXiv},
       eprint = {1610.03460},
 primaryClass = {astro-ph.EP},
       adsurl = {https://ui.adsabs.harvard.edu/abs/2017A&A...598L...5A}
}

@ARTICLE{Venturini_2024_Radius,
       author = {{Venturini}, J. and {Ronco}, M.~P. and {Guilera}, O.~M. and {Haldemann}, J. and {Mordasini}, C. and {Miller Bertolami}, M.},
        title = "{A fading radius valley towards M dwarfs, a persistent density valley across stellar types}",
      journal = {\aap},
         year = 2024,
        month = jun,
       volume = {686},
          eid = {L9},
        pages = {L9},
          doi = {10.1051/0004-6361/202349088},
archivePrefix = {arXiv},
       eprint = {2404.01967},
 primaryClass = {astro-ph.EP},
       adsurl = {https://ui.adsabs.harvard.edu/abs/2024A&A...686L...9V}
}

@ARTICLE{Bitsch_2021_formation,
       author = {{Bitsch}, Bertram and {Raymond}, Sean N. and {Buchhave}, Lars A. and {Bello-Arufe}, Aaron and {Rathcke}, Alexander D. and {Schneider}, Aaron David},
        title = "{Dry or water world? How the water contents of inner sub-Neptunes constrain giant planet formation and the location of the water ice line}",
      journal = {\aap},
         year = 2021,
        month = may,
       volume = {649},
          eid = {L5},
        pages = {L5},
          doi = {10.1051/0004-6361/202140793},
archivePrefix = {arXiv},
       eprint = {2104.11631},
 primaryClass = {astro-ph.EP},
       adsurl = {https://ui.adsabs.harvard.edu/abs/2021A&A...649L...5B}
}

@article{catling2020archean,
  title={The archean atmosphere},
  author={Catling, David C and Zahnle, Kevin J},
  journal={Science advances},
  volume={6},
  number={9},
  pages={eaax1420},
  year={2020},
  publisher={American Association for the Advancement of Science}
}

@ARTICLE{Petraccone_entropy_2024,
       author = {{Petraccone}, Luigi},
        title = "{Planetary entropy production as a thermodynamic constraint for exoplanet habitability}",
      journal = {\mnras},
         year = 2024,
        month = jan,
       volume = {527},
       number = {3},
        pages = {5547-5552},
          doi = {10.1093/mnras/stad3526},
       adsurl = {https://ui.adsabs.harvard.edu/abs/2024MNRAS.527.5547P}
}

@ARTICLE{Glein_2024_K218b,
       author = {{Glein}, Christopher R.},
        title = "{The Geochemical Potential for Metabolic Processes on the Sub-Neptune Exoplanet K2-18b}",
      journal = {\apjl},
         year = 2024,
        month = apr,
       volume = {964},
       number = {2},
          eid = {L19},
        pages = {L19},
          doi = {10.3847/2041-8213/ad3079},
archivePrefix = {arXiv},
       eprint = {2403.03918},
 primaryClass = {astro-ph.EP},
       adsurl = {https://ui.adsabs.harvard.edu/abs/2024ApJ...964L..19G}
}

@article{stetter2006history,
  title={History of discovery of the first hyperthermophiles},
  author={Stetter, Karl O},
  journal={Extremophiles},
  volume={10},
  number={5},
  pages={357--362},
  year={2006},
  publisher={Springer}
}

@article{fey2001thermophilic,
  title={Thermophilic methanogens in rice field soil},
  author={Fey, Axel and Chin, Kuk Jeong and Conrad, Ralf},
  journal={Environmental microbiology},
  volume={3},
  number={5},
  pages={295--303},
  year={2001},
  publisher={Wiley Online Library}
}

@article{deguchi2011microbial,
  title={Microbial growth at hyperaccelerations up to 403,627$\times$ g},
  author={Deguchi, Shigeru and Shimoshige, Hirokazu and Tsudome, Mikiko and Mukai, Sada-atsu and Corkery, Robert W and Ito, Susumu and Horikoshi, Koki},
  journal={Proceedings of the National Academy of Sciences},
  volume={108},
  number={19},
  pages={7997--8002},
  year={2011},
  publisher={National Academy of Sciences}
}

@Article{molecules31020229,
AUTHOR = {Rybacki, Bartosz and Wysocki, Wojciech and Zajkowski, Tomasz and Brodzik, Robert and Krawczyk, Beata},
TITLE = {Effects of Microgravity, Hypergravity, and Ionizing Radiation on the Enzymatic Activity of Proteinase K},
JOURNAL = {Molecules},
VOLUME = {31},
YEAR = {2026},
NUMBER = {2},
ARTICLE-NUMBER = {229},
URL = {https://www.mdpi.com/1420-3049/31/2/229},
ISSN = {1420-3049},
DOI = {10.3390/molecules31020229}
}

@article{santomartino2020no,
  title={No effect of microgravity and simulated Mars gravity on final bacterial cell concentrations on the International Space Station: applications to space bioproduction},
  author={Santomartino, Rosa and Waajen, Annemiek C and De Wit, Wessel and Nicholson, Natasha and Parmitano, Luca and Loudon, Claire-Marie and Moeller, Ralf and Rettberg, Petra and Fuchs, Felix M and Van Houdt, Rob and others},
  journal={Frontiers in microbiology},
  volume={11},
  pages={579156},
  year={2020},
  publisher={Frontiers Media SA}
}

@article{kumar2013pressure,
  title={Pressure and temperature dependence of growth and morphology of Escherichia coli: experiments and stochastic model},
  author={Kumar, Pradeep and Libchaber, Albert},
  journal={Biophysical journal},
  volume={105},
  number={3},
  pages={783--793},
  year={2013},
  publisher={Elsevier}
}

@ARTICLE{Michoud_2016,
       author = {{Michoud}, Gr{\'e}goire and {Jebbar}, Mohamed},
        title = "{High hydrostatic pressure adaptive strategies in an obligate piezophile Pyrococcus yayanosii}",
      journal = {Scientific Reports},
         year = 2016,
        month = jun,
       volume = {6},
          eid = {27289},
        pages = {27289},
          doi = {10.1038/srep27289},
       adsurl = {https://ui.adsabs.harvard.edu/abs/2016NatSR...627289M}
}

@article{futo2024ecophysiological,
  title={Ecophysiological characterisation of a Klebsormidium strain isolated from a cave environment},
  author={Fut{\'o}, P{\'e}ter and Lengyel, Edina and Fut{\'o}, M{\'a}t{\'e} and N{\'e}meth, Zolt{\'a}n and Pirger, Zsolt and Kom{\'a}romy, Andr{\'a}s and Padis{\'a}k, Judit and Felf{\"o}ldi, Tam{\'a}s and Kutasi, J{\'o}zsef and Bern{\'a}t, G{\'a}bor},
  journal={Journal of Applied Phycology},
  volume={36},
  number={3},
  pages={1197--1208},
  year={2024},
  publisher={Springer}
}

@article{giordano2000photosynthesis,
  title={Photosynthesis in the caves of Frasassi (Italy)},
  author={Giordano, Mario and Mobili, Francesco and Pezzoni, Valerio and Hein, Michael K and Davis, Joseph S},
  journal={Phycologia},
  volume={39},
  number={5},
  pages={384--389},
  year={2000},
  publisher={Taylor \& Francis}
}

@ARTICLE{Noack_2016_Water,
       author = {{Noack}, L. and {H{\"o}ning}, D. and {Rivoldini}, A. and {Heistracher}, C. and {Zimov}, N. and {Journaux}, B. and {Lammer}, H. and {Van Hoolst}, T. and {Bredeh{\"o}ft}, J.~H.},
        title = "{Water-rich planets: How habitable is a water layer deeper than on Earth?}",
      journal = {\icarus},
         year = 2016,
        month = oct,
       volume = {277},
        pages = {215-236},
          doi = {10.1016/j.icarus.2016.05.009},
       adsurl = {https://ui.adsabs.harvard.edu/abs/2016Icar..277..215N}
}

@ARTICLE{Seager_2021_Possibilities,
       author = {{Seager}, Sara and {Petkowski}, Janusz J. and {G{\"u}nther}, Maximilian N. and {Bains}, William and {Mikal-Evans}, Thomas and {Deming}, Drake},
        title = "{Possibilities for an Aerial Biosphere in Temperate Sub Neptune-Sized Exoplanet Atmospheres}",
      journal = {Universe},
         year = 2021,
        month = may,
       volume = {7},
       number = {6},
          eid = {172},
        pages = {172},
          doi = {10.3390/universe7060172},
archivePrefix = {arXiv},
       eprint = {2106.07729},
 primaryClass = {astro-ph.EP},
       adsurl = {https://ui.adsabs.harvard.edu/abs/2021Univ....7..172S}
}

@article{hoppe2024photosynthetic,
  title={Photosynthetic light requirement near the theoretical minimum detected in Arctic microalgae},
  author={Hoppe, Clara JM and Fuchs, Niels and Notz, Dirk and Anderson, Philip and Assmy, Philipp and Berge, J{\o}rgen and Bratbak, Gunnar and Guillou, Ga{\"e}l and Kraberg, Alexandra and Larsen, Aud and others},
  journal={Nature Communications},
  volume={15},
  number={1},
  pages={7385},
  year={2024},
  publisher={Nature Publishing Group UK London}
}

@article{qian2018cryo,
  title={Cryo-EM structure of the Blastochloris viridis LH1--RC complex at 2.9 {\AA}},
  author={Qian, Pu and Siebert, C Alistair and Wang, Peiyi and Canniffe, Daniel P and Hunter, C Neil},
  journal={Nature},
  volume={556},
  number={7700},
  pages={203--208},
  year={2018},
  publisher={Nature Publishing Group UK London}
}

@ARTICLE{Mielke_2013_Photo,
       author = {{Mielke}, Steven P. and {Kiang}, Nancy Y. and {Blankenship}, Robert E. and {Mauzerall}, David},
        title = "{Photosystem trap energies and spectrally-dependent energy-storage efficiencies in the Chl d-utilizing cyanobacterium, Acaryochloris marina}",
      journal = {Biochimica et Biophysica Acta (BBA) - Bioenergetics},
         year = 2013,
        month = mar,
       volume = {1827},
       number = {3},
        pages = {255-265},
          doi = {10.1016/j.bbabio.2012.11.002},
       adsurl = {https://ui.adsabs.harvard.edu/abs/2013BBAcB1827..255M}
}

@article{kuhl2005niche,
  title={A niche for cyanobacteria containing chlorophyll d},
  author={K{\"u}hl, Michael and Chen, Min and Ralph, Peter J and Schreiber, Ulrich and Larkum, Anthony WD},
  journal={Nature},
  volume={433},
  number={7028},
  pages={820--820},
  year={2005},
  publisher={Nature Publishing Group UK London}
}

@ARTICLE{Li2025arXiv,
       author = {{Li}, Jie and {Bergin}, Edwin A. and {Hirschmann}, Marc M. and {Blake}, Geoffrey A. and {Ciesla}, Fred J. and {Kempton}, Eliza M.-R.},
        title = "{Soot Planets instead of Water Worlds}",
      journal = {arXiv e-prints},
         year = 2025,
        month = aug,
          eid = {arXiv:2508.16781},
        pages = {arXiv:2508.16781},
          doi = {10.48550/arXiv.2508.16781},
archivePrefix = {arXiv},
       eprint = {2508.16781},
 primaryClass = {astro-ph.EP},
       adsurl = {https://ui.adsabs.harvard.edu/abs/2025arXiv250816781L}
}

@ARTICLE{Werlen_2025_Dry,
       author = {{Werlen}, Aaron and {Dorn}, Caroline and {Burn}, Remo and {Schlichting}, Hilke E. and {Grimm}, Simon L. and {Young}, Edward D.},
        title = "{Sub-Neptunes Are Drier than They Seem: Rethinking the Origins of Water-rich Worlds}",
      journal = {\apjl},
         year = 2025,
        month = sep,
       volume = {991},
       number = {1},
          eid = {L16},
        pages = {L16},
          doi = {10.3847/2041-8213/adff73},
archivePrefix = {arXiv},
       eprint = {2507.00765},
 primaryClass = {astro-ph.EP},
       adsurl = {https://ui.adsabs.harvard.edu/abs/2025ApJ...991L..16W}
}

@ARTICLE{Lin_2025_Carbon,
       author = {{Lin}, Zifan and {Seager}, Sara},
        title = "{Carbon-rich Sub-Neptune Interiors Are Compatible with JWST Observations}",
      journal = {\apjl},
         year = 2025,
        month = sep,
       volume = {990},
       number = {2},
          eid = {L35},
        pages = {L35},
          doi = {10.3847/2041-8213/adfcc8},
archivePrefix = {arXiv},
       eprint = {2508.15117},
 primaryClass = {astro-ph.EP},
       adsurl = {https://ui.adsabs.harvard.edu/abs/2025ApJ...990L..35L}
}

@ARTICLE{Chen_2025_TRAPPIST1,
       author = {{Chen}, Howard and {Clement}, Matthew S. and {Wang}, Le ''Chris'' and {Gu}, Jesse T.},
        title = "{Born Dry or Born Wet? A Palette of Water Growth Histories in TRAPPIST-1 Analogs and Compact Planetary Systems}",
      journal = {\apjl},
         year = 2025,
        month = sep,
       volume = {991},
       number = {1},
          eid = {L11},
        pages = {L11},
          doi = {10.3847/2041-8213/adf282},
archivePrefix = {arXiv},
       eprint = {2510.12794},
 primaryClass = {astro-ph.EP},
       adsurl = {https://ui.adsabs.harvard.edu/abs/2025ApJ...991L..11C}
}

@ARTICLE{Rigby_Ocean_2024,
       author = {{Rigby}, Frances E. and {Madhusudhan}, Nikku},
        title = "{On the ocean conditions of Hycean worlds}",
      journal = {\mnras},
         year = 2024,
        month = mar,
       volume = {529},
       number = {1},
        pages = {409-424},
          doi = {10.1093/mnras/stae413},
archivePrefix = {arXiv},
       eprint = {2402.12330},
 primaryClass = {astro-ph.EP},
       adsurl = {https://ui.adsabs.harvard.edu/abs/2024MNRAS.529..409R}
}

@ARTICLE{kimura2022predicted,
       author = {{Kimura}, Tadahiro and {Ikoma}, Masahiro},
        title = "{Predicted diversity in water content of terrestrial exoplanets orbiting M dwarfs}",
      journal = {Nature Astronomy},
         year = 2022,
        month = nov,
       volume = {6},
        pages = {1296-1307},
          doi = {10.1038/s41550-022-01781-1},
archivePrefix = {arXiv},
       eprint = {2209.14563},
 primaryClass = {astro-ph.EP},
       adsurl = {https://ui.adsabs.harvard.edu/abs/2022NatAs...6.1296K}
}

@ARTICLE{Constantinou_2026_TOI270d,
       author = {{Constantinou}, Savvas and {Madhusudhan}, Nikku and {Holmberg}, M{\r{a}}ns},
        title = "{The atmospheric composition of TOI-270 d}",
      journal = {\aap},
         year = 2026,
        month = jan,
       volume = {705},
          eid = {A25},
        pages = {A25},
          doi = {10.1051/0004-6361/202452192},
archivePrefix = {arXiv},
       eprint = {2511.13830},
 primaryClass = {astro-ph.EP},
       adsurl = {https://ui.adsabs.harvard.edu/abs/2026A&A...705A..25C}
}

@article{gao2023hazy,
  title={The hazy and metal-rich atmosphere of GJ 1214 b constrained by near-and mid-infrared transmission spectroscopy},
  author={Gao, Peter and Piette, Anjali AA and Steinrueck, Maria E and Nixon, Matthew C and Zhang, Michael and Kempton, Eliza M-R and Bean, Jacob L and Rauscher, Emily and Parmentier, Vivien and Batalha, Natasha E and others},
  journal={The Astrophysical Journal},
  volume={951},
  number={2},
  pages={96},
  year={2023},
  publisher={IOP Publishing}
}

@ARTICLE{Leconte_2024_K218b,
       author = {{Leconte}, J{\'e}r{\'e}my and {Spiga}, Aymeric and {Cl{\'e}ment}, No{\'e} and {Guerlet}, Sandrine and {Selsis}, Franck and {Milcareck}, Gwena{\"e}l and {Cavali{\'e}}, Thibault and {Moreno}, Rapha{\"e}l and {Lellouch}, Emmanuel and {Carri{\'o}n-Gonz{\'a}lez}, {\'O}scar and {Charnay}, Benjamin and {Lef{\`e}vre}, Maxence},
        title = "{A 3D picture of moist-convection inhibition in hydrogen-rich atmospheres: Implications for K2-18 b}",
      journal = {\aap},
         year = 2024,
        month = jun,
       volume = {686},
          eid = {A131},
        pages = {A131},
          doi = {10.1051/0004-6361/202348928},
archivePrefix = {arXiv},
       eprint = {2401.06608},
 primaryClass = {astro-ph.EP},
       adsurl = {https://ui.adsabs.harvard.edu/abs/2024A&A...686A.131L}
}

@article{tom2018experimental,
  title={Experimental evolution of UV resistance in a phage},
  author={Tom, Eric F and Molineux, Ian J and Paff, Matthew L and Bull, James J},
  journal={PeerJ},
  volume={6},
  pages={e5190},
  year={2018},
  publisher={PeerJ Inc.}
}

@ARTICLE{Jordan_2025_Albedo,
       author = {{Jordan}, Sean and {Shorttle}, Oliver and {Quanz}, Sascha P.},
        title = "{Planetary Albedo Is Limited by the Above-cloud Atmosphere: Implications for Sub-Neptune Climates}",
      journal = {\apj},
         year = 2025,
        month = nov,
       volume = {993},
       number = {1},
          eid = {86},
        pages = {86},
          doi = {10.3847/1538-4357/ae0192},
archivePrefix = {arXiv},
       eprint = {2504.12030},
 primaryClass = {astro-ph.EP},
       adsurl = {https://ui.adsabs.harvard.edu/abs/2025ApJ...993...86J}
}

@ARTICLE{Innes_2023_hycean,
       author = {{Innes}, Hamish and {Tsai}, Shang-Min and {Pierrehumbert}, Raymond T.},
        title = "{The Runaway Greenhouse Effect on Hycean Worlds}",
      journal = {\apj},
         year = 2023,
        month = aug,
       volume = {953},
       number = {2},
          eid = {168},
        pages = {168},
          doi = {10.3847/1538-4357/ace346},
archivePrefix = {arXiv},
       eprint = {2304.02698},
 primaryClass = {astro-ph.EP},
       adsurl = {https://ui.adsabs.harvard.edu/abs/2023ApJ...953..168I}
}

@article{zablocki2016diversity,
  title={Diversity and ecology of viruses in hyperarid desert soils},
  author={Zablocki, Olivier and Adriaenssens, Evelien M and Cowan, Don},
  journal={Applied and environmental microbiology},
  volume={82},
  number={3},
  pages={770--777},
  year={2016},
  publisher={American Society for Microbiology}
}

@article{hatfull2015dark,
  title={Dark matter of the biosphere: the amazing world of bacteriophage diversity},
  author={Hatfull, Graham F},
  journal={Journal of virology},
  volume={89},
  number={16},
  pages={8107--8110},
  year={2015},
  publisher={American Society for Microbiology 1752 N St., NW, Washington, DC}
}

@article{mushegian2020there,
  title={Are there 1031 virus particles on earth, or more, or fewer?},
  author={Mushegian, AR},
  journal={Journal of bacteriology},
  volume={202},
  number={9},
  pages={10--1128},
  year={2020},
  publisher={American Society for Microbiology 1752 N St., NW, Washington, DC}
}

@article{dion2020phage,
  title={Phage diversity, genomics and phylogeny},
  author={Dion, Mo{\"\i}ra B and Oechslin, Frank and Moineau, Sylvain},
  journal={Nature Reviews Microbiology},
  volume={18},
  number={3},
  pages={125--138},
  year={2020},
  publisher={Nature Publishing Group UK London}
}

@ARTICLE{Suttle2005_Viruses_sea,
       author = {{Suttle}, Curtis A.},
        title = "{Viruses in the sea}",
      journal = {\nat},
         year = 2005,
        month = sep,
       volume = {437},
       number = {7057},
        pages = {356-361},
          doi = {10.1038/nature04160},
       adsurl = {https://ui.adsabs.harvard.edu/abs/2005Natur.437..356S}
}

@article{karczewska2023tackle,
  title={How to tackle bacteriophages: the review of approaches with mechanistic insight},
  author={Karczewska, Monika and Strzelecki, Patryk and Szalewska-Pa{\l}asz, Agnieszka and Nowicki, Dariusz},
  journal={International Journal of Molecular Sciences},
  volume={24},
  number={5},
  pages={4447},
  year={2023},
  publisher={MDPI}
}

@article{hendrix2002bacteriophages,
  title={Bacteriophages: evolution of the majority},
  author={Hendrix, Roger W},
  journal={Theoretical population biology},
  volume={61},
  number={4},
  pages={471--480},
  year={2002},
  publisher={Elsevier}
}

@ARTICLE{2025NatAs.tmp..256R,
       author = {{Roy}, Pierre-Alexis and {Benneke}, Bj{\"o}rn and {Fournier-Tondreau}, Marylou and {Coulombe}, Louis-Philippe and {Piaulet-Ghorayeb}, Caroline and {Lafreni{\`e}re}, David and {Allart}, Romain and {Cowan}, Nicolas B. and {Dang}, Lisa and {Johnstone}, Doug and {Langeveld}, Adam B. and {Pelletier}, Stefan and {Radica}, Michael and {Taylor}, Jake and {Albert}, Lo{\"\i}c and {Doyon}, Ren{\'e} and {Flagg}, Laura and {Jayawardhana}, Ray and {MacDonald}, Ryan J. and {Turner}, Jake D.},
        title = "{Diversity in the haziness and chemistry of temperate sub-Neptunes}",
      journal = {Nature Astronomy},
         year = 2025,
        month = dec,
          doi = {10.1038/s41550-025-02723-3},
archivePrefix = {arXiv},
       eprint = {2512.10876},
 primaryClass = {astro-ph.EP},
       adsurl = {https://ui.adsabs.harvard.edu/abs/2025NatAs.tmp..256R}
}

@ARTICLE{Levy_Lobster_2005,
       author = {{L{\'e}Vy}, M. and {Gavart}, M. and {M{\'e}Mery}, L. and {Caniaux}, G. and {Paci}, A.},
        title = "{A four-dimensional mesoscale map of the spring bloom in the northeast Atlantic (POMME experiment): Results of a prognostic model}",
      journal = {Journal of Geophysical Research (Oceans)},
         year = 2005,
        month = jul,
       volume = {110},
       number = {C7},
          eid = {C07S21},
        pages = {C07S21},
          doi = {10.1029/2004JC002588},
       adsurl = {https://ui.adsabs.harvard.edu/abs/2005JGRC..110.7S21L}
}

@ARTICLE{Long2021_MARBL,
       author = {{Long}, Matthew C. and {Moore}, J. Keith and {Lindsay}, Keith and {Levy}, Michael and {Doney}, Scott C. and {Luo}, Jessica Y. and {Krumhardt}, Kristen M. and {Letscher}, Robert T. and {Grover}, Maxwell and {Sylvester}, Zephyr T.},
        title = "{Simulations With the Marine Biogeochemistry Library (MARBL)}",
      journal = {Journal of Advances in Modeling Earth Systems},
         year = 2021,
        month = dec,
       volume = {13},
       number = {12},
          eid = {e2021MS002647},
        pages = {e2021MS002647},
          doi = {10.1029/2021MS00264710.1002/essoar.10507358.1},
       adsurl = {https://ui.adsabs.harvard.edu/abs/2021JAMES..1302647L}
}

@article{schonheit1980growth,
  title={Growth parameters (K s, $\mu$max, Y s) of Methanobacterium thermoautotrophicum},
  author={Sch{\"o}nheit, Peter and Moll, Johanna and Thauer, Rudolf K},
  journal={Archives of Microbiology},
  volume={127},
  number={1},
  pages={59--65},
  year={1980},
  publisher={Springer}
}

@ARTICLE{Lai_2022_Thermocline,
       author = {{Lai}, Yanhong and {Yang}, Jun},
        title = "{Thermocline Depth on Water-rich Exoplanets}",
      journal = {\apj},
         year = 2022,
        month = jul,
       volume = {933},
       number = {2},
          eid = {152},
        pages = {152},
          doi = {10.3847/1538-4357/ac7221},
archivePrefix = {arXiv},
       eprint = {2206.12049},
 primaryClass = {astro-ph.EP},
       adsurl = {https://ui.adsabs.harvard.edu/abs/2022ApJ...933..152L}
}

@article{franzmann1997methanogenium,
  title={Methanogenium frigidum sp. nov., a psychrophilic, H2-using methanogen from Ace Lake, Antarctica},
  author={Franzmann, Peter D and Liu, Yitai and Balkwill, David L and Aldrich, Henry C and Conway De Macario, Everly and Boone, David R},
  journal={International Journal of Systematic and Evolutionary Microbiology},
  volume={47},
  number={4},
  pages={1068--1072},
  year={1997},
  publisher={Microbiology Society}
}

@ARTICLE{Gause_struggle_1936,
       author = {{Gause}, G.~F.},
        title = "{The Struggle for Existence}",
      journal = {Soil Science},
         year = 1936,
        month = feb,
       volume = {41},
       number = {2},
        pages = {159},
          doi = {10.1097/00010694-193602000-00018},
       adsurl = {https://ui.adsabs.harvard.edu/abs/1936SoilS..41..159G}
}

@article{takai2008cell,
  title={Cell proliferation at 122 C and isotopically heavy CH4 production by a hyperthermophilic methanogen under high-pressure cultivation},
  author={Takai, Ken and Nakamura, Kentaro and Toki, Tomohiro and Tsunogai, Urumu and Miyazaki, Masayuki and Miyazaki, Junichi and Hirayama, Hisako and Nakagawa, Satoshi and Nunoura, Takuro and Horikoshi, Koki},
  journal={Proceedings of the National Academy of Sciences},
  volume={105},
  number={31},
  pages={10949--10954},
  year={2008},
  publisher={National Academy of Sciences}
}

@ARTICLE{Barrier_2025_Hycean,
       author = {{Barrier}, Edouard F.~L. and {Madhusudhan}, Nikku},
        title = "{General circulation models of Hycean worlds}",
      journal = {\mnras},
         year = 2025,
        month = dec,
       volume = {544},
       number = {4},
        pages = {4098-4118},
          doi = {10.1093/mnras/staf1948},
archivePrefix = {arXiv},
       eprint = {2511.07546},
 primaryClass = {astro-ph.EP},
       adsurl = {https://ui.adsabs.harvard.edu/abs/2025MNRAS.544.4098B}
}

@article{marotzke1999convective,
  title={Convective mixing and the thermohaline circulation},
  author={Marotzke, Jochem and Scott, Jeffery R},
  journal={Journal of Physical Oceanography},
  volume={29},
  number={11},
  pages={2962--2970},
  year={1999}
}

@ARTICLE{Si_2022_planet,
       author = {{Si}, Yidongfang and {Yang}, Jun and {Liu}, Yonggang},
        title = "{Planetary climate under extremely high vertical diffusivity}",
      journal = {\aap},
         year = 2022,
        month = feb,
       volume = {658},
          eid = {A33},
        pages = {A33},
          doi = {10.1051/0004-6361/202140778},
archivePrefix = {arXiv},
       eprint = {2111.04947},
 primaryClass = {astro-ph.EP},
       adsurl = {https://ui.adsabs.harvard.edu/abs/2022A&A...658A..33S}
}

@ARTICLE{Marshall_1999_ocean,
       author = {{Marshall}, John and {Schott}, Friedrich},
        title = "{Open-ocean convection: Observations, theory, and models}",
      journal = {Reviews of Geophysics},
         year = 1999,
        month = feb,
       volume = {37},
       number = {1},
        pages = {1-64},
          doi = {10.1029/98RG02739},
       adsurl = {https://ui.adsabs.harvard.edu/abs/1999RvGeo..37....1M}
}

@ARTICLE{Marotzke_1991_ocean,
       author = {{Marotzke}, Jochem},
        title = "{Influence of Convective Adjustment on the Stability of the Thermohaline Circulation}",
      journal = {Journal of Physical Oceanography},
         year = 1991,
        month = jun,
       volume = {21},
       number = {6},
        pages = {903-913},
          doi = {10.1175/1520-0485(1991)021<0903:IOCAOT>2.0.CO;2},
       adsurl = {https://ui.adsabs.harvard.edu/abs/1991JPO....21..903M}
}

@ARTICLE{Waterhouse2014_Ocean,
       author = {{Waterhouse}, Amy F. and {MacKinnon}, Jennifer A. and {Nash}, Jonathan D. and {Alford}, Matthew H. and {Kunze}, Eric and {Simmons}, Harper L. and {Polzin}, Kurt L. and {St. Laurent}, Louis C. and {Sun}, Oliver M. and {Pinkel}, Robert and {Talley}, Lynne D. and {Whalen}, Caitlin B. and {Huussen}, Tycho N. and {Carter}, Glenn S. and {Fer}, Ilker and {Waterman}, Stephanie and {Naveira Garabato}, Alberto C. and {Sanford}, Thomas B. and {Lee}, Craig M.},
        title = "{Global Patterns of Diapycnal Mixing from Measurements of the Turbulent Dissipation Rate}",
      journal = {Journal of Physical Oceanography},
         year = 2014,
        month = jul,
       volume = {44},
       number = {7},
        pages = {1854-1872},
          doi = {10.1175/JPO-D-13-0104.1},
       adsurl = {https://ui.adsabs.harvard.edu/abs/2014JPO....44.1854W}
}

@ARTICLE{Wunsch2004Mixing,
       author = {{Wunsch}, Carl and {Ferrari}, Raffaele},
        title = "{Vertical Mixing, Energy, and the General Circulation of the Oceans}",
      journal = {Annual Review of Fluid Mechanics},
         year = 2004,
        month = jan,
       volume = {36},
        pages = {281-314},
          doi = {10.1146/annurev.fluid.36.050802.122121},
       adsurl = {https://ui.adsabs.harvard.edu/abs/2004AnRFM..36..281W}
}

@ARTICLE{Di_Paolo_2025_Ocean,
       author = {{Di Paolo}, Maria and {Stevens}, David P. and {Joshi}, Manoj and {Hall}, Rob A.},
        title = "{Nonlinear and Nonmonotonic Effect of Ocean Tidal Mixing on Exoplanet Climates and Habitability}",
      journal = {\apjl},
         year = 2025,
        month = apr,
       volume = {982},
       number = {2},
          eid = {L48},
        pages = {L48},
          doi = {10.3847/2041-8213/adbca3},
       adsurl = {https://ui.adsabs.harvard.edu/abs/2025ApJ...982L..48D}
}

@ARTICLE{Munk_Ocean_1998,
       author = {{Munk}, Walter and {Wunsch}, Carl},
        title = "{Abyssal recipes II: energetics of tidal and wind mixing}",
      journal = {Deep Sea Research Part I: Oceanographic Research},
         year = 1998,
        month = dec,
       volume = {45},
       number = {12},
        pages = {1977-2010},
          doi = {10.1016/S0967-0637(98)00070-3},
       adsurl = {https://ui.adsabs.harvard.edu/abs/1998DSRI...45.1977M}
}

@article{zhang2022influence,
  title={Influence of cyanobacterial bloom accumulation and dissipation on underwater light attenuation in a large and shallow lake},
  author={Zhang, Manxue and Zhang, Yunlin and Zhou, Yongqiang and Zhang, Yibo and Shi, Kun and Jiang, Cuiling},
  journal={Environmental Science and Pollution Research},
  volume={29},
  number={52},
  pages={79082--79094},
  year={2022},
  publisher={Springer}
}

@article{pomeroy2007microbial,
  title={The microbial loop},
  author={Pomeroy, Lawrence R and leB. WILLIAMS, PETER J and Azam, Farooq and Hobbie, John E},
  journal={Oceanography},
  volume={20},
  number={2},
  pages={28--33},
  year={2007},
  publisher={JSTOR}
}

@ARTICLE{Azam_MicrobialLoop_1983,
       author = {{Azam}, F. and {Fenchel}, T. and {Field}, JG and {Gray}, JS and {Meyer-Reil}, LA and {Thingstad}, F.},
        title = "{The Ecological Role of Water-Column Microbes in the Sea}",
      journal = {Marine Ecology Progress Series},
         year = 1983,
        month = jan,
       volume = {10},
        pages = {257-263},
          doi = {10.3354/meps010257},
       adsurl = {https://ui.adsabs.harvard.edu/abs/1983MEPS...10..257A}
}

@Inbook{Overmann2013,
author="Overmann, J{\"o}rg
and Garcia-Pichel, Ferran",
editor="Rosenberg, Eugene
and DeLong, Edward F.
and Lory, Stephen
and Stackebrandt, Erko
and Thompson, Fabiano",
title="The Phototrophic Way of Life",
bookTitle="The Prokaryotes: Prokaryotic Communities and Ecophysiology",
year="2013",
publisher="Springer Berlin Heidelberg",
address="Berlin, Heidelberg",
pages="203--257",
isbn="978-3-642-30123-0",
doi="10.1007/978-3-642-30123-0_51",
url="https://doi.org/10.1007/978-3-642-30123-0_51"
}

@article{princiotta2016temperature,
  title={Temperature-dependent phagotrophy and phototrophy in a mixotrophic chrysophyte},
  author={Princiotta, Sarah DeVaul and Smith, Brian T and Sanders, Robert W},
  journal={Journal of phycology},
  volume={52},
  number={3},
  pages={432--440},
  year={2016},
  publisher={Wiley Online Library}
}

@article{das2011enhanced,
  title={Enhanced algae growth in both phototrophic and mixotrophic culture under blue light},
  author={Das, Probir and Lei, Wang and Aziz, Siti Sarah and Obbard, Jeffrey Philip},
  journal={Bioresource technology},
  volume={102},
  number={4},
  pages={3883--3887},
  year={2011},
  publisher={Elsevier}
}

@article{parkin1980effects,
  title={The effects of light quality on the growth of phototrophic bacteria in lakes},
  author={Parkin, TB and Brock, TD},
  journal={Archives of microbiology},
  volume={125},
  number={1},
  pages={19--27},
  year={1980},
  publisher={Springer}
}

@ARTICLE{Ginsburg_Panspermia_2018,
       author = {{Ginsburg}, Idan and {Lingam}, Manasvi and {Loeb}, Abraham},
        title = "{Galactic Panspermia}",
      journal = {\apjl},
         year = 2018,
        month = nov,
       volume = {868},
       number = {1},
          eid = {L12},
        pages = {L12},
          doi = {10.3847/2041-8213/aaef2d},
archivePrefix = {arXiv},
       eprint = {1810.04307},
 primaryClass = {astro-ph.EP},
       adsurl = {https://ui.adsabs.harvard.edu/abs/2018ApJ...868L..12G}
}

@ARTICLE{Kova_2022_Astrobio,
       author = {{Kova{\v{c}}evi{\'c}}, Andjelka B.},
        title = "{On possible life-dispersal patterns beyond the Earth}",
      journal = {International Journal of Astrobiology},
         year = 2022,
        month = apr,
       volume = {21},
       number = {2},
        pages = {78-95},
          doi = {10.1017/S1473550421000379},
archivePrefix = {arXiv},
       eprint = {2202.07347},
 primaryClass = {astro-ph.EP},
       adsurl = {https://ui.adsabs.harvard.edu/abs/2022IJAsB..21...78K}
}

@ARTICLE{Forgan_SRP_2019,
       author = {{Forgan}, Duncan H.},
        title = "{Predator-prey behaviour in self-replicating interstellar probes}",
      journal = {International Journal of Astrobiology},
         year = 2019,
        month = dec,
       volume = {18},
       number = {6},
        pages = {552-561},
          doi = {10.1017/S1473550419000053},
archivePrefix = {arXiv},
       eprint = {1903.00770},
 primaryClass = {physics.pop-ph},
       adsurl = {https://ui.adsabs.harvard.edu/abs/2019IJAsB..18..552F}
}

@article{leger2004new,
  title={A new family of planets?“Ocean-Planets”},
  author={L{\'e}ger, Alain and Selsis, Franck and Sotin, Christophe and Guillot, Tristan and Despois, Didier and Mawet, D and Ollivier, Marc and Lab{\`e}que, A and Valette, C and Brachet, F and others},
  journal={Icarus},
  volume={169},
  number={2},
  pages={499--504},
  year={2004},
  publisher={Elsevier}
}

@ARTICLE{Kasting_HZ_1993,
       author = {{Kasting}, James F. and {Whitmire}, Daniel P. and {Reynolds}, Ray T.},
        title = "{Habitable Zones around Main Sequence Stars}",
      journal = {\icarus},
         year = 1993,
        month = jan,
       volume = {101},
       number = {1},
        pages = {108-128},
          doi = {10.1006/icar.1993.1010},
       adsurl = {https://ui.adsabs.harvard.edu/abs/1993Icar..101..108K}
}

@article{Madhu_PNAS_2025_Sub_JWST,
  title={Exploring the sub-Neptune frontier with JWST},
  author={Madhusudhan, Nikku and Holmberg, M{\aa}ns and Constantinou, Savvas and Cooke, Gregory J.},
  journal={Proceedings of the National Academy of Sciences},
  volume={122},
  number={39},
  pages={e2416194122},
  year={2025},
  doi={10.1073/pnas.2416194122},
  url={https://www.pnas.org/doi/10.1073/pnas.2416194122}
}

@article{vaage2014optimal,
  title={Optimal defense strategies in an idealized microbial food web under trade-off between competition and defense},
  author={V{\aa}ge, Selina and Storesund, Julia E and Giske, Jarl and Thingstad, T Frede},
  journal={PLoS One},
  volume={9},
  number={7},
  pages={e101415},
  year={2014},
  publisher={Public Library of Science San Francisco, USA}
}

@article{vadeboncoeur2021blue,
  title={Blue waters, green bottoms: Benthic filamentous algal blooms are an emerging threat to clear lakes worldwide},
  author={Vadeboncoeur, Yvonne and Moore, Marianne V and Stewart, Simon D and Chandra, Sudeep and Atkins, Karen S and Baron, Jill S and Bouma-Gregson, Keith and Brothers, Soren and Francoeur, Steven N and Genzoli, Laurel and others},
  journal={BioScience},
  volume={71},
  number={10},
  pages={1011--1027},
  year={2021},
  publisher={Oxford University Press}
}

@article{moore2017bio,
  title={Bio-optical properties of cyanobacteria blooms in western Lake Erie},
  author={Moore, Timothy S and Mouw, Colleen B and Sullivan, James M and Twardowski, Michael S and Burtner, Ashley M and Ciochetto, Audrey B and McFarland, Malcolm N and Nayak, Aditya R and Paladino, Danna and Stockley, Nicole D and others},
  journal={Frontiers in Marine Science},
  volume={4},
  pages={300},
  year={2017},
  publisher={Frontiers Media SA}
}

@article{emeliantsev2020quantification,
  title={Quantification of chlorosomal bacteriochlorophylls using absorption spectra of green sulfur bacteria in natural water},
  author={Emeliantsev, PS and Zhiltsova, AA and Krasnova, ED and Voronov, DA and Rymar, VV and Patsaeva, SV},
  journal={Moscow University Physics Bulletin},
  volume={75},
  pages={137--142},
  year={2020},
  publisher={Springer}
}

@article{manske2005physiology,
  title={Physiology and phylogeny of green sulfur bacteria forming a monospecific phototrophic assemblage at a depth of 100 meters in the Black Sea},
  author={Manske, Ann K and Glaeser, Jens and Kuypers, Marcel MM and Overmann, Jörg},
  journal={Applied and environmental microbiology},
  volume={71},
  number={12},
  pages={8049--8060},
  year={2005},
  publisher={American Society for Microbiology}
}

@article{chew2007bacteriochlorophyllide,
  title={Bacteriochlorophyllide c C-82 and C-121 methyltransferases are essential for adaptation to low light in Chlorobaculum tepidum},
  author={Chew, Aline Gomez Maqueo and Frigaard, Niels-Ulrik and Bryant, Donald A},
  journal={Journal of bacteriology},
  volume={189},
  number={17},
  pages={6176--6184},
  year={2007},
  publisher={American Society for Microbiology}
}

@article{laufer2017physiological,
  title={Physiological characterization of a halotolerant anoxygenic phototrophic Fe (II)-oxidizing green-sulfur bacterium isolated from a marine sediment},
  author={Laufer, Katja and Niemeyer, Annika and Nikeleit, Verena and Halama, Maximilian and Byrne, James M and Kappler, Andreas},
  journal={FEMS Microbiology Ecology},
  volume={93},
  number={5},
  pages={fix054},
  year={2017},
  publisher={Oxford University Press}
}

@article{frigaard2003chlorobium,
  title={Chlorobium tepidum: insights into the structure, physiology, and metabolism of a green sulfur bacterium derived from the complete genome sequence},
  author={Frigaard, Niels-Ulrik and Chew, Aline Gomez Maqueo and Li, Hui and Maresca, Julia A and Bryant, Donald A},
  journal={photosynthesis research},
  volume={78},
  pages={93--117},
  year={2003},
  publisher={Springer}
}

@article{vano2006chaos,
  title={Chaos in low-dimensional Lotka--Volterra models of competition},
  author={Vano, JA and Wildenberg, JC and Anderson, MB and Noel, JK and Sprott, JC},
  journal={Nonlinearity},
  volume={19},
  number={10},
  pages={2391},
  year={2006},
  publisher={IOP Publishing}
}

@article{overmann1992extremely,
  title={An extremely low-light adapted phototrophic sulfur bacterium from the Black Sea},
  author={Overmann, J{\"o}rg and Cypionka, Heribert and Pfennig, Norbert},
  journal={Limnology and oceanography},
  volume={37},
  number={1},
  pages={150--155},
  year={1992},
  publisher={Wiley Online Library}
}

@article{kim1980mechanism,
  title={Mechanism of ozone inactivation of bacteriophage f2},
  author={Kim, Chi K and Gentile, David M and Sproul, Otis J},
  journal={Applied and environmental microbiology},
  volume={39},
  number={1},
  pages={210--218},
  year={1980}
}

@article{kozlov2013chaos,
  title={On chaos in Lotka--Volterra systems: an analytical approach},
  author={Kozlov, Vladimir and Vakulenko, Sergey},
  journal={Nonlinearity},
  volume={26},
  number={8},
  pages={2299},
  year={2013},
  publisher={IOP Publishing}
}

@article{duran2002removal,
  title={Removal and inactivation of indicator bacteriophages in fresh waters},
  author={Duran, AE and Muniesa, M and Mendez, X and Valero, F and Lucena, F and Jofre, J},
  journal={Journal of applied microbiology},
  volume={92},
  number={2},
  pages={338--347},
  year={2002},
  publisher={Blackwell Science Ltd Oxford, UK}
}

@article{muller2005inactivation,
  title={Inactivation of bacteriophages by thermal and high-pressure treatment},
  author={M{\"u}ller-Merbach, M and Rauscher, T and Hinrichs, J},
  journal={International Dairy Journal},
  volume={15},
  number={6-9},
  pages={777--784},
  year={2005},
  publisher={Elsevier}
}

@article{luque2025insufficient,
  title={Insufficient evidence for DMS and DMDS in the atmosphere of K2-18 b. From a joint analysis of JWST NIRISS, NIRSpec, and MIRI observations},
  author={Luque, R and Piaulet-Ghorayeb, C and Radica, M and Xue, Q and Zhang, M and Bean, JL and Samra, D and Steinrueck, ME},
  journal={arXiv preprint arXiv:2505.13407},
  year={2025}
}

@ARTICLE{Schmidt2025_K2_18b,
       author = {{Schmidt}, Stephen P. and {MacDonald}, Ryan J. and {Tsai}, Shang-Min and {Radica}, Michael and {Wang}, Le-Chris and {Ahrer}, Eva-Maria and {Bell}, Taylor J. and {Fisher}, Chloe and {Thorngren}, Daniel P. and {Wogan}, Nicholas and {May}, Erin M. and {Ferrari}, Piero and {Bennett}, Katherine A. and {Rustamkulov}, Zafar and {L{\'o}pez-Morales}, Mercedes and {Sing}, David K.},
        title = "{A Comprehensive Reanalysis of K2-18 b's JWST NIRISS+NIRSpec Transmission Spectrum}",
      journal = {arXiv e-prints},
         year = 2025,
        month = jan,
          eid = {arXiv:2501.18477},
        pages = {arXiv:2501.18477},
          doi = {10.48550/arXiv.2501.18477},
archivePrefix = {arXiv},
       eprint = {2501.18477},
 primaryClass = {astro-ph.EP},
       adsurl = {https://ui.adsabs.harvard.edu/abs/2025arXiv250118477S}
}

@ARTICLE{Benneke2024TOI270d,
       author = {{Benneke}, Bj{\"o}rn and {Roy}, Pierre-Alexis and {Coulombe}, Louis-Philippe and {Radica}, Michael and {Piaulet}, Caroline and {Ahrer}, Eva-Maria and {Pierrehumbert}, Raymond and {Krissansen-Totton}, Joshua and {Schlichting}, Hilke E. and {Hu}, Renyu and {Yang}, Jeehyun and {Christie}, Duncan and {Thorngren}, Daniel and {Young}, Edward D. and {Pelletier}, Stefan and {Knutson}, Heather A. and {Miguel}, Yamila and {Evans-Soma}, Thomas M. and {Dorn}, Caroline and {Gagnebin}, Anna and {Fortney}, Jonathan J. and {Komacek}, Thaddeus and {MacDonald}, Ryan and {Raul}, Eshan and {Cloutier}, Ryan and {Acuna}, Lorena and {Lafreni{\`e}re}, David and {Cadieux}, Charles and {Doyon}, Ren{\'e} and {Welbanks}, Luis and {Allart}, Romain},
        title = "{JWST Reveals CH$_4$, CO$_2$, and H$_2$O in a Metal-rich Miscible Atmosphere on a Two-Earth-Radius Exoplanet}",
      journal = {arXiv e-prints},
         year = 2024,
        month = mar,
          eid = {arXiv:2403.03325},
        pages = {arXiv:2403.03325},
          doi = {10.48550/arXiv.2403.03325},
archivePrefix = {arXiv},
       eprint = {2403.03325},
 primaryClass = {astro-ph.EP},
       adsurl = {https://ui.adsabs.harvard.edu/abs/2024arXiv240303325B}
}

@article{luckinbill1974effects,
  title={The effects of space and enrichment on a predator-prey system},
  author={Luckinbill, Leo S},
  journal={Ecology},
  volume={55},
  number={5},
  pages={1142--1147},
  year={1974},
  publisher={Wiley Online Library}
}

@article{persson2001effects,
  title={Effects of enrichment on simple aquatic food webs},
  author={Persson, Anders and Hansson, Lars-Anders and Br{\"o}nmark, Christer and Lundberg, Per and Pettersson, Lars B and Greenberg, Larry and Nilsson, P Anders and Nystr{\"o}m, Per and Romare, Pia and Tranvik, Lars},
  journal={The American Naturalist},
  volume={157},
  number={6},
  pages={654--669},
  year={2001},
  publisher={The University of Chicago Press}
}

@article{vos2004inducible,
  title={Inducible defences and the paradox of enrichment},
  author={Vos, Matthijs and Kooi, Bob W and DeAngelis, Donald L and Mooij, Wolf M},
  journal={Oikos},
  volume={105},
  number={3},
  pages={471--480},
  year={2004},
  publisher={Wiley Online Library}
}

@article{jansen1995regulation,
  title={Regulation of predator-prey systems through spatial interactions: a possible solution to the paradox of enrichment},
  author={Jansen, Vincent AA},
  journal={Oikos},
  pages={384--390},
  year={1995},
  publisher={JSTOR}
}

@ARTICLE{Heath_Mdwarf_1999,
       author = {{Heath}, Martin J. and {Doyle}, Laurance R. and {Joshi}, Manoj M. and {Haberle}, Robert M.},
        title = "{Habitability of Planets Around Red Dwarf Stars}",
      journal = {Origins of Life and Evolution of the Biosphere},
         year = 1999,
        month = aug,
       volume = {29},
       number = {4},
        pages = {405-424},
          doi = {10.1023/A:1006596718708},
       adsurl = {https://ui.adsabs.harvard.edu/abs/1999OLEB...29..405H}
}

@article{toyokawa2017scrounging,
  title={Scrounging by foragers can resolve the paradox of enrichment},
  author={Toyokawa, Wataru},
  journal={Royal Society open science},
  volume={4},
  number={3},
  pages={160830},
  year={2017},
  publisher={The Royal Society Publishing}
}

@article{mougi2007resolution,
  title={A resolution of the paradox of enrichment},
  author={Mougi, Akihiko and Nishimura, Kinya},
  journal={Journal of Theoretical Biology},
  volume={248},
  number={1},
  pages={194--201},
  year={2007},
  publisher={Elsevier}
}

@article{jensen2005paradoxes,
  title={Paradoxes or theoretical failures? The jury is still out},
  author={Jensen, Christopher XJ and Ginzburg, Lev R},
  journal={Ecological Modelling},
  volume={188},
  number={1},
  pages={3--14},
  year={2005},
  publisher={Elsevier}
}

@article{britton1989aggregation,
  title={Aggregation and the competitive exclusion principle},
  author={Britton, Nicholas F},
  journal={Journal of theoretical biology},
  volume={136},
  number={1},
  pages={57--66},
  year={1989},
  publisher={Elsevier}
}

@article{cropp2012mechanisms,
  title={The mechanisms of coexistence and competitive exclusion in complex plankton ecosystem models},
  author={Cropp, Roger and Norbury, John},
  journal={Ecosystems},
  volume={15},
  pages={200--212},
  year={2012},
  publisher={Springer}
}

@article{armstrong1980competitive,
  title={Competitive exclusion},
  author={Armstrong, Robert A and McGehee, Richard},
  journal={The American Naturalist},
  volume={115},
  number={2},
  pages={151--170},
  year={1980},
  publisher={University of Chicago Press}
}

@article{hardin1960competitive,
  title={The competitive exclusion principle: an idea that took a century to be born has implications in ecology, economics, and genetics.},
  author={Hardin, Garrett},
  journal={science},
  volume={131},
  number={3409},
  pages={1292--1297},
  year={1960},
  publisher={American Association for the Advancement of Science}
}

@article{roy2007towards,
  title={Towards a resolution of ‘the paradox of the plankton’: A brief overview of the proposed mechanisms},
  author={Roy, Shovonlal and Chattopadhyay, J},
  journal={Ecological complexity},
  volume={4},
  number={1-2},
  pages={26--33},
  year={2007},
  publisher={Elsevier}
}

@article{hallegraeff2003harmful,
  title={Harmful algal blooms: a global overview},
  author={Hallegraeff, Gustaaf M},
  journal={Manual on harmful marine microalgae},
  volume={33},
  pages={1--22},
  year={2003},
  publisher={Unesco Paris}
}

@article{gilpin1972enriched,
  title={Enriched predator-prey systems: theoretical stability},
  author={Gilpin, Michael E},
  journal={Science},
  volume={177},
  number={4052},
  pages={902--904},
  year={1972},
  publisher={American Association for the Advancement of Science}
}

@article{may1972limit,
  title={Limit cycles in predator-prey communities},
  author={May, Robert M},
  journal={Science},
  volume={177},
  number={4052},
  pages={900--902},
  year={1972},
  publisher={American Association for the Advancement of Science}
}

@ARTICLE{1971Sci...171..385R,
       author = {{Rosenzweig}, Michael L.},
        title = "{Paradox of Enrichment: Destabilization of Exploitation Ecosystems in Ecological Time}",
      journal = {Science},
         year = 1971,
        month = jan,
       volume = {171},
       number = {3969},
        pages = {385-387},
          doi = {10.1126/science.171.3969.385},
       adsurl = {https://ui.adsabs.harvard.edu/abs/1971Sci...171..385R}
}

@article{eiler2004composition,
  title={Composition of freshwater bacterial communities associated with cyanobacterial blooms in four Swedish lakes},
  author={Eiler, Alexander and Bertilsson, Stefan},
  journal={Environmental Microbiology},
  volume={6},
  number={12},
  pages={1228--1243},
  year={2004},
  publisher={Wiley Online Library}
}

@article{paerl2013harmful,
  title={Harmful cyanobacterial blooms: causes, consequences, and controls},
  author={Paerl, Hans W and Otten, Timothy G},
  journal={Microbial ecology},
  volume={65},
  pages={995--1010},
  year={2013},
  publisher={Springer}
}

@article{huisman2018cyanobacterial,
  title={Cyanobacterial blooms},
  author={Huisman, Jef and Codd, Geoffrey A and Paerl, Hans W and Ibelings, Bas W and Verspagen, Jolanda MH and Visser, Petra M},
  journal={Nature Reviews Microbiology},
  volume={16},
  number={8},
  pages={471--483},
  year={2018},
  publisher={Nature Publishing Group UK London}
}

@article{scheffer2003plankton,
  title={Why plankton communities have no equilibrium: solutions to the paradox},
  author={Scheffer, Marten and Rinaldi, Sergio and Huisman, Jef and Weissing, Franz J},
  journal={Hydrobiologia},
  volume={491},
  pages={9--18},
  year={2003},
  publisher={Springer}
}

@article{hutchinson1961paradox,
  title={The paradox of the plankton},
  author={Hutchinson, G Evelyn},
  journal={The American Naturalist},
  volume={95},
  number={882},
  pages={137--145},
  year={1961},
  publisher={Science Press}
}

@incollection{turchin2013complex,
  title={Complex population dynamics: a theoretical/empirical synthesis (MPB-35)},
  author={Turchin, Peter},
  booktitle={Complex population dynamics},
  year={2013},
  publisher={Princeton university press}
}

@ARTICLE{Holmberg2024Hycean,
       author = {{Holmberg}, M{\r{a}}ns and {Madhusudhan}, Nikku},
        title = "{Possible Hycean conditions in the sub-Neptune TOI-270 d}",
      journal = {\aap},
         year = 2024,
        month = mar,
       volume = {683},
          eid = {L2},
        pages = {L2},
          doi = {10.1051/0004-6361/202348238},
archivePrefix = {arXiv},
       eprint = {2403.03244},
 primaryClass = {astro-ph.EP},
       adsurl = {https://ui.adsabs.harvard.edu/abs/2024A&A...683L...2H}
}

@article{zhiltsova2022comparative,
  title={Comparative analysis of spectral methods for determining bacteriochlorophyll d concentration in green sulfur bacteria in water},
  author={Zhiltsova, AA and Filippova, OA and Krasnova, ED and Voronov, DA and Patsaeva, SV},
  journal={Atmospheric and Oceanic Optics},
  volume={35},
  number={5},
  pages={562--568},
  year={2022},
  publisher={Springer}
}

@ARTICLE{Madhu2023CarbonBearing,
       author = {{Madhusudhan}, Nikku and {Sarkar}, Subhajit and {Constantinou}, Savvas and {Holmberg}, M{\r{a}}ns and {Piette}, Anjali A.~A. and {Moses}, Julianne I.},
        title = "{Carbon-bearing Molecules in a Possible Hycean Atmosphere}",
      journal = {\apjl},
         year = 2023,
        month = oct,
       volume = {956},
       number = {1},
          eid = {L13},
        pages = {L13},
          doi = {10.3847/2041-8213/acf577},
archivePrefix = {arXiv},
       eprint = {2309.05566},
 primaryClass = {astro-ph.EP},
       adsurl = {https://ui.adsabs.harvard.edu/abs/2023ApJ...956L..13M}
}

@ARTICLE{2025AtmRe.31607940S,
       author = {{Szymszov{\'a}}, Simona and {L{\'a}ska}, Kamil and {Kim}, Seong-Joong and {Park}, Sang-Jong},
        title = "{Variability of solar radiation and cloud cover in the Antarctic Peninsula region}",
      journal = {Atmospheric Research},
         year = 2025,
        month = apr,
       volume = {316},
          eid = {107940},
        pages = {107940},
          doi = {10.1016/j.atmosres.2025.107940},
       adsurl = {https://ui.adsabs.harvard.edu/abs/2025AtmRe.31607940S}
}

@article{otte1991comparative,
  title={A comparative study of the optical characteristics of intact cells of photosynthetic green sulfur bacteria containing bacteriochlorophyll c, d or e},
  author={Otte, Stephan CM and van der Heiden, Jos C and Pfennig, Norbert and Amesz, Jan},
  journal={Photosynthesis research},
  volume={28},
  pages={77--87},
  year={1991},
  publisher={Springer}
}

@article{kharcheva2016quantification,
  title={Quantification of two forms of green sulfur bacteria in their natural habitat using bacteriochlorophyll fluorescence spectra},
  author={Kharcheva, Anastasia V and Zhiltsova, Anna A and Lunina, Olga N and Savvichev, Alexander S and Patsaeva, Svetlana V},
  journal={Saratov Fall Meeting 2015: Third International Symposium on Optics and Biophotonics and Seventh Finnish-Russian Photonics and Laser Symposium (PALS)},
  volume={9917},
  pages={159--166},
  year={2016},
  organization={SPIE}
}

@article{lichtenthaler2001chlorophylls,
  title={Chlorophylls and carotenoids: Measurement and characterization by UV-VIS spectroscopy},
  author={Lichtenthaler, Hartmut K and Buschmann, Claus},
  journal={Current protocols in food analytical chemistry},
  volume={1},
  number={1},
  pages={F4--3},
  year={2001},
  publisher={Wiley Online Library}
}

@article{costa2017potential,
  title={Potential of Rhodobacter capsulatus grown in anaerobic-light or aerobic-dark conditions as bioremediation agent for biological wastewater treatments},
  author={Costa, Stefania and Ganzerli, Saverio and Rugiero, Irene and Pellizzari, Simone and Pedrini, Paola and Tamburini, Elena},
  journal={Water},
  volume={9},
  number={2},
  pages={108},
  year={2017},
  publisher={MDPI}
}

@article{musat2008single,
  title={A single-cell view on the ecophysiology of anaerobic phototrophic bacteria},
  author={Musat, Niculina and Halm, Hannah and Winterholler, B{\"a}rbel and Hoppe, Peter and Peduzzi, Sandro and Hillion, Francois and Horreard, Francois and Amann, Rudolf and J{\o}rgensen, Bo B and Kuypers, Marcel MM},
  journal={Proceedings of the National Academy of Sciences},
  volume={105},
  number={46},
  pages={17861--17866},
  year={2008},
  publisher={National Academy of Sciences}
}

@article{folt1989field,
  title={Field study comparing growth and viability of a population of phototrophic bacteria},
  author={Folt, Carol L and Wevers, Mary Jo and Yoder-Williams, Michael P and Howmiller, Richard P},
  journal={Applied and environmental microbiology},
  volume={55},
  number={1},
  pages={78--85},
  year={1989}
}

@article{eichler1988new,
  title={A new purple sulfur bacterium from stratified freshwater lakes, Amoebobacter purpureus sp. nov.},
  author={Eichler, Barbara and Pfennig, Norbert},
  journal={Archives of Microbiology},
  volume={149},
  pages={395--400},
  year={1988},
  publisher={Springer}
}

@article{schmidt1970variable,
  title={Variable cellular composition of Chromatium in growing cultures},
  author={Schmidt, Gregory L and Kamen, Martin D},
  journal={Archiv f{\"u}r Mikrobiologie},
  volume={73},
  pages={1--18},
  year={1970},
  publisher={Springer}
}

@article{knoblauch1999psychrophilic,
  title={Psychrophilic sulfate-reducing bacteria isolated from permanently cold Arctic marine sediments: description of Desulfofrigus oceanense gen. nov., sp. nov., Desulfofrigus fragile sp. nov., Desulfofaba gelida gen. nov., sp. nov., Desulfotalea psychrophila gen. nov., sp. nov. and Desulfotalea arctica sp. nov},
  author={Knoblauch, Christian and Sahm, Kerstin and J{\o}rgensen, Bo B},
  journal={International Journal of Systematic and Evolutionary Microbiology},
  volume={49},
  number={4},
  pages={1631--1643},
  year={1999},
  publisher={Microbiology Society}
}

@article{scholze2021psychrophilic,
  title={Psychrophilic properties of sulfate-reducing bacteria in Arctic marine sediments},
  author={Scholze, Caroline and J{\o}rgensen, Bo Barker and R{\o}y, Hans},
  journal={Limnology and Oceanography},
  volume={66},
  pages={S293--S302},
  year={2021},
  publisher={Wiley Online Library}
}

@article{biebl1978growth,
  title={Growth yields of green sulfur bacteria in mixed cultures with sulfur and sulfate reducing bacteria},
  author={Biebl, Hanno and Pfennig, Norbert},
  journal={Archives of microbiology},
  volume={117},
  pages={9--16},
  year={1978},
  publisher={Springer}
}

@article{virtanen2022chlorophyll,
  title={Chlorophyll does not reflect green light--how to correct a misconception},
  author={Virtanen, Olli and Constantinidou, Emanuella and Tyystj{\"a}rvi, Esa},
  journal={Journal of Biological Education},
  volume={56},
  number={5},
  pages={552--559},
  year={2022},
  publisher={Taylor \& Francis}
}

@article{bricaud1998variations,
  title={Variations of light absorption by suspended particles with chlorophyll a concentration in oceanic (case 1) waters: Analysis and implications for bio-optical models},
  author={Bricaud, Annick and Morel, Andr{\'e} and Babin, Marcel and Allali, Karima and Claustre, Herv{\'e}},
  journal={Journal of Geophysical Research: Oceans},
  volume={103},
  number={C13},
  pages={31033--31044},
  year={1998},
  publisher={Wiley Online Library}
}

@article{nelson1993phytoplankton,
  title={Phytoplankton light absorption and the package effect in California coastal waters},
  author={Nelson, Norman B and Pr{\'e}zelin, Barbara B and Bidigare, Robert R},
  journal={Marine Ecology Progress Series},
  pages={217--227},
  year={1993},
  publisher={JSTOR}
}

@article{weissing1994growth,
  title={Growth and competition in a light gradient},
  author={Weissing, Franz J and Huisman, Jef},
  journal={Journal of theoretical biology},
  volume={168},
  number={3},
  pages={323--336},
  year={1994},
  publisher={Elsevier}
}

@article{huisman1994light,
  title={Light-limited growth and competition for light in well-mixed aquatic environments: an elementary model},
  author={Huisman, Jef and Weissing, Franz J},
  journal={Ecology},
  volume={75},
  number={2},
  pages={507--520},
  year={1994},
  publisher={Wiley Online Library}
}

@article{huisman1995competition,
  title={Competition for nutrients and light in a mixed water column: a theoretical analysis},
  author={Huisman, Jef and Weissing, Franz J},
  journal={The American Naturalist},
  volume={146},
  number={4},
  pages={536--564},
  year={1995},
  publisher={University of Chicago Press}
}

@article{kostakis2020development,
  title={Development of a bio-optical model for the Barents Sea to quantitatively link glider and satellite observations},
  author={Kostakis, Ina and R{\"o}ttgers, R{\"u}diger and Orkney, Andrew and Bouman, Heather A and Porter, Marie and Cottier, Finlo and Berge, J{\o}rgen and McKee, David},
  journal={Philosophical Transactions of the Royal Society A},
  volume={378},
  number={2181},
  pages={20190367},
  year={2020},
  publisher={The Royal Society Publishing}
}

@ARTICLE{2017AN....338..413S,
       author = {{Sato}, S. and {Wang}, Zh. and {Cuntz}, M.},
        title = "{Climatological and ultraviolet-based habitability of possible exomoons in F-star systems}",
      journal = {Astronomische Nachrichten},
         year = 2017,
        month = may,
       volume = {338},
       number = {4},
        pages = {413-427},
          doi = {10.1002/asna.201613279},
archivePrefix = {arXiv},
       eprint = {1503.02560},
 primaryClass = {astro-ph.SR},
       adsurl = {https://ui.adsabs.harvard.edu/abs/2017AN....338..413S}
}

@ARTICLE{Madhu_DMS_2025,
       author = {{Madhusudhan}, Nikku and {Constantinou}, Savvas and {Holmberg}, M{\r{a}}ns and {Sarkar}, Subhajit and {Piette}, Anjali A.~A. and {Moses}, Julianne I.},
        title = "{New Constraints on DMS and DMDS in the Atmosphere of K2-18 b from JWST MIRI}",
      journal = {\apjl},
         year = 2025,
        month = apr,
       volume = {983},
       number = {2},
          eid = {L40},
        pages = {L40},
          doi = {10.3847/2041-8213/adc1c8},
archivePrefix = {arXiv},
       eprint = {2504.12267},
 primaryClass = {astro-ph.EP},
       adsurl = {https://ui.adsabs.harvard.edu/abs/2025ApJ...983L..40M}
}

@ARTICLE{Hu_2025_K218b,
       author = {{Hu}, Renyu and {Bello-Arufe}, Aaron and {Tokadjian}, Armen and {Yang}, Jeehyun and {Damiano}, Mario and {Roy}, Pierre-Alexis and {Coulombe}, Louis-Philippe and {Madhusudhan}, Nikku and {Constantinou}, Savvas and {Benneke}, Bj{\"o}rn},
        title = "{A water-rich interior in the temperate sub-Neptune K2-18 b revealed by JWST}",
      journal = {arXiv e-prints},
         year = 2025,
        month = jul,
          eid = {arXiv:2507.12622},
        pages = {arXiv:2507.12622},
          doi = {10.48550/arXiv.2507.12622},
archivePrefix = {arXiv},
       eprint = {2507.12622},
 primaryClass = {astro-ph.EP},
       adsurl = {https://ui.adsabs.harvard.edu/abs/2025arXiv250712622H}
}

@ARTICLE{Kopparapu_HZ_2013,
       author = {{Kopparapu}, Ravi Kumar and {Ramirez}, Ramses and {Kasting}, James F. and {Eymet}, Vincent and {Robinson}, Tyler D. and {Mahadevan}, Suvrath and {Terrien}, Ryan C. and {Domagal-Goldman}, Shawn and {Meadows}, Victoria and {Deshpande}, Rohit},
        title = "{Habitable Zones around Main-sequence Stars: New Estimates}",
      journal = {\apj},
         year = 2013,
        month = mar,
       volume = {765},
       number = {2},
          eid = {131},
        pages = {131},
          doi = {10.1088/0004-637X/765/2/131},
archivePrefix = {arXiv},
       eprint = {1301.6674},
 primaryClass = {astro-ph.EP},
       adsurl = {https://ui.adsabs.harvard.edu/abs/2013ApJ...765..131K}
}

@article{o2009perturbations,
  title={Perturbations to trophic interactions and the stability of complex food webs},
  author={O'Gorman, Eoin J and Emmerson, Mark C},
  journal={Proceedings of the National Academy of Sciences},
  volume={106},
  number={32},
  pages={13393--13398},
  year={2009},
  publisher={National Academy of Sciences}
}

@article{griffin2011predator,
  title={Predator diversity stabilizes and strengthens trophic control of a keystone grazer},
  author={Griffin, John N and Silliman, Brian R},
  journal={Biology letters},
  volume={7},
  number={1},
  pages={79--82},
  year={2011},
  publisher={The Royal Society}
}

@article{bengil2016bio,
  title={A bio-optical model for integration into ecosystem models for the Ligurian Sea},
  author={Bengil, Fethi and McKee, David and Be{\c{s}}iktepe, S{\"u}kr{\"u} T and Calzado, Violeta Sanjuan and Trees, Charles},
  journal={Progress in Oceanography},
  volume={149},
  pages={1--15},
  year={2016},
  publisher={Elsevier}
}

@article{rosenzweig1963graphical,
  title={Graphical representation and stability conditions of predator-prey interactions},
  author={Rosenzweig, Michael L and MacArthur, Robert H},
  journal={The American Naturalist},
  volume={97},
  number={895},
  pages={209--223},
  year={1963},
  publisher={Science Press}
}

@article{vogl2012bacteriochlorophyll,
  title={Bacteriochlorophyll f: properties of chlorosomes containing the “forbidden chlorophyll”},
  author={Vogl, Kajetan and Tank, Marcus and Orf, Gregory S and Blankenship, Robert E and Bryant, Donald A},
  journal={Frontiers in Microbiology},
  volume={3},
  pages={298},
  year={2012},
  publisher={Frontiers Media SA}
}

@ARTICLE{2023NatAs...7.1070Y,
       author = {{Yang}, Jun and {Zhang}, Yixiao and {Fu}, Zuntao and {Yan}, Mingyu and {Song}, Xinyi and {Wei}, Mengyu and {Liu}, Jiachen and {Ding}, Feng and {Tan}, Zhihong},
        title = "{Cloud behaviour on tidally locked rocky planets from global high-resolution modelling}",
      journal = {Nature Astronomy},
         year = 2023,
        month = sep,
       volume = {7},
        pages = {1070-1080},
          doi = {10.1038/s41550-023-02015-8},
archivePrefix = {arXiv},
       eprint = {2306.12186},
 primaryClass = {astro-ph.EP},
       adsurl = {https://ui.adsabs.harvard.edu/abs/2023NatAs...7.1070Y}
}

@article{treguier2023mixed,
  title={The mixed-layer depth in the Ocean Model Intercomparison Project (OMIP): impact of resolving mesoscale eddies},
  author={Treguier, Anne Marie and de Boyer Mont{\'e}gut, Clement and Bozec, Alexandra and Chassignet, Eric P and Fox-Kemper, Baylor and McC. Hogg, Andy and Iovino, Doroteaciro and Kiss, Andrew E and Le Sommer, Julien and Li, Yiwen and others},
  journal={Geoscientific Model Development},
  volume={16},
  number={13},
  pages={3849--3872},
  year={2023},
  publisher={Copernicus Publications G{\"o}ttingen, Germany}
}

@ARTICLE{2017A&A...601A.120B,
       author = {{Boutle}, Ian A. and {Mayne}, Nathan J. and {Drummond}, Benjamin and {Manners}, James and {Goyal}, Jayesh and {Hugo Lambert}, F. and {Acreman}, David M. and {Earnshaw}, Paul D.},
        title = "{Exploring the climate of Proxima B with the Met Office Unified Model}",
      journal = {\aap},
         year = 2017,
        month = may,
       volume = {601},
          eid = {A120},
        pages = {A120},
          doi = {10.1051/0004-6361/201630020},
archivePrefix = {arXiv},
       eprint = {1702.08463},
 primaryClass = {astro-ph.EP},
       adsurl = {https://ui.adsabs.harvard.edu/abs/2017A&A...601A.120B}
}

@ARTICLE{2013ApJ...771L..45Y,
       author = {{Yang}, Jun and {Cowan}, Nicolas B. and {Abbot}, Dorian S.},
        title = "{Stabilizing Cloud Feedback Dramatically Expands the Habitable Zone of Tidally Locked Planets}",
      journal = {\apjl},
         year = 2013,
        month = jul,
       volume = {771},
       number = {2},
          eid = {L45},
        pages = {L45},
          doi = {10.1088/2041-8205/771/2/L45},
archivePrefix = {arXiv},
       eprint = {1307.0515},
 primaryClass = {astro-ph.EP},
       adsurl = {https://ui.adsabs.harvard.edu/abs/2013ApJ...771L..45Y}
}

@ARTICLE{2020ApJ...894...84S,
       author = {{Sergeev}, Denis E. and {Lambert}, F. Hugo and {Mayne}, Nathan J. and {Boutle}, Ian A. and {Manners}, James and {Kohary}, Krisztian},
        title = "{Atmospheric Convection Plays a Key Role in the Climate of Tidally Locked Terrestrial Exoplanets: Insights from High-resolution Simulations}",
      journal = {\apj},
         year = 2020,
        month = may,
       volume = {894},
       number = {2},
          eid = {84},
        pages = {84},
          doi = {10.3847/1538-4357/ab8882},
archivePrefix = {arXiv},
       eprint = {2004.03007},
 primaryClass = {astro-ph.EP},
       adsurl = {https://ui.adsabs.harvard.edu/abs/2020ApJ...894...84S}
}

@ARTICLE{2021ApJ...913..101L,
       author = {{Lef{\`e}vre}, Maxence and {Turbet}, Martin and {Pierrehumbert}, Raymond},
        title = "{3D Convection-resolving Model of Temperate, Tidally Locked Exoplanets}",
      journal = {\apj},
         year = 2021,
        month = jun,
       volume = {913},
       number = {2},
          eid = {101},
        pages = {101},
          doi = {10.3847/1538-4357/abf2c1},
archivePrefix = {arXiv},
       eprint = {2104.05559},
 primaryClass = {astro-ph.EP},
       adsurl = {https://ui.adsabs.harvard.edu/abs/2021ApJ...913..101L}
}

@ARTICLE{2024ApJ...970....7S,
       author = {{Sergeev}, Denis E. and {Boutle}, Ian A. and {Lambert}, F. Hugo and {Mayne}, Nathan J. and {Bendall}, Thomas and {Kohary}, Krisztian and {Olivier}, Enrico and {Shipway}, Ben},
        title = "{The Impact of the Explicit Representation of Convection on the Climate of a Tidally Locked Planet in Global Stretched-mesh Simulations}",
      journal = {\apj},
         year = 2024,
        month = jul,
       volume = {970},
       number = {1},
          eid = {7},
        pages = {7},
          doi = {10.3847/1538-4357/ad4ecd},
archivePrefix = {arXiv},
       eprint = {2402.19277},
 primaryClass = {astro-ph.EP},
       adsurl = {https://ui.adsabs.harvard.edu/abs/2024ApJ...970....7S}
}

@ARTICLE{2003JApMe..42.1421P,
       author = {{Pfister}, G. and {McKenzie}, R.~L. and {Liley}, J.~B. and {Thomas}, A. and {Forgan}, B.~W. and {Long}, C.~N.},
        title = "{Cloud Coverage Based on All-Sky Imaging and Its Impact on Surface Solar Irradiance.}",
      journal = {Journal of Applied Meteorology},
         year = 2003,
        month = oct,
       volume = {42},
       number = {10},
        pages = {1421-1434},
          doi = {10.1175/1520-0450(2003)042<1421:CCBOAI>2.0.CO;2},
       adsurl = {https://ui.adsabs.harvard.edu/abs/2003JApMe..42.1421P}
}

@article{perez2015reconstruction,
  title={Reconstruction of long-term direct solar irradiance data series using a model based on the Cloud Modification Factor},
  author={P{\'e}rez-Burgos, Ana and Rom{\'a}n, Roberto and Bilbao, Julia and de Miguel, Argimiro and Oteiza, Pilar},
  journal={Renewable energy},
  volume={77},
  pages={115--124},
  year={2015},
  publisher={Elsevier}
}

@ARTICLE{2008ACP.....8.2493S,
       author = {{Staiger}, H. and {den Outer}, P.~N. and {Bais}, A.~F. and {Feister}, U. and {Johnsen}, B. and {Vuilleumier}, L.},
        title = "{Hourly resolved cloud modification factors in the ultraviolet}",
      journal = {Atmospheric Chemistry \& Physics},
         year = 2008,
        month = may,
       volume = {8},
       number = {9},
        pages = {2493-2508},
          doi = {10.5194/acp-8-2493-200810.5194/acpd-8-181-2008},
       adsurl = {https://ui.adsabs.harvard.edu/abs/2008ACP.....8.2493S}
}

@ARTICLE{2020ApJ...896L..16S,
       author = {{Salazar}, Andrea M. and {Olson}, Stephanie L. and {Komacek}, Thaddeus D. and {Stephens}, Haynes and {Abbot}, Dorian S.},
        title = "{The Effect of Substellar Continent Size on Ocean Dynamics of Proxima Centauri b}",
      journal = {\apjl},
         year = 2020,
        month = jun,
       volume = {896},
       number = {1},
          eid = {L16},
        pages = {L16},
          doi = {10.3847/2041-8213/ab94c1},
archivePrefix = {arXiv},
       eprint = {2005.14185},
 primaryClass = {astro-ph.EP},
       adsurl = {https://ui.adsabs.harvard.edu/abs/2020ApJ...896L..16S}
}

@article{platt1983photosynthetically,
  title={Photosynthetically-competent phytoplankton from the aphotic zone of the deep ocean.},
  author={Platt, Trevor and Rao, Subba and Smith, JC and Li, WK and Irwin, B and Horne, EPW and Sameoto, DD},
  journal={Marine ecology progress series. Oldendorf},
  volume={10},
  number={2},
  pages={105--110},
  year={1983}
}

@article{guo2018insight,
  title={Insight into the pico-and nano-phytoplankton communities in the deepest biosphere, the Mariana Trench},
  author={Guo, Ruoyu and Liang, Yantao and Xin, Yu and Wang, Long and Mou, Shanli and Cao, Chunjie and Xie, Ruize and Zhang, Chuanlun and Tian, Jiwei and Zhang, Yongyu},
  journal={Frontiers in Microbiology},
  volume={9},
  pages={2289},
  year={2018},
  publisher={Frontiers Media SA}
}

@article{agusti2015ubiquitous,
  title={Ubiquitous healthy diatoms in the deep sea confirm deep carbon injection by the biological pump},
  author={Agust{\'\i}, Susana and Gonz{\'a}lez-Gordillo, J Ignacio and Vaqu{\'e}, Dolors and Estrada, Marta and Cerezo, Mar{\'\i}a Isabel and Salazar, Guillem and Gasol, Josep M and Duarte, Carlos M},
  journal={Nature communications},
  volume={6},
  number={1},
  pages={7608},
  year={2015},
  publisher={Nature Publishing Group UK London}
}

@article{lee2007euphotic,
  title={Euphotic zone depth: Its derivation and implication to ocean-color remote sensing},
  author={Lee, ZhongPing and Weidemann, Alan and Kindle, John and Arnone, Robert and Carder, Kendall L and Davis, Curtiss},
  journal={Journal of Geophysical Research: Oceans},
  volume={112},
  number={C3},
  year={2007},
  publisher={Wiley Online Library}
}

@ARTICLE{Abrevaya2025_EXO_UV,
       author = {{Abrevaya}, Ximena C. and {Odert}, Petra and {Oppezzo}, Oscar J. and {Leizinger}, Martin and {Luna}, Gerardo Juan M. and {Patel}, Manish R. and {Hanslmeier}, Arnold},
        title = "{The EXO-UV program: lastest advances of experimental studies to investigate the biological impact of UV radiation on exoplanets}",
      journal = {arXiv e-prints},
         year = 2025,
        month = feb,
          eid = {arXiv:2502.08752},
        pages = {arXiv:2502.08752},
          doi = {10.48550/arXiv.2502.08752},
archivePrefix = {arXiv},
       eprint = {2502.08752},
 primaryClass = {astro-ph.EP},
       adsurl = {https://ui.adsabs.harvard.edu/abs/2025arXiv250208752A}
}

@article{balaji2022methods,
  title={Methods to control harmful algal blooms: A review},
  author={Balaji-Prasath, Barathan and Wang, Ying and Su, Yu Ping and Hamilton, David P and Lin, Hong and Zheng, Luwei and Zhang, Yong},
  journal={Environmental Chemistry Letters},
  volume={20},
  number={5},
  pages={3133--3152},
  year={2022},
  publisher={Springer}
}

@incollection{calbet2024paradox,
  title={The Paradox of the Plankton: The Kill the Winner Hypothesis},
  author={Calbet, Albert},
  booktitle={Plankton in a Changing World: The Impact of Global Change on Marine Ecosystems},
  pages={87--90},
  year={2024},
  publisher={Springer}
}

@ARTICLE{2019AnRFM..51..275P,
       author = {{Pierrehumbert}, Raymond T. and {Hammond}, Mark},
        title = "{Atmospheric Circulation of Tide-Locked Exoplanets}",
      journal = {Annual Review of Fluid Mechanics},
         year = 2019,
        month = jan,
       volume = {51},
       number = {1},
        pages = {275-303},
          doi = {10.1146/annurev-fluid-010518-040516},
       adsurl = {https://ui.adsabs.harvard.edu/abs/2019AnRFM..51..275P}
}

@ARTICLE{2017CeMDA.129..509B,
       author = {{Barnes}, Rory},
        title = "{Tidal locking of habitable exoplanets}",
      journal = {Celestial Mechanics and Dynamical Astronomy},
         year = 2017,
        month = dec,
       volume = {129},
       number = {4},
        pages = {509-536},
          doi = {10.1007/s10569-017-9783-7},
archivePrefix = {arXiv},
       eprint = {1708.02981},
 primaryClass = {astro-ph.EP},
       adsurl = {https://ui.adsabs.harvard.edu/abs/2017CeMDA.129..509B}
}

@article{macarthur1958population,
  title={Population ecology of some warblers of northeastern coniferous forests},
  author={MacArthur, Robert H},
  journal={Ecology},
  volume={39},
  number={4},
  pages={599--619},
  year={1958},
  publisher={JSTOR}
}

@article{gause1934experimental,
  title={Experimental analysis of Vito Volterra's mathematical theory of the struggle for existence},
  author={Gause, Georgy F},
  journal={Science},
  volume={79},
  number={2036},
  pages={16--17},
  year={1934},
  publisher={American Association for the Advancement of Science}
}

@book{maclulich1937fluctuations,
  author    = {MacLulich, Duncan Alexander},
  title     = {Fluctuations in the Numbers of the Varying Hare (Lepus americanus)},
  series    = {University of Toronto Studies, Biological Series},
  volume    = {43},
  publisher = {University of Toronto Press},
  address   = {Toronto},
  year      = {1937}
}

@article{nedorezov2016dynamics,
  title={The dynamics of the lynx--hare system: an application of the Lotka--Volterra model},
  author={Nedorezov, LV},
  journal={Biophysics},
  volume={61},
  number={1},
  pages={149--154},
  year={2016},
  publisher={Springer}
}

@ARTICLE{2020ApJ...895...19O,
       author = {{Olson}, Stephanie L. and {Jansen}, Malte and {Abbot}, Dorian S.},
        title = "{Oceanographic Considerations for Exoplanet Life Detection}",
      journal = {\apj},
         year = 2020,
        month = may,
       volume = {895},
       number = {1},
          eid = {19},
        pages = {19},
          doi = {10.3847/1538-4357/ab88c9},
archivePrefix = {arXiv},
       eprint = {1909.02928},
 primaryClass = {astro-ph.EP},
       adsurl = {https://ui.adsabs.harvard.edu/abs/2020ApJ...895...19O}
}

@article{butala2023unique,
  title={Unique relationships between phages and endospore-forming hosts},
  author={Butala, Matej and Drago{\v{s}}, Anna},
  journal={Trends in microbiology},
  volume={31},
  number={5},
  pages={498--510},
  year={2023},
  publisher={Elsevier}
}

@article{schwartz2022phage,
  title={Phage-encoded sigma factors alter bacterial dormancy},
  author={Schwartz, DA and Lehmkuhl, BK and Lennon, JT},
  journal={Msphere},
  volume={7},
  number={4},
  pages={e00297--22},
  year={2022},
  publisher={American Society for Microbiology 1752 N St., NW, Washington, DC}
}

@article{bi2023bio,
  title={Bio-geo-optical modelling of natural waters},
  author={Bi, Shun and Hieronymi, Martin and R{\"o}ttgers, R{\"u}diger},
  journal={Frontiers in Marine Science},
  volume={10},
  pages={1196352},
  year={2023},
  publisher={Frontiers Media SA}
}

@article{dobric2018temperature,
  title={Temperature affects lag period and growth of bacteria in soil according to a Ratkowsky (square root) model after a drying/rewetting episode},
  author={Dobri{\'c}, Sanja and B{\aa}{\aa}th, Erland},
  journal={Soil Biology and Biochemistry},
  volume={124},
  pages={32--37},
  year={2018},
  publisher={Elsevier}
}

@article{zwietering1991modeling,
  title={Modeling of bacterial growth as a function of temperature},
  author={Zwietering, MH and De Koos, JT and Hasenack, BE and De Witt, JC and Van't Riet, K},
  journal={Applied and environmental microbiology},
  volume={57},
  number={4},
  pages={1094--1101},
  year={1991}
}

@article{ratkowsky1983model,
  title={Model for bacterial culture growth rate throughout the entire biokinetic temperature range},
  author={Ratkowsky, DA and Lowry, RK and McMeekin, TA and Stokes, AN and Chandler, RE217594},
  journal={Journal of bacteriology},
  volume={154},
  number={3},
  pages={1222--1226},
  year={1983}
}

@article{knoblauch1999effect,
  title={Effect of temperature on sulphate reduction, growth rate and growth yield in five psychrophilic sulphate-reducing bacteria from Arctic sediments},
  author={Knoblauch, Christian and J{\o}rgensen, Bo Barker},
  journal={Environmental Microbiology},
  volume={1},
  number={5},
  pages={457--467},
  year={1999},
  publisher={Wiley Online Library}
}

@article{fischetti2008bacteriophage,
  title={Bacteriophage lysins as effective antibacterials},
  author={Fischetti, Vincent A},
  journal={Current opinion in microbiology},
  volume={11},
  number={5},
  pages={393--400},
  year={2008},
  publisher={Elsevier}
}

@article{maffei2024phage,
  title={Phage Paride can kill dormant, antibiotic-tolerant cells of Pseudomonas aeruginosa by direct lytic replication},
  author={Maffei, Enea and Woischnig, Anne-Kathrin and Burkolter, Marco R and Heyer, Yannik and Humolli, Dorentina and Th{\"u}rkauf, Nicole and Bock, Thomas and Schmidt, Alexander and Manfredi, Pablo and Egli, Adrian and others},
  journal={Nature Communications},
  volume={15},
  number={1},
  pages={175},
  year={2024},
  publisher={Nature Publishing Group UK London}
}

@ARTICLE{Claudi2020Life,
       author = {{Claudi}, Riccardo and {Alei}, Eleonora and {Battistuzzi}, Mariano and {Cocola}, Lorenzo and {Erculiani}, Marco Sergio and {Pozzer}, Anna Caterina and {Salasnich}, Bernardo and {Simionato}, Diana and {Squicciarini}, Vito and {Poletto}, Luca and {La Rocca}, Nicoletta},
        title = "{Super-Earths, M Dwarfs, and Photosynthetic Organisms: Habitability in the Lab}",
      journal = {Life},
         year = 2020,
        month = dec,
       volume = {11},
       number = {1},
        pages = {10},
          doi = {10.3390/life11010010},
archivePrefix = {arXiv},
       eprint = {2101.04448},
 primaryClass = {astro-ph.EP},
       adsurl = {https://ui.adsabs.harvard.edu/abs/2020Life...11...10C}
}

@ARTICLE{Jung2023Light,
       author = {{Jung}, Patrick and {Harion}, Felix and {Wu}, Shujie and {N{\"u}rnberg}, Dennis J. and {Bellamoli}, Francesco and {Guillen}, Antonio and {Leira}, Manuel and {Lakatos}, Michael},
        title = "{Dark Blue-Green: Cave-inhabiting Cyanobacteria as a Model for Astrobiology}",
      journal = {Frontiers in Astronomy and Space Sciences},
         year = 2023,
        month = feb,
       volume = {10},
          eid = {5},
        pages = {5},
          doi = {10.3389/fspas.2023.1107371},
       adsurl = {https://ui.adsabs.harvard.edu/abs/2023FrASS..1007371J}
}

@ARTICLE{Kaltenegger2010Spectra,
       author = {{Kaltenegger}, Lisa and {Selsis}, Frank and {Fridlund}, Malcolm and {Lammer}, Helmut and {Beichman}, Charles and {Danchi}, William and {Eiroa}, Carlos and {Henning}, Thomas and {Herbst}, Tom and {L{\'e}ger}, Alain and {Liseau}, Ren{\'e} and {Lunine}, Jonathan and {Paresce}, Francesco and {Penny}, Alan and {Quirrenbach}, Andreas and {R{\"o}ttgering}, Huub and {Schneider}, Jean and {Stam}, Daphne and {Tinetti}, Giovanna and {White}, Glenn J.},
        title = "{Deciphering Spectral Fingerprints of Habitable Exoplanets}",
      journal = {Astrobiology},
         year = 2010,
        month = jan,
       volume = {10},
       number = {1},
        pages = {89-102},
          doi = {10.1089/ast.2009.0381},
archivePrefix = {arXiv},
       eprint = {0906.2263},
 primaryClass = {astro-ph.EP},
       adsurl = {https://ui.adsabs.harvard.edu/abs/2010AsBio..10...89K}
}

@ARTICLE{Cooke2023Degenerate,
       author = {{Cooke}, G.~J. and {Marsh}, D.~R. and {Walsh}, C. and {Youngblood}, A.},
        title = "{Degenerate Interpretations of O$_{3}$ Spectral Features in Exoplanet Atmosphere Observations Due to Stellar UV Uncertainties: A 3D Case Study with TRAPPIST-1 e}",
      journal = {\apj},
         year = 2023,
        month = dec,
       volume = {959},
       number = {1},
          eid = {45},
        pages = {45},
          doi = {10.3847/1538-4357/ad0381},
archivePrefix = {arXiv},
       eprint = {2309.15239},
 primaryClass = {astro-ph.EP},
       adsurl = {https://ui.adsabs.harvard.edu/abs/2023ApJ...959...45C}
}

@ARTICLE{Coelho2024Purple,
       author = {{Coelho}, L{\'\i}gia Fonseca and {Kaltenegger}, Lisa and {Zinder}, Stephen and {Philpot}, William and {Price}, Taylor L. and {Hamilton}, Trinity L.},
        title = "{Purple is the new green: biopigments and spectra of Earth-like purple worlds}",
      journal = {\mnras},
         year = 2024,
        month = may,
       volume = {530},
       number = {2},
        pages = {1363-1368},
          doi = {10.1093/mnras/stae601},
archivePrefix = {arXiv},
       eprint = {2404.10105},
 primaryClass = {astro-ph.EP},
       adsurl = {https://ui.adsabs.harvard.edu/abs/2024MNRAS.530.1363C}
}

@INPROCEEDINGS{Kaltenegger2019DarkWaters,
       author = {{Kaltenegger}, Lisa},
        title = "{Dark water oceans on exoplanets orbiting cool stars}",
    booktitle = {AAS/Division for Extreme Solar Systems Abstracts},
         year = 2019,
       series = {AAS/Division for Extreme Solar Systems Abstracts},
       volume = {51},
        month = aug,
          eid = {502.05},
        pages = {502.05},
       adsurl = {https://ui.adsabs.harvard.edu/abs/2019ESS.....450205K}
}

@ARTICLE{2018ApJ...864...75K,
       author = {{Kite}, Edwin S. and {Ford}, Eric B.},
        title = "{Habitability of Exoplanet Waterworlds}",
      journal = {\apj},
         year = 2018,
        month = sep,
       volume = {864},
       number = {1},
          eid = {75},
        pages = {75},
          doi = {10.3847/1538-4357/aad6e0},
archivePrefix = {arXiv},
       eprint = {1801.00748},
 primaryClass = {astro-ph.EP},
       adsurl = {https://ui.adsabs.harvard.edu/abs/2018ApJ...864...75K}
}

@ARTICLE{2019AsBio..19.1075C,
       author = {{Chan}, Marjorie A. and {Hinman}, Nancy W. and {Potter-McIntyre}, Sally L. and {Schubert}, Keith E. and {Gillams}, Richard J. and {Awramik}, Stanley M. and {Boston}, Penelope J. and {Bower}, Dina M. and {Des Marais}, David J. and {Farmer}, Jack D. and {Jia}, Tony Z. and {King}, Penelope L. and {Hazen}, Robert M. and {L{\'e}veill{\'e}}, Richard J. and {Papineau}, Dominic and {Rempfert}, Kaitlin R. and {S{\'a}nchez-Rom{\'a}n}, M{\'o}nica and {Spear}, John R. and {Southam}, Gordon and {Stern}, Jennifer C. and {Cleaves}, Henderson James},
        title = "{Deciphering Biosignatures in Planetary Contexts}",
      journal = {Astrobiology},
         year = 2019,
        month = sep,
       volume = {19},
       number = {9},
        pages = {1075-1102},
          doi = {10.1089/ast.2018.1903},
       adsurl = {https://ui.adsabs.harvard.edu/abs/2019AsBio..19.1075C}
}

@ARTICLE{2019IJAsB..18..204S,
       author = {{Stevenson}, David S. and {Large}, Sean},
        title = "{Evolutionary exobiology: towards the qualitative assessment of biological potential on exoplanets}",
      journal = {International Journal of Astrobiology},
         year = 2019,
        month = jun,
       volume = {18},
       number = {3},
        pages = {204-208},
          doi = {10.1017/S1473550417000349},
       adsurl = {https://ui.adsabs.harvard.edu/abs/2019IJAsB..18..204S}
}

@ARTICLE{2022ApJ...927L..30W,
       author = {{Wright}, Jason T. and {Haqq-Misra}, Jacob and {Frank}, Adam and {Kopparapu}, Ravi and {Lingam}, Manasvi and {Sheikh}, Sofia Z.},
        title = "{The Case for Technosignatures: Why They May Be Abundant, Long-lived, Highly Detectable, and Unambiguous}",
      journal = {\apjl},
         year = 2022,
        month = mar,
       volume = {927},
       number = {2},
          eid = {L30},
        pages = {L30},
          doi = {10.3847/2041-8213/ac5824},
archivePrefix = {arXiv},
       eprint = {2203.10899},
 primaryClass = {astro-ph.EP},
       adsurl = {https://ui.adsabs.harvard.edu/abs/2022ApJ...927L..30W}
}

@ARTICLE{2023ApJ...943...27L,
       author = {{Lingam}, Manasvi and {Haqq-Misra}, Jacob and {Wright}, Jason T. and {Huston}, Macy J. and {Frank}, Adam and {Kopparapu}, Ravi},
        title = "{Technosignatures: Frameworks for Their Assessment}",
      journal = {\apj},
         year = 2023,
        month = jan,
       volume = {943},
       number = {1},
          eid = {27},
        pages = {27},
          doi = {10.3847/1538-4357/acaca0},
       adsurl = {https://ui.adsabs.harvard.edu/abs/2023ApJ...943...27L}
}

@article{kara2000optimal,
  title={An optimal definition for ocean mixed layer depth},
  author={Kara, A Birol and Rochford, Peter A and Hurlburt, Harley E},
  journal={Journal of Geophysical Research: Oceans},
  volume={105},
  number={C7},
  pages={16803--16821},
  year={2000},
  publisher={Wiley Online Library}
}

@ARTICLE{2018MNRAS.477..727N,
       author = {{Nicholson}, Arwen E. and {Wilkinson}, David M. and {Williams}, Hywel T.~P. and {Lenton}, Timothy M.},
        title = "{Gaian bottlenecks and planetary habitability maintained by evolving model biospheres: the ExoGaia model}",
      journal = {\mnras},
         year = 2018,
        month = jun,
       volume = {477},
       number = {1},
        pages = {727-740},
          doi = {10.1093/mnras/sty658},
archivePrefix = {arXiv},
       eprint = {1803.08063},
 primaryClass = {astro-ph.EP},
       adsurl = {https://ui.adsabs.harvard.edu/abs/2018MNRAS.477..727N}
}

@ARTICLE{2025MNRAS.tmp...96M,
       author = {{Mitchell}, Emily G. and {Madhusudhan}, Nikku},
        title = "{Prospects for biological evolution on Hycean worlds}",
      journal = {\mnras},
         year = 2025,
        month = jan,
          doi = {10.1093/mnras/staf094},
archivePrefix = {arXiv},
       eprint = {2502.07872},
 primaryClass = {astro-ph.EP},
       adsurl = {https://ui.adsabs.harvard.edu/abs/2025MNRAS.tmp...96M}
}

@ARTICLE{2024AsBio..24.1187W,
       author = {{Wordsworth}, R. and {Cockell}, C.},
        title = "{Self-Sustaining Living Habitats in Extraterrestrial Environments}",
      journal = {Astrobiology},
         year = 2024,
        month = dec,
       volume = {24},
       number = {12},
        pages = {1187-1195},
          doi = {10.1089/ast.2024.0080},
archivePrefix = {arXiv},
       eprint = {2409.14477},
 primaryClass = {astro-ph.EP},
       adsurl = {https://ui.adsabs.harvard.edu/abs/2024AsBio..24.1187W}
}

@article{rankin1999chemical,
  title={The chemical stratification and microbial communities of Ace Lake, Antarctica: a review of the characteristics of a marine-derived meromictic lake},
  author={Rankin, Lynne M and Gobson, JAE and Franzmann, PD and Burton, HR},
  journal={Polarforschung},
  volume={66},
  number={1/2},
  pages={33--52},
  year={1999},
  publisher={Alfred Wegener Institute for Polar and Marine Research \& German Society of~…}
}

@ARTICLE{Abrevaya2024,
       author = {{Abrevaya}, X.~C. and {Odert}, P. and {Oppezzo}, O.~J. and {Leitzinger}, M. and {Luna}, G.~J.~M. and {Guenther}, E. and {Patel}, M.~R. and {Hanslmeier}, A.},
        title = "{An experimental study of the biological impact of a superflare on the TRAPPIST-1 planets}",
      journal = {\mnras},
         year = 2024,
        month = dec,
       volume = {535},
       number = {2},
        pages = {1616-1624},
          doi = {10.1093/mnras/stae2433},
       adsurl = {https://ui.adsabs.harvard.edu/abs/2024MNRAS.535.1616A}
}

@ARTICLE{Kuzucan2025Ecoli,
       author = {{Kuzucan}, Asena and {Bolmont}, Emeline and {Chaverot}, Guillaume and {Ferreira}, Jaqueline Quirino and {Ibelings}, Bastiaan Willem and {Bhatnagar}, Siddharth and {McGinnis}, Daniel Frank},
        title = "{The Role of Atmospheric Composition in Defining the Habitable Zone Limits and Supporting E. coli Growth}",
      journal = {Life},
         year = 2025,
        month = jan,
       volume = {15},
       number = {1},
          eid = {79},
        pages = {79},
          doi = {10.3390/life15010079},
archivePrefix = {arXiv},
       eprint = {2501.05297},
 primaryClass = {astro-ph.EP},
       adsurl = {https://ui.adsabs.harvard.edu/abs/2025Life...15...79K}
}

@ARTICLE{Seager2020Lab,
       author = {{Seager}, S. and {Huang}, J. and {Petkowski}, J.~J. and {Pajusalu}, M.},
        title = "{Laboratory studies on the viability of life in H$_{2}$-dominated exoplanet atmospheres}",
      journal = {Nature Astronomy},
         year = 2020,
        month = may,
       volume = {4},
        pages = {802-806},
          doi = {10.1038/s41550-020-1069-4},
archivePrefix = {arXiv},
       eprint = {2005.01668},
 primaryClass = {astro-ph.EP},
       adsurl = {https://ui.adsabs.harvard.edu/abs/2020NatAs...4..802S}
}

@ARTICLE{Armstrong2014ObliqHabit,
       author = {{Armstrong}, J.~C. and {Barnes}, R. and {Domagal-Goldman}, S. and {Breiner}, J. and {Quinn}, T.~R. and {Meadows}, V.~S.},
        title = "{Effects of Extreme Obliquity Variations on the Habitability of Exoplanets}",
      journal = {Astrobiology},
         year = 2014,
        month = apr,
       volume = {14},
       number = {4},
        pages = {277-291},
          doi = {10.1089/ast.2013.1129},
archivePrefix = {arXiv},
       eprint = {1404.3686},
 primaryClass = {astro-ph.EP},
       adsurl = {https://ui.adsabs.harvard.edu/abs/2014AsBio..14..277A}
}

@article{wangersky1978lotka,
  title={Lotka-Volterra population models},
  author={Wangersky, Peter J},
  journal={Annual Review of Ecology and Systematics},
  volume={9},
  pages={189--218},
  year={1978},
  publisher={JSTOR}
}

@ARTICLE{2011AsBio..11.1041S,
       author = {{Schulze-Makuch}, Dirk and {M{\'e}ndez}, Abel and {Fair{\'e}n}, Alberto G. and {von Paris}, Philip and {Turse}, Carol and {Boyer}, Grayson and {Davila}, Alfonso F. and {Ant{\'o}nio}, Marina Resendes De Sousa and {Catling}, David and {Irwin}, Louis N.},
        title = "{A Two-Tiered Approach to Assessing the Habitability of Exoplanets}",
      journal = {Astrobiology},
         year = 2011,
        month = dec,
       volume = {11},
       number = {10},
        pages = {1041-1052},
          doi = {10.1089/ast.2010.0592},
       adsurl = {https://ui.adsabs.harvard.edu/abs/2011AsBio..11.1041S}
}

@article{martino2021virus,
  title={Virus and bacteria inactivation using ultraviolet light-emitting diodes},
  author={Martino, Vincent and Ochsner, Kyra and Peters, Paige and Zitomer, Daniel H and Mayer, Brooke K},
  journal={Environmental Engineering Science},
  volume={38},
  number={6},
  pages={458--468},
  year={2021},
  publisher={Mary Ann Liebert, Inc., publishers 140 Huguenot Street, 3rd Floor New~…}
}

@article{jerlov1951optical,
  author  = {Jerlov, Nils Gunnar},
  title   = {Optical studies of ocean water},
  journal = {Reports of the Swedish Deep-Sea Expedition},
  volume  = {3},
  pages   = {1--59},
  year    = {1951}
}

@article{overmann1989pelodictyon,
  title={Pelodictyon phaeoclathratiforme sp. nov., a new brown-colored member of the Chlorobiaceae forming net-like colonies},
  author={Overmann, J{\"o}rg and Pfennig, Norbert},
  journal={Archives of microbiology},
  volume={152},
  number={4},
  pages={401--406},
  year={1989},
  publisher={Springer}
}

@article{powell2005ecology,
  title={Ecology of a novel Synechococcus clade occurring in dense populations in saline Antarctic lakes},
  author={Powell, LM and Bowman, JP and Skerratt, JH and Franzmann, PD and Burton, HR},
  journal={Marine Ecology Progress Series},
  volume={291},
  pages={65--80},
  year={2005}
}

@article{yang2023trade,
  title={Trade-offs between competitive ability and resistance to top-down control in marine microbes},
  author={Yang, Jinny Wu and Chang, Feng-Hsun and Yeh, Yi-Chun and Tsai, An-Yi and Chiang, Kuo-Ping and Shiah, Fuh-Kwo and Gong, Gwo-Ching and Hsieh, Chih-hao},
  journal={Msystems},
  volume={8},
  number={2},
  pages={e01017--22},
  year={2023},
  publisher={American Society for Microbiology 1752 N St., NW, Washington, DC}
}

@article{suttle2007marine,
  title={Marine viruses—major players in the global ecosystem},
  author={Suttle, Curtis A},
  journal={Nature reviews microbiology},
  volume={5},
  number={10},
  pages={801--812},
  year={2007},
  publisher={Nature Publishing Group UK London}
}

@ARTICLE{Mendonca_2014_AstroEco,
       author = {{Mendon{\c{c}}a}, Milton de Souza},
        title = "{Spatial ecology goes to space: Metabiospheres}",
      journal = {\icarus},
         year = 2014,
        month = may,
       volume = {233},
        pages = {348-351},
          doi = {10.1016/j.icarus.2014.01.027},
       adsurl = {https://ui.adsabs.harvard.edu/abs/2014Icar..233..348M}
}

@ARTICLE{Meurer_2024_Astroecology,
       author = {{Meurer}, Juliana Campos and {Haqq-Misra}, Jacob and {Mendon{\c{c}}a}, Jr., Milton de Souza},
        title = "{Astroecology: bridging the gap between ecology and astrobiology}",
      journal = {International Journal of Astrobiology},
         year = 2024,
        month = jan,
       volume = {23},
          eid = {e3},
        pages = {e3},
          doi = {10.1017/S1473550423000265},
       adsurl = {https://ui.adsabs.harvard.edu/abs/2024IJAsB..23....3M}
}

@ARTICLE{2025PSJ.....6....5B,
       author = {{Braam}, Marrick and {Palmer}, Paul I. and {Decin}, Leen and {Mayne}, Nathan J. and {Manners}, James and {Rugheimer}, Sarah},
        title = "{Earth-like Exoplanets in Spin{\textendash}Orbit Resonances: Climate Dynamics, 3D Atmospheric Chemistry, and Observational Signatures}",
      journal = {\psj},
         year = 2025,
        month = jan,
       volume = {6},
       number = {1},
          eid = {5},
        pages = {5},
          doi = {10.3847/PSJ/ad9565},
archivePrefix = {arXiv},
       eprint = {2410.19108},
 primaryClass = {astro-ph.EP},
       adsurl = {https://ui.adsabs.harvard.edu/abs/2025PSJ.....6....5B}
}

@ARTICLE{2021ApJ...921...25C,
       author = {{Colose}, Christopher M. and {Haqq-Misra}, Jacob and {Wolf}, Eric T. and {Del Genio}, Anthony D. and {Barnes}, Rory and {Way}, Michael J. and {Ruedy}, Reto},
        title = "{Effects of Spin-Orbit Resonances and Tidal Heating on the Inner Edge of the Habitable Zone}",
      journal = {\apj},
         year = 2021,
        month = nov,
       volume = {921},
       number = {1},
          eid = {25},
        pages = {25},
          doi = {10.3847/1538-4357/ac135c},
archivePrefix = {arXiv},
       eprint = {2012.07996},
 primaryClass = {astro-ph.EP},
       adsurl = {https://ui.adsabs.harvard.edu/abs/2021ApJ...921...25C}
}

@ARTICLE{2024MNRAS.532.4511L,
       author = {{Liu}, Binghan and {Marsh}, Daniel R. and {Walsh}, Catherine and {Cooke}, Greg and {Sainsbury-Martinez}, Felix},
        title = "{Eccentric orbits may enhance the habitability of Earth-like exoplanets}",
      journal = {\mnras},
         year = 2024,
        month = aug,
       volume = {532},
       number = {4},
        pages = {4511-4523},
          doi = {10.1093/mnras/stae1758},
       adsurl = {https://ui.adsabs.harvard.edu/abs/2024MNRAS.532.4511L}
}

@ARTICLE{Jernigan2023Superhabitability,
       author = {{Jernigan}, Jonathan and {Lafl{\`e}che}, {\'E}milie and {Burke}, Angela and {Olson}, Stephanie},
        title = "{Superhabitability of High-obliquity and High-eccentricity Planets}",
      journal = {\apj},
         year = 2023,
        month = feb,
       volume = {944},
       number = {2},
          eid = {205},
        pages = {205},
          doi = {10.3847/1538-4357/acb81c},
archivePrefix = {arXiv},
       eprint = {2303.02188},
 primaryClass = {astro-ph.EP},
       adsurl = {https://ui.adsabs.harvard.edu/abs/2023ApJ...944..205J}
}

@article{fel1992exponential,
  title={Exponential model of the seasonal thermocline},
  author={Fel'zenbaum, AI},
  journal={Soviet journal of physical oceanography},
  volume={3},
  number={1},
  pages={75--79},
  year={1992},
  publisher={Springer}
}

@ARTICLE{Espinoza_highlights_2025,
       author = {{Espinoza}, N{\'e}stor and {Perrin}, Marshall D.},
        title = "{Highlights from Exoplanet Observations by the James Webb Space Telescope}",
      journal = {arXiv e-prints},
         year = 2025,
        month = may,
          eid = {arXiv:2505.20520},
        pages = {arXiv:2505.20520},
          doi = {10.48550/arXiv.2505.20520},
archivePrefix = {arXiv},
       eprint = {2505.20520},
 primaryClass = {astro-ph.EP},
       adsurl = {https://ui.adsabs.harvard.edu/abs/2025arXiv250520520E}
}

@ARTICLE{Leger_2004_Ocean,
       author = {{L{\'e}ger}, A. and {Selsis}, F. and {Sotin}, C. and {Guillot}, T. and {Despois}, D. and {Mawet}, D. and {Ollivier}, M. and {Lab{\`e}que}, A. and {Valette}, C. and {Brachet}, F. and {Chazelas}, B. and {Lammer}, H.},
        title = "{A new family of planets? ``Ocean-Planets''}",
      journal = {\icarus},
         year = 2004,
        month = jun,
       volume = {169},
       number = {2},
        pages = {499-504},
          doi = {10.1016/j.icarus.2004.01.001},
archivePrefix = {arXiv},
       eprint = {astro-ph/0308324},
 primaryClass = {astro-ph},
       adsurl = {https://ui.adsabs.harvard.edu/abs/2004Icar..169..499L}
}

@misc{jerlov1976marine,
  title={Marine Optics' Elsevier Scientific Publishing Company},
  author={Jerlov, NG},
  year={1976},
  publisher={Oxford) p-20}
}

@book{jerlov1968optical,
  author    = {Jerlov, Nils Gunnar},
  title     = {Optical Oceanography},
  series    = {Elsevier Oceanography Series},
  volume    = {5},
  publisher = {Elsevier},
  address   = {Amsterdam},
  year      = {1968}
}

@article{williamson2023depth,
  title={Depth profiles of Jerlov water types},
  author={Williamson, Craig A and Hollins, Richard C},
  journal={Limnology and Oceanography Letters},
  volume={8},
  number={5},
  pages={781--788},
  year={2023},
  publisher={Wiley Online Library}
}

@ARTICLE{Modi_2023_Impact,
       author = {{Modi}, Ashini and {Estrela}, Raissa and {Valio}, Adriana},
        title = "{Impact of M-dwarf stellar wind and photoevaporation on the atmospheric evolution of small planets}",
      journal = {\mnras},
         year = 2023,
        month = nov,
       volume = {525},
       number = {4},
        pages = {5168-5179},
          doi = {10.1093/mnras/stad2557},
archivePrefix = {arXiv},
       eprint = {2309.10942},
 primaryClass = {astro-ph.EP},
       adsurl = {https://ui.adsabs.harvard.edu/abs/2023MNRAS.525.5168M}
}

@ARTICLE{Zahnle_Cosmic_2017,
       author = {{Zahnle}, Kevin J. and {Catling}, David C.},
        title = "{The Cosmic Shoreline: The Evidence that Escape Determines which Planets Have Atmospheres, and what this May Mean for Proxima Centauri B}",
      journal = {\apj},
         year = 2017,
        month = jul,
       volume = {843},
       number = {2},
          eid = {122},
        pages = {122},
          doi = {10.3847/1538-4357/aa7846},
archivePrefix = {arXiv},
       eprint = {1702.03386},
 primaryClass = {astro-ph.EP},
       adsurl = {https://ui.adsabs.harvard.edu/abs/2017ApJ...843..122Z}
}

@ARTICLE{2025arXiv250419872J,
       author = {{Ji}, Xuan and {Chatterjee}, Richard D. and {Park Coy}, Brandon and {Kite}, Edwin S.},
        title = "{The Cosmic Shoreline Revisited: A Metric for Atmospheric Retention Informed by Hydrodynamic Escape}",
      journal = {arXiv e-prints},
         year = 2025,
        month = apr,
          eid = {arXiv:2504.19872},
        pages = {arXiv:2504.19872},
          doi = {10.48550/arXiv.2504.19872},
archivePrefix = {arXiv},
       eprint = {2504.19872},
 primaryClass = {astro-ph.EP},
       adsurl = {https://ui.adsabs.harvard.edu/abs/2025arXiv250419872J}
}

@ARTICLE{Schwieterman_Biosignatures_2022,
       author = {{Schwieterman}, Edward W. and {Olson}, Stephanie L. and {Pidhorodetska}, Daria and {Reinhard}, Christopher T. and {Ganti}, Ainsley and {Fauchez}, Thomas J. and {Bastelberger}, Sandra T. and {Crouse}, Jaime S. and {Ridgwell}, Andy and {Lyons}, Timothy W.},
        title = "{Evaluating the Plausible Range of N$_{2}$O Biosignatures on Exo-Earths: An Integrated Biogeochemical, Photochemical, and Spectral Modeling Approach}",
      journal = {\apj},
         year = 2022,
        month = oct,
       volume = {937},
       number = {2},
          eid = {109},
        pages = {109},
          doi = {10.3847/1538-4357/ac8cfb},
archivePrefix = {arXiv},
       eprint = {2210.01669},
 primaryClass = {astro-ph.EP},
       adsurl = {https://ui.adsabs.harvard.edu/abs/2022ApJ...937..109S}
}

@article{gilpin1975limit,
  title={Limit cycles in competition communities},
  author={Gilpin, Michael E},
  journal={The American Naturalist},
  volume={109},
  number={965},
  pages={51--60},
  year={1975},
  publisher={University of Chicago Press}
}

@article{lu2002two,
  title={Two limit cycles in three-dimensional Lotka-Volterra systems},
  author={Lu, Zhengyi and Luo, YONG},
  journal={Computers \& Mathematics with Applications},
  volume={44},
  number={1-2},
  pages={51--66},
  year={2002},
  publisher={Elsevier}
}

% Don't change these lines
\bsp	% typesetting comment
\label{lastpage}
\end{document}